\documentclass[10pt,nofootinbib,superscriptaddress,prd]{revtex4-2}
\usepackage{graphicx}
\usepackage{amsmath,stix}
\usepackage{dsfont}
\usepackage{soul}
\usepackage{color}
\usepackage{hyperref}
\usepackage{tcolorbox}
\usepackage{multirow}

\usepackage{tikz,xcolor}
\definecolor{lime}{HTML}{A6CE39}
\DeclareRobustCommand{\orcidicon}{%
	\begin{tikzpicture}
	\draw[lime, fill=lime] (0,0) 
	circle [radius=0.16] 
	node[white] {{\fontfamily{qag}\selectfont \tiny ID}};
	\draw[white, fill=white] (-0.0625,0.095) 
	circle [radius=0.007];
	\end{tikzpicture}
	\hspace{-2mm}
}
\foreach \x in {A, ..., Z}{%
	\expandafter\xdef\csname orcid\x\endcsname{\noexpand\href{https://orcid.org/\csname orcidauthor\x\endcsname}{\noexpand\orcidicon}}
}

\begin{document}
\date{\today}
\title{Introduction to gravitational redshift of quantum photons propagating in curved spacetime}
\author{Luis Adri\'an Alan\'is Rodr\'iguez\orcidA{}}
\affiliation{Institute for Theoretical Physics, Faculty of Mathematics and Natural Sciences, University of Cologne, Z\"ulpicherstrasse 77, 50937 Cologne, Germany }
\affiliation{Institute for Quantum Computing Analytics (PGI-12), Forschungszentrum J\"ulich, 52425 J\"ulich, Germany}

\author{Andreas Wolfgang Schell\orcidB{}}
\affiliation{Institute of Semiconductor and Solid State Physics, Johannes Kepler Universtiy Linz, Altenberger Stra\ss e 69, 4040 Linz, Austria}
\author{David Edward Bruschi\orcidC{}}
\email{david.edward.bruschi@posteo.net}
\affiliation{Institute for Quantum Computing Analytics (PGI-12), Forschungszentrum J\"ulich, 52425 J\"ulich, Germany}

\begin{abstract}
Gravitational redshift is discussed in the context of quantum photons propagating in curved spacetime. A brief introduction to modelling realistic photons is first presented and the effect of gravity on the spectrum computed for photons largely confined along the direction of propagation. It is then shown that redshift-induced transformations on photon operators with sharp momenta are not unitary, while a unitary transformation can be constructed for realistic photons with finite bandwidth. The unitary transformation obtained is then characterized as a multimode mixing operation, which is a generalized rotation of the Hilbert-space basis. Finally, applications of these results are discussed with focus on performance of quantum communication protocols, exploitation of the effects for quantum metrology and sensing, as well as potential for tests of fundamental science.
\end{abstract}

\maketitle

\section*{Introduction}
Gravitational redshift is one of the main predictions of general relativity \cite{Einstein:1916,Earman:Glymour:1980,Wald:1984,Carroll:2019}. In general it occurs when two observers, that exchange photons or pulses of light, are located in a curved spacetime and are subject to different local gravitational potentials. Light signals sent by the source at a given initial frequency are detected by the receiver with a different frequency as measured locally. More than one century after the formulation of the theory \cite{Einstein:1916,Einstein:1915}, gravitational redshift has been unequivocally established by experiments on Earth \cite{Pound:Rebka:1959,Pound:Snider:1964,Mueller:Wold:2006,Chou:Hume:2010,Mueller:Peters:2010,Clifford:2014,Herrmann:Finke:2018,Litvinov:Rudenko:2018,Delva:Puchades:2018,DiPumpo:Ufrecht:2021,Bothwell:Kennedy:2022} as well as via astrophysical observations \cite{Adams:1925,vonHippel:1996,Schunck:Liddle:1997,Cottam:Pearels:2002,Falcon:Winget:2010,Pasquini:Melo:2011,Abuter:Amorim:2018,Chandra:Hwang:2020,El-Badry:2022}. Different proposals have now been put forward for its exploitation in astrophysics \cite{Nunez:Nowakowski:2010}, for quantum information related tasks 
\cite{Peres:Terno:2004}, and for testing novel and fundamental theories of Nature \cite{Hohensee:Chu:2011}. Gravitational redshift remains to date an interesting and intriguing physical phenomenon with applications to many fields of physics.

Classical gravitational redshift, in the sense of a theory where matter is classical, can be well understood already from fundamental principles. In particular, the Einstein Equivalence Principle (EEP) implies that a nontrivial redshift between the frequencies of two observers due to a difference in the gravitational potential at each location should be expected \cite{Einstein:1916,Florides:2002,Carroll:2019}. While the effect 
has been experimentally measured, there has been some debate on the interpretation of the nature 
of the effects itself. In particular, the question that can be asked is if the effect is witnessed by the photon during propagation, or it is due to a mismatch in the basic properties of the measuring devices \cite{Okun:2000,Okun:Selivanov:2000}. Here we will consider that it is the reception and measurement of the photon that allows the experimenter to detect the effect and we do not participate in the debate that, although interesting conceptually, remains out of the scope of this work.

Given the success of general relativity in explaining 
gravitational redshift, it is natural to ask the question: \textit{how does the gravitational redshift affect realistic photons, that is,  quantum excitations of the electromagnetic field with a finite bandwidth and extension}? Seemingly straightforward and therefore easy to answer, this question has not yet been addressed systematically. This is the current state of the art regardless of the fact that such question is of foundational interest since we do not have a comprehensive theory of Nature that naturally accounts for the quantum and relativistic features of physical systems. The problem of unifying general relativity and quantum mechanics has been tackled in the past century leading to a variety of different theories \cite{Kiefer:2012}, such as String Theory 
\cite{Green:Schwarz:Witten:2012,Polchinski:1998} or Loop Quantum Gravity \cite{Rovelli:1998}, each with its own varying degree of success. Nevertheless, recent developments in the field of quantum information \cite{Nielsen:Chuang:2010}, and relativistic quantum information in particular 
\cite{Hu:Lin:Louko:2012}, have fuelled novel approaches to the study of physics at the overlap of general relativity and quantum mechanics without the need for a complete Theory of Everything. Many new models and experimental proposals are now being put forward to deepen our understanding of this key area of physics: these range from collapse models 
\cite{Bassi:Lochan:2013} and gravitationally induced decoherence of a quantum state \cite{Bassi:Grossart:2017,Howl:Penrose:2019} to testing the quantum nature of gravity with tabletop experiments \cite{Carney:Stamp:2019} or to modelling spacetime as a quantum channel for propagating quantum systems \cite{Downes:Ralph:2013,Bruschi:Ralph:2014,Bruschi:Datta:2014,Berera:2020,Berera:Brahma:2021,Berera:Calderon:2022}, from stochastic gravity 
\cite{Hu:Verdaguer:2008} and semiclassical approaches to self gravitation of quantum systems \cite{Bruschi:Wilhelm:2020} to gravitational quantum time dilation \cite{Smith:Ahmadi:2019,Smith:Ahmadi:2020} and to developing setups geared at detecting the quantumness of gravity 
\cite{Bose:Mazumdar:2017,Marletto:Vedral:2017,Bose:Mazumdar:2022} or testing gravitationally-induced effects on the interferometric visibility \cite{Williams:Chiow:2016,Tino:2021}.

In this work we review an approach recently developed to answer the question of gravitational redshift affecting the quantum state of light \cite{Bruschi:Ralph:2014,Bruschi:Datta:2014,Kohlrus:Bruschi:2017,Bruschi:Chatzinotas:2021,Bruschi:Schell:2023}. So far, the idea has been to model the photon as a wavepacket of a quantum field that propagates in curved spacetime and to obtain a relation between the wavepacket generated by the sender, usually called Alice, and the one detected by the receiver, usually called Bob. The transformation of the wavepackets can be interpreted as a transformation of the mode structure of the field, which in turn can be seen as a change of basis in the Hilbert space of the photon. Gravitational redshift can be therefore reinterpreted as a unitary rotation of the Hilbert space or, equivalently, as a \textit{multimode mixing} operation on the field operators 
\cite{Pan:Chen:2012}. This latter point of view naturally connects the predicted transformation to the field of quantum optics \cite{Scully:Zubairy:1997}, where multimode mixing is a common \textit{passive operation} between modes of light (i.e., one that preserves the total number of excitations).

We extend the original approach \cite{Bruschi:Ralph:2014}, which is founded on a simplified $1+1$-dimensional version of the realistic $3+1$-dimensional setup. In the original effort, a toy model was developed in order to obtain the wavepacket transformation without solving complicated equations. Here we show that the approach proposed can be generalized to a $3+1$ scenario and can be therefore justified by properly modelling realistic photon propagation in the full theory. The main conclusion is that realistic photons that are strongly confined along the direction of propagation can be effectively modelled as $1$-dimensional photons with only a frequency degree of freedom. This is complementary to recent work that has considered the deformation of photonic 
wavepackets during propagation in a background curved spacetime in the dimensions perpendicular to that of propagation \cite{Exirifard:Culf:2021,Exirifard:Karimi:2022}.
We also extend the discussion to include potential applications and outlook of the formalism.

This work is organized as follows. In Section~\ref{section:tools}, we introduce the mathematical tools necessary to the work, where we briefly go over the relevant topics from general relativity, quantum field theory and quantum optics. In Section~\ref{section:GravRedshiftRealPhotons} we employ these tools to study the gravitational redshift of real photons as they propagate in curved spacetime to then map the induced effect to a unitary transformation on the photon operator. Section~\ref{section:Applications} is devoted to give an overview of the possible applications this work may have in different fields of physics. We then put our results into perspective and give an outlook in Section~\ref{section:ConsiderationsnOutlook}. Lastly, we provide concluding remarks for this work in Section~\ref{section:Conclusions}.

\section{Tools}\label{section:tools}
Here we present the mathematical tools necessary to describe the setups considered in this work, and obtain the desired results. A detailed introduction to each topic can be found in the references provided.
In this work we assume that the metric has signature $(-,+,+,+)$, use Einstein's summation convention, employ natural units $c=\hbar=G_\textrm{N}=1$ unless explicitly stated, and work in the \textit{Heisenberg picture}\footnote{After German physicist Werner Heisenberg (5 December 1901 -- 1 February 1976).}.

\subsection{General relativity}
We start by recollecting a few notions from general relativity \cite{Misner:Thorne:1973,Wald:1984,Carroll:2019}. Spacetime is a $3+1$- dimensional manifold $\mathcal{M}$ with coordinates $x^\rho$. Derivatives with respect to a coordinate $x^\rho$ are written as $\partial_\mu\equiv\partial/\partial x^\mu$ and are defined by their action on real smooth functions over the manifold.

In a curved spacetime one can introduce \textit{curves}, which are smooth functions $\gamma:I\subset\mathbb{R}\rightarrow \mathcal{M}$. The basic structure of the manifold can be used to give coordinates $x^\mu(\lambda)$ to the curve which are thereby parametrized by $\lambda$. A vector $V$ tangent to the curve $\gamma$ is defined as a derivative operator $V\equiv\frac{d}{d\lambda}\equiv \frac{d x^\mu}{d\lambda} \partial_\mu$, and therefore via its action on functions $f$ defined over the manifold. We have $V(f)=V^\mu\partial_\mu f$, where the components $V^\mu$ of the vector in the coordinate basis chose read $V^\mu:=d x^\mu/d\lambda$. 
In general, any vector $V$ can be used as a basis for vectors, which might not be ``aligned'' along a direction $x^\mu$. Among all possible choices for basis vectors, $\hat{e}_\mu=\partial_\mu$ is a particularly (often) convenient one. \textit{One-forms} are introduced as dual objects to vectors, i.e., linear maps acting on vectors, and they can be written as $\omega=\omega_\mu dx^\mu$, with the defining property that $dx^\mu(\hat{e}_\nu)=\delta^\mu{}_\nu$. More in general, one can have an arbitrary basis element $\theta^\mu$ for the dual space following the same argument as for the vector case.  Vectors are particular cases of tensors, which are linear maps that act on tensorial products of copies of the vector space and the dual space. A \textit{tensor} $T$ of rank $(n,m)$ is multilinear map (i.e., linear on all entries) defined as the map $T:T^*_p\otimes...\otimes T^*_p\otimes T_p\otimes...\otimes T_p\rightarrow \mathbb{R}$ at each point $p$ in its domain, where $T_p$ is the tangent space at point $p$ and $T^*_p$ is the dual space to $T_p$. The tensor product is taken over $n$ copies of $T^*_p$ and $m$ copies of $T_p$. We can write that $T=T^{\mu_1...\mu_n}{}_{\nu_1...\nu_m}\hat{e}_{\mu_1}\otimes...\otimes\hat{e}_{\mu_n}\otimes\theta^{\nu_1}\otimes\theta^{\nu_m}$, where $T^{\mu_1...\mu_n}{}_{\nu_1...\nu_m}$ are the components of the tensor. We will refer to both $T$ and $T^{\mu_1...\mu_n}{}_{\nu_1...\nu_m}$ equivalently as the tensor in the following. Note that, strictly speaking, $T^{\mu_1...\mu_n}{}_{\nu_1...\nu_m}$ are real functions over the manifold and we will rely on context for clarity. We note here that this abuse of notation is common throughout the literature.
Recall that tensorial objects  are constructed such that they remain invariant under coordinate transformations $x^\mu\rightarrow x^{\mu'}(x^\mu)$. In this case we have that: vectors transform as $\hat{e}_\mu\rightarrow\hat{e}_{\mu'}=\frac{\partial x^\mu}{\partial x^{\mu'}}\hat{e}_{\mu}$, one-forms transform as $\theta^\mu\rightarrow\theta^{\mu'}=\frac{\partial x^{\mu'}}{\partial x^\mu}\theta^{\mu}$, and tensors transform as  $T^{\mu_1...\mu_n}{}_{\nu_1...\nu_m}\rightarrow T^{\mu_1'...\mu_n'}{}_{\nu_1'...\nu_m'}=\frac{\partial x^{\mu_1'}}{\partial x^{\mu_1}}...\frac{\partial x^{\mu_n'}}{\partial x^{\mu_n}}\frac{\partial x^{\nu_1}}{\partial x^{\nu_1'}}...\frac{\partial x^{\nu_n}}{\partial x^{\nu_n'}}T^{\mu_1...\mu_n}{}_{\nu_1...\nu_m}$. Finally, we define $T_{(\mu\nu)}:=\frac{1}{2}(T_{\mu\nu}+T_{\nu\mu})$ and $T_{[\mu\nu]}:=\frac{1}{2}(T_{\mu\nu}-T_{\nu\mu})$. The tensor $T_{\mu\nu}$ is \textit{symmetric} if $T_{\mu\nu}=T_{(\mu\nu)}$ (or equivalently $T_{\mu\nu}=T_{\nu\mu}$) and it is \textit{antisymmetric} if $T_{\mu\nu}=T_{[\mu\nu]}$ (or equivalently $T_{\mu\nu}=-T_{\nu\mu}$). Note that one can always write $T_{\mu\nu}=T_{(\mu\nu)}+T_{[\mu\nu]}$, and this is true \textit{solely} for two spacetime indices. These concepts can be generalized to tensors of higher degree but they are unnecessary for our purposes.

The spacetime is endowed with a metric, which is a $(0,2)$-type, non degenerate, symmetric tensor $\mathfrak{g}$ with components $g_{\mu\nu}(x^\rho)$. Flat spacetime is characterized by the metric $\eta_{\mu\nu}=\text{diag}(-1,1,1,1)$, also known as the \textit{Minkowski metric}\footnote{After German mathematician Hermann Minkowski (22 June 1864 -- 12 January 1909).}. Since partial derivatives $\partial_\mu$ of tensorial components do not give rise to the components of another tensor (e.g., $\partial_\mu V^\nu$ is not a tensor), we also introduce covariant derivatives $\nabla_\mu$ in the direction $\mu$ that map tensors to other tensors. Their definition requires the \textit{Leibniz rule}\footnote{After German polymath Gottfried Wilhelm von Leibniz (1 July 1646  -- 14 November 1716).} $\nabla_\mu(T)=\nabla_\mu(T^{\mu_1...\mu_n}{}_{\nu_1...\nu_m})\hat{e}_{\mu_1}\otimes...\otimes\hat{e}_{\mu_n}\otimes\theta^{\nu_1}\otimes\theta^{\nu_m}+T^{\mu_1...\mu_n}{}_{\nu_1...\nu_m}(\nabla_\mu\hat{e}_{\mu_1})\otimes...\otimes\hat{e}_{\mu_n}\otimes\theta^{\nu_1}\otimes\theta^{\nu_m}+...$, as well as the following basic actions: $\nabla_\mu f\equiv\partial_\mu f$, $\nabla_\mu\hat{e}_\nu=\Gamma^{\rho}{}_{\mu\nu} \hat{e}_\rho$ and $\nabla_\mu\theta^\nu=-\Gamma^{\nu}{}_{\mu\rho} \theta^\rho$. The quantities $\Gamma^\mu{}_{\nu\rho}$ are \textit{not} tensors and they are used to define the \textit{connection}, i.e., the way vectors are parallel transported along curves. Note that there are many ways one can choose a connection. While the $\Gamma$-symbols are not tensors per se, the difference of two $\Gamma$-symbols is a tensor. We choose to employ the \textit{Levi-Civita connection}\footnote{After Italian mathematician Tullio Levi-Civita (29 March 1873 -- 29 December 1941).} that is \textit{torsion free} ($\Gamma^\mu{}_{\nu\rho}=\Gamma^\mu{}_{\rho\nu}$) and \textit{metric compatible} (i.e.,  $\nabla_\mu g_{\nu\rho}=0$). This allows us to have \textit{one} set of $\Gamma$-symbols that is \textit{uniquely} determined by the metric. In this case, the coefficients $\Gamma^\mu{}_{\nu\rho}$ are called the \textit{Christoffel symbols}\footnote{After German mathematician and physicist Elwin Bruno Christoffel (10 November 1829 -- 15 March 1900).}. We have 
$\Gamma^\mu{}_{\nu\rho}= \frac{1}{2}g^{\mu\alpha}\left ( g_{\alpha\nu,\rho}+g_{\rho\alpha,\nu}-g_{\rho\nu,\alpha}\right )$. Given the metric, we can 
raise, lower, and contract indices in the standard way. We can classify tangent vectors $V^\mu$ by using the metric to compute their length, or norm, $||V||:=V_\mu V^\mu$. The vector $V^\mu$ is: \textit{timelike}, if $V^\mu V_\mu<0$; \textit{spacelike}, if $V^\mu V_\mu>0$; \textit{null}, if $V^\mu V_\mu=0$.

Geodesics are curves that satisfy $\nabla_{U}U=0$, where $U$ is the vector tangent to the curve and $\nabla_U:=U^\mu\nabla_\mu$ is the covariant derivative along the direction $U$. Given the spacetime of interest we can in principle compute all of the timelike geodesics, i.e., those followed for ideal pointlike massive test particles, and null geodesics, i.e., those followed by ideal pointlike photons. A \textit{Killing vector}\footnote{After German mathematician Wilhelm Karl Joseph Killing (10 May 1847 -- 11 February 1923).} $K\equiv d/d\xi= K^\mu\partial_\mu$ with $K^\mu:=d x^\mu/d\xi$ is a vector that satisfies the defining constraint $\nabla_{(\mu}K_{\nu)}=0$ and therefore enjoys the property $\nabla_K(K^\rho K_\rho)=0$. This means that its magnitude $||K||=\sqrt{-K^\rho K_\rho}$ is constant along the trajectory to which it is tangent. 

Let us now assume that $P:=d/d\lambda$ is the tangent vector to a (null) geodesic and $K$ a Killing vector field. We therefore know that $\nabla_P P=0$ and that the Killing vector satisfies the defining equation $\nabla_{(\mu}  K_{\nu)}=0$. It is possible to use the metric compatibility $\nabla_\rho g_{\mu\nu}=0$ to show that $P^\mu  K_\mu$ is conserved along the geodesic with tangent vector $P$, see \cite{Wald:1984,Carroll:2019}. Concretely, this can be cast as $\nabla_P (P^\mu  K_\mu)=0$, which means that the inner product $P^\mu  K_\mu$ between the tangent vector to the geodesic and the Killing vector remains constant. Thus, since $P^\mu  K_\mu$ is a 
function, we also have that $\nabla_P(P^\mu  K_\mu)=d/d\lambda(P^\mu  K_\mu)=0$ which means that $P^\mu  K_\mu|_{\lambda_{\text{i}}}=P^\mu  K_\mu|_{\lambda}$ at any latter hypersurface $\Sigma$ labelled by $\lambda$ and perpendicular to the Killing vector $P$. Notice that such foliation might not (and, in general, will not) exist across all spacetime.


\subsection{Quantum field theory in curved spacetime}
Photons are excitations of the electromagnetic field. Therefore, the natural choice of the theory to use would be (free) quantum electrodynamics in curved spacetime \cite{Birrell:Davies:1982,Wald:1995,Srednicki:2007}. This, however, would also be an overcomplication  largely unnecessary for the purposes of our work. In fact, at this stage we are not interested in obtaining the quantitatively correct magnitude of a particular effect to be compared with the result of an experiment but we are interested instead in proving that certain transformations are expected to occur in the first place. Therefore, it shall be implicitly assumed that future work must be dedicated to bringing the mathematical predictions provided here to a level where a concrete experiment can be proposed.
A detailed introduction to quantum field theory in curved spacetime can be found in any of the dedicated monographs and it is left to the interested reader \cite{Birrell:Davies:1982,Wald:1995}.

For the purposes of our work, and without loss of generality, we model photons as the excitations of a massless scalar quantum field $\hat{\phi}(x^\mu)$ propagating on classical (curved) $3+1$ background with coordinates $x^\mu$ and metric $g_{\mu\nu}$. Such field can be employed to model one polarization of the electromagnetic field in the regimes considered here \cite{Srednicki:2007}. The classical field $\phi(x^\mu)$ will satisfy the \textit{Klein-Gordon equation}\footnote{After Swedish physicist Oskar Benjamin Klein (15 September 1894 -- 5 February 1977) and German physicist Walter Gordon (13 August 1893 -- 24 December 1939).} 
\begin{equation}\label{equation:one}
\left(\left(\sqrt{-g}\right)^{-1}\partial_\mu \left(g^{\mu\nu} \sqrt{-g}\,\partial_\nu\right)\right)\phi(x^\mu)=0,
\end{equation}
which can also be written conveniently as $\square\phi(x^\mu)=0$ where $\square:=(\sqrt{-g})^{-1}\partial_\mu \bigl(g^{\mu\nu} \sqrt{-g} \partial_\nu\bigr)$. Here $g$ is the determinant of the metric.

Finding solutions to Equation~(\ref{equation:one}) is very difficult since, in a general spacetime, there is no preferred notion of time \cite{Misner:Thorne:1973,Birrell:Davies:1982}. When a notion of time exists, for example the spacetime has a global timelike Killing vector field $K\equiv d/d\xi$, it is possible to meaningfully foliate the spacetime in spacelike hypersurfaces orthogonal to $K$ and separate variables for the energy part and solve the Klein-Gordon equation. Upon quantization one finally obtains
\begin{equation}\label{field:expansion}
\hat{\phi}(x^\mu)=\int d^3k\,\left[\phi_{\boldsymbol{k}}(x^\mu)\,\hat{a}_{\boldsymbol{k}}+\phi_{\boldsymbol{k}}^*(x^\mu)\,\hat{a}_{\boldsymbol{k}}^\dag\right],
\end{equation}
where the mode solutions $\phi_{\boldsymbol{k}}(x^\mu)$ are labelled by the quantum numbers $\boldsymbol{k}$, satisfy $\square\phi_{\boldsymbol{k}}(x^\mu)=0$ and are normalized by $(\phi_{\boldsymbol{k}},\phi_{\boldsymbol{k}'})=\delta^{3}(\boldsymbol{k}-\boldsymbol{k}')$ given the appropriate inner product $(\cdot,\cdot)$. The annihilation and creation operators $\hat{a}_{\boldsymbol{k}},\hat{a}^\dag_{\boldsymbol{k}}$ satisfy the canonical commutation relations $[\hat{a}_{\boldsymbol{k}},\hat{a}_{\boldsymbol{k}'}^\dag]=\delta^{3}(\boldsymbol{k}-\boldsymbol{k}')$, while all other commutators vanish. Note that, in the most general cases, some quantum numbers might be continuous and others discrete. For example, in flat spacetimes we expect to have $\boldsymbol{k}\equiv(k_x,k_y,k_z)\in\mathbb{R}^3$. In any case, we consider spacetimes with no compactified dimensions, and therefore we expect all quantum numbers to be a continuous variables. This also implies that the commutation relations will have Dirac deltas and no \textit{Kronecker deltas}\footnote{After German mathematician Leopold Kronecker (7 December 1823 -- 29 December 1891).}. We do not attempt to be extremely formal in this regard and we leave it to specific cases to study cases where some dimensions are compact. 
The mode solutions also satisfy $i\,\partial_\xi\phi_{\boldsymbol{k}}(x^\mu)=\omega_{\boldsymbol{k}}\,\phi_{\boldsymbol{k}}(x^\mu)$, where $K=d/d\xi$ is a timelike (Killing) vector, which guarantees a consistent notion of particle in time. The frequency $\omega_{\boldsymbol{k}}$ is a function of $\boldsymbol{k}$ that can be obtained by separation of variables in (\ref{equation:one}). For example, in flat spacetime one has $\omega_{\boldsymbol{k}}=|\boldsymbol{k}|$.

If the notion of time is not easily obtainable, in the sense that there is no preferred timelike Killing vector to choose, one can still try to look for consistent ways to propagate solutions from one foliation of the spacetime to the next, but this becomes extremely difficult from an algebraic perspective. Some work in this direction has been performed with a reasonable degree of success \cite{Barbado:Baez-Camargo:2020,Barbado:Baez-Camargo:2021}.

The annihilation operators $\hat{a}_{\boldsymbol{k}}$ define the vacuum state $|0\rangle$ of the theory through $\hat{a}_{\boldsymbol{k}}|0\rangle=0$ for all $\boldsymbol{k}$. The one-particle state with sharp momentum $\boldsymbol{k}$ is the defined by $|1_{\boldsymbol{k}}\rangle:=\hat{a}^\dag_{\boldsymbol{k}}|0\rangle$, and the sharp momentum many-particle state is immediately obtained through standard procedure. It is important to recall that the states $|1_{\boldsymbol{k}}\rangle$ are not properly normalized, in the sense that $\langle1_{\boldsymbol{k}}|1_{\boldsymbol{k}'}\rangle=\delta^3(\boldsymbol{k}-\boldsymbol{k}')$. This can be seen as the consequence of the fact that the particle has nonzero physical support over all of spacetime, and that there is no natural localization due to \textit{Lorentz invariance}\footnote{After Dutch physicist Hendrik Antoon Lorentz (18 July 1853 -- 4 February 1928).}. Such particles, while naturally arising from the theory and therefore useful for immediate understanding of certain important features of particle physics, are not a good model of realistic particles. We will deal with constructing more realistic particles below.

\subsection{Modelling realistic photons}
A physical (realistic) photon is characterized by a finite spatial extension and frequency bandwidth instead of an (infinitely) sharp momentum 
\cite{Bruschi:Chatzinotas:2021,Bruschi:Schell:2023}. Photons with only one frequency do not exist, and what is usually meant by this characterization is that the frequency distribution of the photon is sharply peaked around the red or blue wavelengths. In fact, one can argue from first principles that a photon with an infinitely sharp momentum \textit{must} be spread across the whole of spacetime. In a realistic scenario, instead, we consider a $3+1$-dimensional spacetime within which we expect to find a localized photon propagating in a given direction \cite{Maybee:Hodgson:2019,Hodgson:Southall:2022}. Such photon will be characterized by a spatial extension along the direction of propagation as well as an extension in the directions perpendicular to it. We will refer to such photon in the following as a \textit{realistic photon} with the understanding that polarization is omitted without loss of generality and that it can be readily included when necessary by considering spin-$1$ fields. 

Given the mode functions $\phi_{\boldsymbol{k}}(x^\mu)$ we introduce the \textit{bandwidth function} $F_{\boldsymbol{k}_0}(\boldsymbol{k})$ which is peaked around $\boldsymbol{k}_0$ and has a certain (three dimensional) width. Such function is used to construct the ``bandwidth'' of the photon in momentum space. Note that we can generalize the bandwidth functions to those with multiple peaks. There is no a priori way to determine a specific form for the bandwidth function, nor are we aware of fundamental principles that can inform us on a specific choice. Photons can be engineered in the laboratory to have different bandwidth profiles \cite{Keller:Lange:2004,Nisbet-Jones:Dilley:2011,Chi:Wang:2021}, which seems to suggest that a theory that predicts a preferred such function should not be expected.

We can construct a ``shape profile'' $F^{\text{s}}_{\boldsymbol{k}_0}(x^\mu)$ as
\begin{equation}
F^{\text{s}}_{\boldsymbol{k}_0}(x^\mu):= \int d^3k F_{\boldsymbol{k}_0}(\boldsymbol{k})\phi_{\boldsymbol{k}}(x^\mu),
\end{equation}
and it is immediate to check that it satisfies the Klein-Gordon equation $\square F^{\text{s}}_{\boldsymbol{k}_0}(x^\mu)=0$. We can also say that $F^{\text{s}}_{\boldsymbol{k}_0}(x^\mu)$ is nothing more than a different mode function that can be used as a particular element of a basis for the field expansion (\ref{field:expansion}). An example of a new mode function would be a particular \textit{Unruh mode}\footnote{After Canadian physicist William George Unruh.}~\cite{Unruh:1976,Bruschi:Louko:2010}. Note that, since the bandwidth function $F_{\boldsymbol{k}_0}(\boldsymbol{k})$ has an extension in three momentum dimensions, we also expect the shape function $F^{\text{s}}_{\boldsymbol{k}_0}(x^\mu)$ to have extension in three dimensions.

We now move to the construction of a \textit{realistic photon operator}, that is, one that satisfies the usual canonical commutation relations. We define the annihilation operator $\hat{A}_{\boldsymbol{k}_0}(x^\rho)$ through the expression
\begin{equation}\label{fat:operator}
\hat{A}_{\boldsymbol{k}_0}(x^\mu):=\int d^3k F_{\boldsymbol{k}_0}(\boldsymbol{k})\phi_{\boldsymbol{k}}(x^\mu)\,\hat{a}_{\boldsymbol{k}},
\end{equation}
and constrain it to a hypersurface $\Sigma$ by setting $\hat{A}_{\boldsymbol{k}_0}(x^\mu)|_\Sigma$, which is taken to signify that $\phi_{\boldsymbol{k}}(x^\mu)$ is evaluated at $\Sigma$. We sometimes just write $\hat{A}_{\boldsymbol{k}_0}$ in place of $\hat{A}_{\boldsymbol{k}_0}(x^\mu)|_\Sigma$. It is immediate to verify that $[\hat{A}_{\boldsymbol{k}_0},\hat{A}_{\boldsymbol{k}_0}^\dag]=1$ if and only if $\int d^3k |F_{\boldsymbol{k}_0,x|_\Sigma}(\boldsymbol{k})|^2=1$, where we have 
introduced the function $F_{\boldsymbol{k}_0,x|_\Sigma}(\boldsymbol{k}):=F_{\boldsymbol{k}_0}(\boldsymbol{k})\phi_{\boldsymbol{k}}(x^\mu)|_\Sigma$ for convenience of presentation.  This condition guarantees that there must be peaks because such function is an element of $\mathcal{L}^2$ and therefore vanishes at infinity (there are directions that are not compact and thus of infinite support). We will assume normalization of these functions from now on.

Let us introduce the inner product $\langle F,G\rangle:=\int d^3k  F^*(\boldsymbol{k})G(\boldsymbol{k})$ between functions $F(\boldsymbol{k})$ and $G(\boldsymbol{k})$. We can therefore write that $\langle F_{\boldsymbol{k}_0,x|_\Sigma},F_{\boldsymbol{k}_0,x|_\Sigma}\rangle=1$ as our given normalization condition. Furthermore, we note that the \textit{Hilbert space}\footnote{After German mathematician David Hilbert (23 January 1862 -- 14 February 1943).} $\mathcal{H}$ of the (scalar) photon is infinite dimensional and therefore we need to introduce the set of functions $F_{\underline{\lambda},x|_\Sigma}(\boldsymbol{k})$ determined by a set of parameters $\underline{\lambda}$ such that, together with $F_{\boldsymbol{k}_0,x|_\Sigma}(\boldsymbol{k})$, they form an orthonormal basis. In practice this means that $\langle F_{\boldsymbol{k}_0,x|_\Sigma},F_{\underline{\lambda},x|_\Sigma}\rangle=0$ for all $\underline{\lambda}$, while $\langle F_{\underline{\lambda},x|_\Sigma},F_{\underline{\lambda}',x|_\Sigma}\rangle=\delta(\underline{\lambda}-\underline{\lambda}')$. The delta here is a function of the variables $\underline{\lambda}$ that label the new basis. Field operators associated with the modes $F_{\underline{\lambda},x|_\Sigma}(\boldsymbol{k})$ can then be defined as $\hat{A}_{\underline{\lambda}}:=\int d^3k\,F_{\underline{\lambda},x|_\Sigma}(\boldsymbol{k})\,\hat{a}_\omega$ and therefore $[\hat{A}_{\boldsymbol{k}_0},\hat{A}^\dag_{\underline{\lambda}}]=0$ while $[\hat{A}^\dag_{\underline{\lambda}},\hat{A}^\dag_{\underline{\lambda}'}]=\delta(\underline{\lambda}-\underline{\lambda}')$. The explicit construction of the basis $\{F_{\underline{\lambda}}\}$ might be very difficult, if not impossible, to obtain in practice. In general, however, the expression of each element basis will not be needed.

\subsection{Gravitational redshift}
Gravitational redshift is one of the key predictions of general relativity \cite{Misner:Thorne:1973}. It has been unequivocally confirmed experimentally \cite{Pound:Rebka:1959,Pound:Snider:1964,Mueller:Wold:2006,Chou:Hume:2010,Mueller:Peters:2010,Clifford:2014,Herrmann:Finke:2018,Litvinov:Rudenko:2018,Delva:Puchades:2018,DiPumpo:Ufrecht:2021,Bothwell:Kennedy:2022}, and it is also accounted for in existing technology such as the global positioning system (GPS) \cite{Ashby:2003}. Regardless of these experimental successes, as well as its theoretical standing, it is fair to say that to date this 
effect lacks a conclusive explanation \cite{Okun:2000,Wilhelm:Bhola:2014}. It remains unclear if it is a fundamental effect witnessed by the photons due to their propagation in curved spacetime, or a consequence of the effects of such curvature on local emitting and measuring devices. If one assumes the second point of view, then the gravitational redshift is not viewed as a ``change in frequency of the photon'', but rather as a mismatch in the frequencies of the constituents forming, for example, the detecting devices of the sender and receiver respectively. 

Here we take the approach that a frequency is what a (localized) observer measures with his (local) clock 
\cite{Mieling:2021,Hu:Lin:Louko:2012}. With this in mind, we consider two (ideal and pointlike) observers Alice and Bob that move along paths in curved spacetime with tangent four-vectors $U_{\text{A}}$ and $U_{\text{B}}$ respectively. Alice measures proper time $\tau_{\textrm{A}}$ and frequency $\omega_{\textrm{A}}$ locally using her clock, while Bob measures proper time $\tau_{\textrm{B}}$ and frequency $\omega_{\textrm{B}}$ locally using his clock. Alice generates a pulse of light at a location $x_{\text{A}}$ along her path, and this pulse travels through spacetime to be received by Bob at location $x_{\text{B}}$ along his.
The generic expression for the \textit{redshift} \cite{Wald:1984}, denoted by $z$, is therefore given~by
\begin{equation}\label{general:redshift}
(1+z)\equiv\chi^2:=\frac{\omega_{\textrm{B}}}{\omega_{\textrm{A}}}=\frac{(P_\mu\,U_{\text{B}}^\mu)|_{x_{\text{B}}}}{(P_\mu\,U_{\text{A}}^\mu)|_{x_{\text{A}}}}.
\end{equation}
In this work we also use the nonnegative parameter $\chi$ for consistency with recent literature 
\cite{Bruschi:Chatzinotas:2021,Bruschi:Schell:2023}.   

\vspace{0.2cm}

\begin{tcolorbox}[colback=orange!3!white,colframe=orange!85!black,title= \textbf{Box 1:} Static observers in Schwarzschild spacetime,label=example:Schwarzschild]
As an example we consider the case of \textit{Schwarzschild spacetime}\addtocounter{footnote}{1}\footnote{After German physicist Karl Schwarzschild (9 October 1873 -- 11 May 1916).} that can be used to model the metric outside a spherical nonrotating object of mass $M$, see \cite{Wald:1984,Carroll:2019}. Schwarzschild spacetime is spherically symmetric with coordinates $x^\mu\equiv(t,r,\theta,\varphi)$, is static, and is given the line element
\begin{align*}
ds^2=-f(r)dt^2+\frac{dr^2}{f(r)}+r^2d\theta^2+r^2\sin^2\theta d\varphi^2.
\end{align*}
Here $f(r):=1-r_{\text{S}}/r$ and $r_{\text{S}}:=(2G_{\text{N}}M)/c^2$ is the \textit{Schwarzschild radius} of the massive object. For reference, the Earth has a Schwarzschild radius of approximatively $r_{\text{S}}\approx9$ mm. 
The spacetime is endowed with the timelike Killing vector $K\equiv\partial_t=(1,0,0,0)$, where the norm squared $||K||^2:=K^\mu K_\mu=-f(r)$ is negative, as required for timelike vectors, as long as we remain outside the event horizon located at $r_{\text{S}}$.

We consider two observers Alice and Bob who are located at constant radii $r_{\text{A}}$ and $r_{\text{B}}$ (i.e., $\theta$ and $\phi$ are constant) and have four-velocities $U_{\text{A}}\equiv1/\sqrt{f(r_{\text{A}})}(1,0,0,0)$ and $U_{\text{B}}\equiv1/\sqrt{f(r_{\text{B}})}(1,0,0,0)$ respectively. A photon propagating from Alice to Bob has four momentum $P^\mu=(f^{-1}(r),-1,0,0)$, which is null. We can therefore compute the gravitational redshift (\ref{general:redshift:Killing:path:general}) witnessed when they exchange photons. We find
\begin{equation*}
\chi^2=\frac{\omega_{\textrm{B}}}{\omega_{\textrm{A}}}=\frac{\sqrt{f(r_{\text{A}})}}{\sqrt{f(r_{\text{B}})}},
\end{equation*}
which is the well 
known formula from the literature. Notice that Alice and Bob are \textit{not} following geodesics since they do not have angular momentum and therefore need to use a propulsion mechanism to remain at a constant distance from the planet. In fact, we can compute the proper acceleration $A:=\nabla_U U$, which is zero if and only if the observer follows a geodesic. In our case, it turns 
out that the only nonzero component is $A^1\equiv A^r=1/2\partial_r g_{00}$ obtained by noting that $A^r=(U^0)^2\Gamma^r{}_{00}$. We have $A^r=r_{\text{S}}/(2r^2)=G_{\text{N}}M/r^2$,
where the radius $r$ is evaluated at the location of Alice or Bob.
This is precisely the acceleration necessary to maintain the observer fixed at location $r$ as predicted by Newtonian mechanics.
\end{tcolorbox}

\vspace{0.2cm}

Gravitational redshift is a particular instance of change of frequency as measured by an emitter and an observer. The existence of purely gravitational redshift, and its dependence on the parameters of the system, depends on the scenario adopted and dominates when considering observers that are static with respect to each other. In general, the expression (\ref{general:redshift}) contains a kinematic contribution that we can associate to the \textit{Doppler shift}\footnote{After Austrian mathematician and physicist Christian Andreas Doppler (29 November 1803 -- 17 March 1853).}, which occurs as a consequence of relative motion of observers, as well as a purely gravitational contribution. The situation becomes even worse in the case of dynamical spacetimes, since it can occur that the emitter and observer are located in regions of spacetime endowed with a (asymptotic) timelike Killing vector but in between the spacetime is dynamic. In this case we expect that, when considering quantum mechanical effects during the propagation of the photon, there might be additional particle creation phenomena occurring \cite{Birrell:Davies:1982,Wald:1995}. An example would be particle creation due to an expanding universe. In this case, the universe is flat in the asymptotic past and future but expands at a certain rate in between. Particles are created as a result from the quantum vacuum, and signals sent from a past observer are distorted in a complicated way from the perspective of a future one \cite{Parker:1969,Parker:1971}.

Le us now assume that there is a timelike Killing vector $K$ in some region of spacetime. We can impose to the two observers to follow two paths whose tangent vectors $U^\mu$ are aligned to the \textit{same} Killing vector field \cite{Wald:1984,Carroll:2019}. The idea is that observers following such paths are static with respect to each other. In this case, we would have $U^\mu=K^\mu/(\sqrt{-K^\rho K_\rho})$, since we must have $U^\mu U_\mu=-1$ for the trajectory of a physical observer. This in turn implies that
\begin{equation}\label{general:redshift:Killing:path:general}
\chi^2=\frac{\sqrt{-K^\rho K_\rho}|_{x_{\text{A}}}}{\sqrt{-K^\rho K_\rho}|_{x_{\text{B}}}}\frac{(P_\mu\,K_{\text{B}}^\mu)|_{x_{\text{B}}}}{(P_\mu\,K_{\text{A}}^\mu)|_{x_{\text{A}}}}=\frac{\sqrt{-K^\rho K_\rho}|_{x_{\text{A}}}}{\sqrt{-K^\rho K_\rho}|_{x_{\text{B}}}}
\end{equation}
since $P^\mu K_\mu$ is conserved along the null geodesic followed by the photon \cite{Wald:1984,Carroll:2019}.

An observer following the path with tangent vector $U$ witnesses the passage of proper time $\tau$ as 
measured by his local clock. This means that we have $U^\mu=d x^\mu/d\tau$ and $K^\mu=(d x^\mu/d \xi)$, and the proper time $\tau$ is related to the parameter $\xi$ by the relation $d \tau=||K||d\xi$.

Notice that if the Killing vector $K$ is also tangent to a geodesic it must satisfy $\nabla_K K=0$, and it is immediate to see that $\nabla_P(K^\mu K_\mu)=-2P_\sigma \nabla_K K^\sigma=0$, which means that the magnitude $||K||^2=-K^\mu K_\mu$ of the Killing vector $K$ is preserved along the null geodesic followed by the pulse of light.  It would follow that $(1+z)=\chi^2=1$ which would imply 
that there would be no gravitational redshift. 
This is to be expected since there can be no redshift between two observers following the \textit{same} geodesic, which is an equivalent statement 
to saying that there is no Doppler effect between two inertial observers following the same inertial trajectory. An example of this scenario would be two observers free falling straight towards the Earth along the \textit{same} radial direction. The situation would change if Alice and Bob were to follow \textit{different} geodesics. An example of this second scenario would be an observer free falling straight towards the Earth, with the other one orbiting around the planet. 
Even two observers free falling towards the Earth along \textit{different} directions would witness a Doppler effect due to relative motion. The second to last  example is explained in detail in 
\hyperref[example:Schwarzschild]{Box 2}.

\vspace{0.2cm}

\begin{tcolorbox}[colback=orange!3!white,colframe=orange!85!black,title=\textbf{Box 2:} Observers following different geodesics in Schwarzschild spacetime,label=example:Schwarzschild]
As a second example we consider again the case of Schwarzschild spacetime.
The spacetime is endowed with a spacelike rotational Killing vector $R:=\partial_\varphi=(0,0,0,1)$. We can use $R$ together with $K$ to construct the new Killing vectors $J_\pm=K\pm\Omega R$, where $\Omega>0$ is an angular constant that we fix below. Assuming we lie on the equatorial $\theta=\pi/2$ plane we have $||J||^2=-(f(r)-\Omega^2r^2)$. We can construct the timelike vectors $V_\pm=1/\sqrt{f(r)-\Omega^2r^2}(K\pm\Omega R)$ that are normalizied by $V_\pm{}_\mu V_\pm^\mu=-1$. If we require that $V_\pm^\mu$ are also geodesics followed by Bob, then they satisfy $\nabla_{V_\pm} V_\pm^\mu=0$, which implies that Bob's proper acceleration $A_\pm:=\nabla_{V_\pm}V_\pm$ also vanishes. This in turn implies that $0=\Gamma^\mu{}_{00}(V_\pm^0)^2+\Gamma^\mu{}_{33}(V_\pm^3)^2+2\Gamma^\mu{}_{03}V_\pm^0V_\pm^3$. Using the explicit expressions for these coefficients and the components of $V_\pm$ we find that the geodesic equation implies $\Omega=\sqrt{G_{\text{N}}M/r^3}$, i.e.,  exactly the angular parameter that is found in Newtonian gravity for a stable circular orbit. We therefore have $V_\pm=1/\sqrt{1-3G_{\text{N}}M/r}\,\bigl(1,0,0,\pm\sqrt{G_{\text{N}}M/r^3}\bigr)$. The two solutions can be interpreted as Bob moving in orbit clockwise or counterclockwise.  

Alice, who is located at a constant point in space, has four momentum $U^\mu=1/\sqrt{f(r)}(1,0,0,0)$. We can therefore compute the gravitational redshift (\ref{general:redshift}) witnessed when Alice and Bob exchange photons. We find,
\begin{equation*}
\chi^2=\frac{\omega_{\textrm{B}}}{\omega_{\textrm{A}}}=\frac{\sqrt{1-2G_{\text{N}}M/r_{\text{A}}}}{\sqrt{1-3G_{\text{N}}M/r_{\text{B}}}},
\end{equation*}
which coincides as expected with the solution found in the literature. 
\end{tcolorbox}

\subsection{Quantum optics\label{section:quantum:optics}}

Electrodynamics in flat or curved spacetime can be fully described using quantum field theory \cite{Srednicki:2007}. The theory requires the four-vector potential $A^\mu$ that is used to define the \textit{Faraday tensor}\footnote{After English scientist Michael Farady (22 September 1791 -- 25 August 1867).} $F_{\mu\nu}:=\nabla_\mu A_\nu-\nabla_\nu A_\mu$ obeying \textit{Maxwell's equations}\footnote{After Scottish scientist James Clerk Maxwell (13 June 1831 -- 5 November 1879).} 
\begin{align*}
\left\{
\begin{matrix}
\nabla_\mu F^{\mu\nu}&=&J^\nu\nonumber\\
\nabla_{[\rho} F_{\mu\nu]}&=&0
\end{matrix}
\right.
\end{align*}
in tensor form, where $J^\mu$ is the four-current. Note that in flat spacetime one has the identification $A^\mu\equiv(\varphi,\boldsymbol{A})$ and $J^\mu\equiv(\rho,\boldsymbol{j})$ in terms of charge density $\rho$, electric current $\boldsymbol{j}$, scalar potential $\phi$ and vector potential $\boldsymbol{A}$. This absolute distinction becomes meaningless in (strongly) curved spacetime.

Solving Maxwell's equations can require significant effort especially when sources are present (i.e., when $J^\mu\neq0$). Furthermore, when spacetime is curved the complexity increases dramatically, leaving little hope for analytical solutions. In flat spacetime when no charges are present one can derive the field equations $\square A^\mu=0$ for the four-potential and a gauge must be chosen in order to obtain concrete solutions. Since the field equations are linear, one expects to obtain the full field expression as a linear superposition of plane waves solutions similar to the expression (\ref{field:expansion}) presented above for the scalar field. Choosing the \textit{Coulomb gauge}\footnote{After French engineer and physicist Charles-Augustin de Coulomb (14 June 1736 -- 23 August 1806).} $\nabla_j A^j=0$, it is easy to show that the $A_0$ component is not dynamical and the remaining three degrees of freedom $A_k$ are the ones that will be present in the kinematics. The difference with the scalar field case will be that additional degrees of freedom are present, such as the spin $s=\pm1$. Therefore, the mode structure must be upgraded from the functions $\exp[i k_\mu x^\mu]$ to the quantities $\exp[i k_\mu x^\mu]\varepsilon_\sigma(\boldsymbol{k})$, where $\varepsilon_\sigma(\boldsymbol{k})$ are three-dimensional vectors \cite{Srednicki:2007}. One then obtains
\begin{equation}\label{field:expansion:electromagnetism}
\boldsymbol{A}(x^\mu)=\sum_\sigma\int d^3k\,\left[\varepsilon_\sigma(\boldsymbol{k})e^{i k_\mu x^\mu}\,\hat{a}_{\sigma,\boldsymbol{k}}+\varepsilon^*_\sigma(\boldsymbol{k})e^{-i k_\mu x^\mu}\,\hat{a}_{\sigma,\boldsymbol{k}}^\dag\right].
\end{equation}
Note that there are only two independent polarization degrees of freedom in this expression, since the Coulomb gauge implies the constraint $\boldsymbol{k}\cdot\varepsilon_\sigma(\boldsymbol{k})=0$ and we have $[\hat{a}_{\sigma,\boldsymbol{k}},\hat{a}_{\sigma',\boldsymbol{k}'}^\dag]=\delta_{\sigma,\sigma'}\delta^3(\boldsymbol{k}-\boldsymbol{k}')$ while all other commutators vanish.

Solving Maxwell's equations inside a medium, or in curved spacetime, does not lead to a simple expression as the one in (\ref{field:expansion:electromagnetism}). For example, when a laser propagates through a crystal it would be an extremely taxing task to model accurately the interaction of the atoms and free charged particles with the field itself. Instead, if it is possible to reduce the whole process by showing that there are effectively few degrees of freedom that are interacting, this can lead to a great deal of formal simplification and greater experimental control over the system.

Quantum optics is the field of physics that has been developed to simplify the complexity of quantum electrodynamics when a coherent source of light is considered (i.e., a laser) \cite{Scully:Zubairy:1997}. In this case, interaction with cavities or atoms leads to the situation where few field modes to play a significant role while all others can be effectively ignored \cite{Scully:Zubairy:1997}. This approach leads to a simplified, elegant yet powerful formalism that allows to model the physics that are studied in everyday modern quantum optics laboratories employing a few degrees of freedom only. The idea is that a finite collection 
of second-quantized field operators $\hat{a}_n$ satisfying the canonical commutation relations $[\hat{a}_n,\hat{a}_m^\dag]=\delta_{nm}$ will be involved in the definition of the quantum states of the system, and will be used to define the Hamiltonian that 
governs its dynamics. 

These operators are used to define the vacuum state $\hat{a}_n|0\rangle=0$ for all $n$, and normalized \textit{Fock states}\footnote{After Soviet physicist Vladimir Aleksandrovich Fock (22 December 1898 -- 27 December 1974).} are defined by the standard second-quantized expression $|n_1n_2...\rangle:=\prod_k(\hat{a}^\dag_{k})^{n_k}/\sqrt{n_k!}|0\rangle$. Examples of the power of this approach are the ability to account for effects such as \textit{parametric down-conversion} \cite{Christ:Brecht:2013,Couteau:2018}  and the \textit{Hong-Ou-Mandel}\footnote{After Korean physicist Chung Ki Hong, Chinese-American physicist Zhe-Yu Jeff Ou, and American physicist Leonard Mandel (9 May 1927 -- 9 February 2001).} effect 
\cite{Hong:Ou:1987, Bouchard:Sit:2021}. 
Among all possible quantum states that can be realized in the laboratory, the ones that are most commonly considered in this field are listed in \hyperref[boxList]{Box 3}.

\vspace{0.2cm}

\begin{tcolorbox}[colback=orange!3!white,colframe=orange!85!black,title=\textbf{Box 3:} List of prominent quantum optical states,label=boxList]
In the table below we list some of the most prominent states used in quantum optics. Note that these states can be slightly generalized by adding relative phases in the appropriate place. We choose to set them to zero for simplicity of presentation. To give a perspective on the meaning of the parameters we can compute the average number $\langle N\rangle:=\text{Tr}(\hat{\rho}\hat{N})$ of particles in each state as a benchmark, where $\hat{N}:=\sum_k\hat{a}^\dag_k\hat{a}_k$ is the particle number operator, $k=1,...,N$, and $N$ is the number of modes. We have: $\langle N\rangle=|\alpha|^2$ for the coherent state, $\langle N\rangle=(e^{(\hbar\omega)/(k_{\text{B}}T)}-1)^{-1}$ for the thermal state, $\langle N\rangle=2\sinh^2s$ for the single-mode squeezed state, $\langle N\rangle=2\sinh^2r$ for the two-mode squeezed state, $\langle N\rangle=N$ for the N00N state. 

\begin{center}
\begin{tabular}{ |p{2.2cm}||p{0.9cm}|p{2.2cm}|p{2.6cm}|p{4.8cm}|  }
 \hline
 \multicolumn{5}{|c|}{List of prominent quantum optical states} \\
 \hline
 State Name & Type & State operator & Operator & Fock-state representation\\
 \hline
  \hline
 Coherent state   &   Pure  & 
 $|\alpha\rangle=\hat{U}(\alpha)|0\rangle $ & $\hat{U}(\alpha)=e^{\alpha\hat{a}^\dag-\alpha^*\hat{a}}$ &  $|\alpha\rangle=e^{-|\alpha|^2/2}\sum_n\frac{(\hat{a}^\dag)^n}{\sqrt{n!}}|0\rangle$\\
  \hline
 Thermal state & Mixed & $\hat{\rho}_T$ &  & $\hat{\rho}_T=\bigl(1-e^{-\frac{\hbar\omega}{k_{\text{B}}T}}\bigr)\sum_n e^{-\frac{n\hbar\omega}{k_{\text{B}}T}}|n\rangle\langle n|$ \\
  \hline
 Squeezed state: single mode & Pure & $|s\rangle:=\hat{U}(s)|0\rangle$  & $\hat{U}(s)=e^{s(\hat{a}^{\dag2}-\hat{a}^2)}$ & $|s\rangle=\sum_n \sqrt{\frac{(2n)!}{2^n n!}}\frac{\tanh^n (2s)}{\sqrt{\cosh (2s)}}|2n\rangle$\\
 \hline
 Squeezed state: two modes & Pure & $|r\rangle:=\hat{U}(r)|0\rangle$  & $\hat{U}(s)=e^{r(\hat{a}^{\dag}\hat{b}^\dag-\hat{a}\hat{b})}$ & $|s\rangle=\sum_n \frac{\tanh^n r}{\cosh r}|n,n\rangle$\\
 \hline
  N00N state & Pure & $|\psi_{N00N}\rangle$  &  & $|\psi_{N00N}\rangle=\frac{1}{\sqrt{2}}[|N0\rangle+|0N\rangle]$\\
 \hline
\end{tabular}
\end{center}
Interestingly, we note that the vacuum state $|0\rangle$ is the only pure thermal state.
\end{tcolorbox}

\vspace{0.2cm}

Although quantum optics provides the tools to study the physics that occur in a laboratory where field modes are manipulated through linear optical gates and ultimately interact with single atoms, clouds of atoms  \cite{Kockum:2021,Soro:Kockum:2022} or crystals \cite{Drummond:Corney:2006,Lukishova:2014}, in the past years it has become evident that its principles, language and techniques can be used to describe a myriad of phenomena seemingly pertaining to completely disconnected fields. For example, quantum squeezed states naturally arise in the Unruh effect \cite{Unruh:1976} and \textit{Hawking effect}\footnote{After English physicist Stephen William Hawking (8 January 1942 -- 14 March 2018).} 
\cite{Hawking:1974,Grishchuk:Sidorov:1990}, in particle creation phenomena due to an expanding universe 
\cite{Hu:Kang:Matacz:1994}, as well as particle creation due to moving boundary conditions \cite{Dodonov:2010,Bruschi:Fuentes:2012,Bruschi:Lee:2013}.
In this sense, quantum optics can also be viewed as a set of tools and concepts that can be applied to better understand, characterize and extract information from many physical systems regardless of their concrete incarnation.

\subsection{Linear dynamics\label{section:liner:dynamics}}
Among all possible dynamics allowed in Nature, we can restrict ourselves to the regime of \textit{linear dynamics}. By linear in this context we mean the fact that the Hamiltonian is quadratic in  the annihilation and creation operators (or, equivalently, the quadrature operators 
\cite{Scully:Zubairy:1997}). This is not related to the fundamental linearity of quantum mechanics. Linear dynamics are paramount in quantum optical laboratories. We tackle linear dynamics using the \textit{symplectic formalism} to map unitary operators to (low-dimensional) matrices. An extensive review can be found in the literature \cite{Adesso:Ragy:2014}. 

We consider a system of $N$ bosonic (quantum harmonic oscillators) modes with annihilation and creation operators $\hat{a}_n,\hat{a}_n^\dag$ that satisfy the canonical commutation relations $[\hat{a}_n,\hat{a}_m]=\delta_{nm}$, while all others vanish. It is convenient to collect all of the operators in the operator vector $\hat{\mathbb{X}}:=(\hat{a}_1,\hat{a}_2,...,\hat{a}_N,\hat{a}_1^\dag,\hat{a}_2^\dag,...,\hat{a}_N^\dag)^{\text{Tp}}$, where Tp stands for transpose. The canonical commutation relations can be recast as $[\hat{X}_n,\hat{X}_m^\dag]=i\Omega_{nm}$, where the matrix $\boldsymbol{\Omega}:=\text{diag}(-i,...-i,i,...,i)$ is called the \textit{symplectic form} and $\hat{X}_n$ is the $n$-th element of the vector $\hat{\mathbb{X}}$.
Any linear unitary evolution $\hat{U}(t)$ of our system can be represented by a $2N\times2N$ \textit{symplectic} matrix $\boldsymbol{S}(t)$ through the fundamental equation 
\begin{equation}\label{time:evolution:operator:symplectic:representation}
\hat{\mathbb{X}}(t)=\hat{U}(t)^\dag\,\hat{\mathbb{X}}(0)\,\hat{U}(t)=\boldsymbol{S}(t)\,\hat{\mathbb{X}}(0).
\end{equation}
The defining property of a symplectic matrix $\boldsymbol{S}$ is that it satisfies $\boldsymbol{S}\,\boldsymbol{\Omega}\,\boldsymbol{S}^\dag=\boldsymbol{S}^\dag\,\boldsymbol{\Omega}\,\boldsymbol{S}=\boldsymbol{\Omega}$. 

Any quadratic Hamiltonian $\hat{H}$ can be put in a matrix form $\boldsymbol{H}$ via the relation $\hat{H}=(\hbar/2)\hat{\mathbb{X}}^\dag\,\boldsymbol{H}\,\hat{\mathbb{X}}$. Given the choice of ordering of the operators in the vector $\hat{\mathbb{X}}$, the matrices $\boldsymbol{S}$ and $\boldsymbol{H}$ have the expression
\begin{equation}
\boldsymbol{H}
=
\begin{pmatrix}
\boldsymbol{U} & \boldsymbol{V} \\
\boldsymbol{V}^* & \boldsymbol{U}^*
\end{pmatrix},\quad\quad
\boldsymbol{S}
=
\begin{pmatrix}
\boldsymbol{\alpha} & \boldsymbol{\beta}\\
\boldsymbol{\beta}^* & \boldsymbol{\alpha}^*
\end{pmatrix},
\end{equation}
where $\boldsymbol{U}$ and $\boldsymbol{V}$ satisfy $\boldsymbol{U}=\boldsymbol{U}^\dag$ and $\boldsymbol{V}=\boldsymbol{V}^T$. Notice that the defining property of the symplectic matrix $\boldsymbol{S}$ is equivalent to the well-known \textit{Bogoliubov identities}\footnote{After Soviet mathematician and physicist Nikolay Nikolayevich Bogoliubov (21 August 1909 -- 13 February 1992).},
 which in matrix form read $\boldsymbol{\alpha}\,\boldsymbol{\alpha}^\dag-\boldsymbol{\beta}\,\boldsymbol{\beta}^\dag=\mathds{1}$ and $\boldsymbol{\alpha}\,\boldsymbol{\beta}^{\text{Tp}}-\boldsymbol{\beta}\,\boldsymbol{\alpha}^{\text{Tp}}=0$. Thus Bogoliubov transformations are symplectic transformations (and viceversa).

We conclude that the action \eqref{time:evolution:operator:symplectic:representation} of the time evolution operator $\hat{U}(t)$ on the (vector of) creation and annihilation operators implies that it has the \textit{symplectic representation} 
\begin{equation}
\boldsymbol{S}(t)=\overset{\leftarrow}{\mathcal{T}}\,\exp\left[\boldsymbol{\Omega}\,\int_0^t\,dt'\,\boldsymbol{H}(t')\right].
\end{equation}
The symbol $\overset{\leftarrow}{\mathcal{T}}$ stands for the time-ordering operator.

In the context of interest to this work, one applies the main techniques to Gaussian states of light that will allow for analytical results. The ambition is to characterize the effects and understand the physical principles that lie below them rather than to provide a complete concrete description of a realistic implementation.

\subsection{Covariance Matrix Formalism}
Among all possible states in the Hilbert space, we can choose to restrict ourselves to the class of \textit{Gaussian states}\footnote{After German mathematician Johann Carl Friedrich Gauss (30 April 1777 -- 23 February 1855).}
. Gaussian states are those quantum states with a Gaussian \textit{Wigner function}\footnote{After Hungarian physicist Eugene Paul Wigner (17 November 1902 -- 1 January 1995).}. These states are prominent across many areas of physics \cite{Adesso:Ragy:2014}, and quantum optics in particular. When considered in conjunction with \textit{linear dynamics}, they allow for a full description and characterisation of the whole physical system using the covariance matrix formalism \cite{Bruschi:Lee:2013,Brown:Martin-Martinez:2013,Bruschi:Xuereb:2018}. Note that, while the analytical solution obtained for linear dynamics can be obtained independently of the initial state, the covariance matrix formalism can be employed only when considering Gaussian states. A full introduction to this topic is left to the literature \cite{Adesso:Ragy:2014}. 

 Any Gaussian state $\hat{\rho}_\text{G}$ of $N$ bosonic modes \textit{fully} characterised by the $2N$-dimensional vector  $d$ of first moments and the $2N\times 2N$ covariance matrix of second moments $\boldsymbol{\sigma}$ defined by the elements $d_n:=\langle \hat{X}_n\rangle_{\hat{\rho}_\text{G}}$ and $\sigma_{nm}:=\langle\hat{X}_n\hat{X}_m^\dag+\hat{X}_m^\dag\hat{X}_n\rangle_{\hat{\rho}_\text{G}}-2\,\langle \hat{X}_n\rangle_{\hat{\rho}_\text{G}}\langle \hat{X}_m^\dag\rangle_{\hat{\rho}_\text{G}}$. Here, $\langle\hat{A}\rangle_{\hat{\rho}_\text{G}}:=\text{Tr}(\hat{A}\hat{\rho}_\text{G})$ is the average of the operator $\hat{A}$ with respect to the state $\hat{\rho}_\text{G}$.

Given the above, we see that the \textit{von Neumann equation}\footnote{After Hungarian-American mathematician John von Neumann (28 December 1903 -- 8 February 1957).} $\hat{\rho}_\text{G}(t)=\hat{U}(t)\,\hat{\rho}_\text{G}(0)\,\hat{U}^\dag(t)$ takes the form
\begin{equation*}\label{covariance:matrix:time:evolution}
\boldsymbol{\sigma}(t) = \boldsymbol{S}(t)\,\boldsymbol{\sigma}(0)\,\boldsymbol{S}^\dag(t)\quad \text{and} \quad d(t)=\boldsymbol{S}\,d(0).
\end{equation*}
\textit{Williamson's theorem}\footnote{After Scottish mathematician John Williamson (23 May 1901 – ? 1949).} guarantees that any $2\,N\times2\,N$ matrix, such as the covariance matrix $\boldsymbol{\sigma}$, can be put in diagonal form as
$\boldsymbol{\sigma} = \boldsymbol{s}\,\boldsymbol{\nu}_\oplus\,\boldsymbol{s}^\dag$ by an appropriate symplectic matrix $\boldsymbol{s}$, see \cite{Williamson:1923}. The diagonal matrix $\boldsymbol{\nu}_\oplus$ is called the \textit{Williamson form} of the covariance matrix $\boldsymbol{\sigma}$ and has the expression $\boldsymbol{\nu}_\oplus=\text{diag}(\nu_1,...,\nu_N,\nu_1,...,\nu_N)$, where $\nu_n\geq1$ are called the \textit{symplectic eigenvalues} of $\boldsymbol{\sigma}$ and are found as the absolute value of the spectrum of $i\,\boldsymbol{\Omega}\,\boldsymbol{\sigma}$. The general expression for such eigenvalues is $\nu_n=\coth\bigl(\frac{\hbar\,\omega_n}{2\,k_\text{B}\,T_n}\bigr)$, where $T_n$ is a local temperature of each subsystem. This is equivalent to the statement that Gaussian states are locally (i.e., in terms of single subsystems) equivalent to thermal states (i.e., up to local unitary transformations). Clearly, when $T_n=0$ for all $n$ one has $\boldsymbol{\nu}_\oplus\equiv\mathds{1}$, i.e., the state is pure.
Finally, we note that in this formalism tracing over a subsystem is performed by deleting the corresponding rows and columns in the covariance matrix.

\vspace{0.2cm}

\begin{tcolorbox}[colback=orange!3!white,colframe=orange!85!black,title= \textbf{Box 4:} Examples of Gaussian states in the covariance matrix formalism,label=exampleCovarianceMatrix]
In this formalism, we can conveniently write unitary operators induced by quadratic Hamiltonians in matrix form. For example, we can consider a \textit{single-mode squeezing} operation $\hat{U}_{\textrm{SMS}}(s)=\exp[s((\hat{a}^\dag)^2-\hat{a}^2)]$, which is represented by the $2\times2$ matrix $\boldsymbol{S}_{\textrm{SMS}}(s)$ of the form
\begin{equation*}
    \boldsymbol{S}_{\textrm{SMS}}(s)= \begin{pmatrix}
\cosh s & \sinh s\\
\sinh s & \cosh s
\end{pmatrix},
\end{equation*}
where $s$ is the squeezing parameter.
We can also look at two-mode operations such as \textit{beam-splitting} $\hat{U}_{\textrm{BS}}(\theta)=\exp[\theta(\hat{a}^\dag\hat{b}-\hat{a}\hat{b}^\dag)]$ and \textit{two-mode squeezing} $\hat{U}_{\textrm{TMS}}(s)=\exp[r(\hat{a}^\dag\hat{b}^\dag-\hat{a}\hat{b})]$. Here $\theta$ is the beam-splitting angle and $r$ is the squeezing parameter. The espective matrix forms $\boldsymbol{S}_{\textrm{BS}}(\theta)$ and $\boldsymbol{S}_{\textrm{SMS}}(r)$ read
\begin{equation*}
    \boldsymbol{S}_{\textrm{BS}}(\theta)= 
    \begin{pmatrix}
\cos\theta & \sin\theta & 0 & 0\\
-\sin\theta & \cos\theta & 0 & 0\\
0 & 0 & \cos\theta & \sin\theta\\
0 & 0 & -\sin\theta & \cos\theta
\end{pmatrix},
\quad
\boldsymbol{S}_{\textrm{TMS}}(r)= 
\begin{pmatrix}
\cosh r & 0 & 0 & \sinh r\\
0 & \cosh r & \sinh r & 0\\
0 & \sinh r & \cosh r & 0\\
\sinh r & 0 & 0 & \cosh r
\end{pmatrix}.
\end{equation*}
Note that, in the literature, one often uses the transmittivity $\tau$ of the beam-splitter, defined by a rotation of $\theta=\arccos(\tau)$ in phase space, instead of the angle $\theta$. It is also convenient, for later purposes, to write $\boldsymbol{S}_{\textrm{BS}}(\theta)=\boldsymbol{R}(\theta)\oplus\boldsymbol{R}(\theta)$, where $\boldsymbol{R}(\theta)$ is the orthogonal matrix that appears in both diagonal blocks of $\boldsymbol{S}_{\textrm{BS}}(\theta)$.
\end{tcolorbox}

\subsection{Quantum metrology}
Parameter estimation is a key endeavour of physical sciences. The predictions of a theory, for example, must be tested against experimental measurements. To estimate parameters with high precision, a reduction in the statistical error is not only desired but necessary. Such a reduction can be achieved classically by employing $N$ independent measurements and averaging the outcomes, thereby resulting in an error scaling $\propto N^{-1/2}$ as stated by the central limit theorem. A natural question is what would the role of quantum features be in estimating a parameter of interest. 

 Quantum metrology encompasses techniques and strategies to employ genuine quantum features, such as coherence and entanglement, to design successful strategies for error reduction. Within this field, it has been shown that the precision can be enhanced if quantum properties such as squeezing and entanglement are exploited, and optimal estimation strategies for the measurement of the final sate are performed. This allows for a scaling of the error $\propto N^{-1}$, usually referred to as the \textit{Heisenberg limit} \cite{Giovannetti:Lloyd:2011}. Applications of the techniques are manifold. For instance, we can consider a quantum state undergoing a unitary transformation that encodes a parameter of interest that is not an observable of the system, such as time, temperature \cite{Hovhannisyan:Jorgensen:2021}, acceleration, an unknown phase shift between different field modes \cite{Wang:Wu:2019}, or small perturbations due to spacetime changes \cite{Ahmadi:Bruschi:2014}. Among the plethora of physical systems that can be used for this purpose, photons are usually the most appropriate quantum systems to be employed due to the relative ease of their generation, manipulation, and detection \cite{Polino:Valeri:2020}. In recent developments, studies on the estimation of spacetime parameters of the Earth using photons as a realization for quantum metrology tasks have been proposed \cite{Bruschi:Datta:2014,Kohlrus:Bruschi:Fuentes:2019}.

Let us now work with a unitary channel parametrized by $\Theta$, the parameter which we intend to measure. The channel is implemented by a unitary operator $\hat{U}_{\Theta}$, which maps an initial state $\hat{\rho}_0$ to the state $\hat{\rho}_{\Theta}=\hat{U}_{\Theta}\hat{\rho}_0\hat{U}^\dag_{\Theta}$. Our ambition is to bound the mean error $\langle (\Delta\hat{\Theta})^2\rangle$ on our random variable $\Theta$. The strategy that we need to employ requires us to distinguish between two states $\hat{\rho}_{\Theta}$ and $\hat{\rho}_{\Theta+d\Theta}$ that differ by an infinitesimal change $d\Theta$ of the parameter. We can quantify the distinguishability of these two states by means of the operational measure called the \textit{Fisher information}\footnote{After British polymath Sir Ronald Aylmer Fisher (17 February 1890 -- 29 July 1962).}, which gives a lower bound to the mean-square error via the \textit{Cram\'er-Rao bound}\footnote{After Swedish mathematician Harald Cramér (25 September 1893 -- 5 October 1985) and Indian mathematician Calyampudi Radhakrishna Rao.} as 
\begin{align}
\langle (\Delta\hat{\Theta})^2\rangle\geq (N\mathcal{F}(\Theta))^{-1},
\end{align}
where $N$ is the number of input probes, the quantity $\mathcal{F}(\Theta)=\int d\lambda p(\lambda | \Theta)(d\ln[p(\lambda | \Theta)]/d\lambda)^2$ is called the \textit{Fisher information}, and $p(\lambda | \Theta)$ defines the likelihood function with respect to a chosen positive operator valued measurement (POVM) $\{\hat{\Theta}_\lambda\}$, with $\Sigma_\lambda\hat{\mathcal{O}}_\lambda=\mathbb{1}$ \cite{Giovannetti:Lloyd:2011}. It can be shown that an even stronger bound can be achieved by optimizing all possible quantum measurements, thus obtaining $(\Delta\hat\Theta)^2\geq1/[N\mathcal{F}(\Theta)]\geq1/[N\mathcal{H}(\Theta)]$, where $\mathcal{H}(\Theta)$ is the \textit{quantum Fisher information} (QFI) \cite{Braunstein:Caves:1994}. It is worth noting that, even though the optimal measurements for which the Cramér-Rao bound becomes asymptotically tight can be readily computed, implementing them in the laboratory might be extremely difficult if not impossible, a problem that requires devising suboptimal strategies such as homodyne or heterodyne detection \cite{Vidrighin:Donati:2014}.

The QFI can be computed by means of the \textit{fidelity} $\mathcal{F}(\hat{\rho},\hat{\rho}'):=\textrm{Tr}\bigl(\bigl(\sqrt{\hat{\rho}}\hat{\rho}'\sqrt{\hat{\rho}}\bigr)^{-1/2}\bigr)$ of two quantum states. In our case, the relation between the QFI and the Fisher information $\mathcal{F}(\hat{\rho}_{\Theta},\hat{\rho}_{\Theta+d\Theta})$ reads
\begin{equation}
    \mathcal{H}(\Theta)=\lim_{d\Theta\rightarrow0}\frac{8\left(1-\sqrt{\mathcal{F}\left(\hat{\rho}_{\Theta},\hat{\rho}_{\Theta+d\Theta}\right)}\right)}{d\Theta^2}.
\end{equation}
Here we refrain from using this generic formalism and we exploit instead the covariance matrix formalism by restricting ourselves to Gaussian states of light with vanishing first moments. This is a good way to obtain first answers to the problem at hand without the need of arduous algebraic computations necessary in the more general cases. The states $\hat{\rho}_{\Theta}$ and $\hat{\rho}_{\Theta+d\Theta}$ will be replaced by their covariance matrices $\boldsymbol{\sigma}_{\Theta}$ and $\boldsymbol{\sigma}_{\Theta+d\Theta}$, and we focus on the case of no initial first moments. The fidelity then reads 
\begin{equation}
    \label{eq:Fidelity}\mathcal{F}\left(\boldsymbol{\sigma}_{\Theta},\boldsymbol{\sigma}_{\Theta+d\Theta}\right)=4\left(\sqrt{\gamma}+\sqrt{\lambda}-\sqrt{\left ( \sqrt{\gamma}-\sqrt{\lambda} \right )^2-\eta}\right)^{-1},
\end{equation}
where we have defined $\gamma :=\det\left(\mathbb{1}+i\boldsymbol{\Omega}\boldsymbol{\sigma}_{\Theta}i\boldsymbol{\Omega}\boldsymbol{\sigma}_{\Theta+d\Theta}\right)$, $\lambda :=\det\left(\mathbb{1}+i\boldsymbol{\Omega}\boldsymbol{\sigma}_{\Theta}\right) \det\left(\mathbb{1}+i\boldsymbol{\Omega}\boldsymbol{\sigma}_{\Theta+d\Theta}\right)$ and $\eta :=\det\left(\boldsymbol{\Omega}\boldsymbol{\sigma}_{\Theta}+\boldsymbol{\Omega}\boldsymbol{\sigma}_{\Theta+d\Theta}\right)$ for simplicity of presentation \cite{Marian:Marian:2008,Marian:Marian:2012,Vidrighin:Donati:2014}.

These expressions will be of fundamental practical importance when attempting to calculate the errors on parameter estimation in concrete schemes. They have already been used successfully in the literature \cite{Bruschi:Datta:2014,Kohlrus:Bruschi:2017}. We will briefly report on their concrete use later on in this work.

\section{Gravitational redshift of realistic photons}\label{section:GravRedshiftRealPhotons}
Here we study the gravitational redshift of realistic photons. The overall idea is to be able to account for the propagation of the photon in the case where it is mostly confined along the direction of propagation, and then to map the overall effect induced by the gravitational redshift to a unitary transformation acting on the photon operator. Thus, propagation from A to B is indistinguishable from an appropriate unitary transformation acting on the photon at the receiver's location.

\subsection{Effective propagation of realistic photons}
We now would like to describe the propagation of realistic photons in curved spacetime. We assume that there is a timelike Killing vector $K$ with $||K||:=\sqrt{-K^\rho K_\rho}$, and that the photon propagates along a geodesic given by the solution to $\nabla_P P=0$. Here $P\equiv d/d\lambda$ is a null vector with $||P||^2\equiv P_\mu P^\mu=0$.

\begin{figure}[ht!]
\centering
\includegraphics[width=0.60\linewidth]{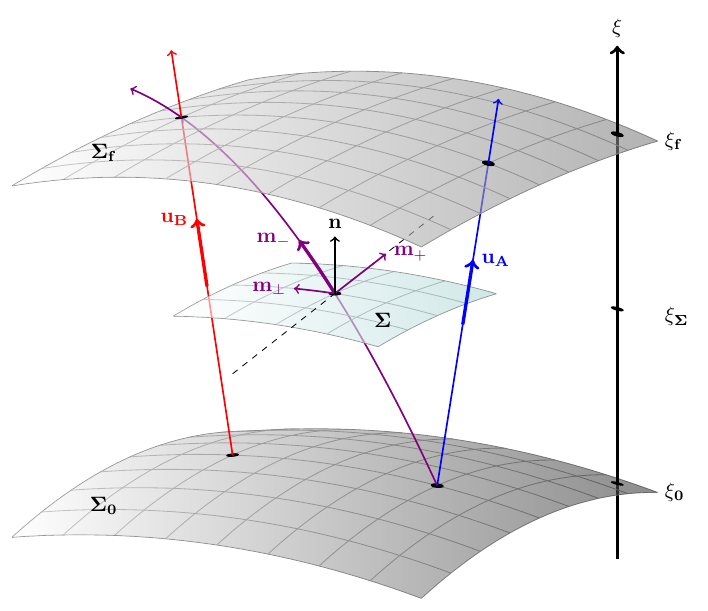}
\caption{A pictorial representation of the generic scheme considered in this work. Note that, in general, the notion of perpendicularity in $3+1$-dimensional curved spacetime cannot be faithfully reproduced. The figure should be taken as a tool to help visualize the foliation induced by the Killing vector $K$ as well as the four-vector decomposition \eqref{n:mplus:mminus:decomposition}.}\label{figure:paths:configuration}
\end{figure}

Let us introduce the vector $n:=K/||K||$, which is tangent to the path followed by static observers.
We can therefore construct two null vectors $m_\pm\equiv d/d\lambda_\pm$ via
\begin{equation}\label{vector:plus:minus}
m_{\pm}{}:=\frac{1}{2}(n\pm m_\perp{}),
\end{equation}
where $m_\perp\equiv d/d\lambda_\perp$ is an appropriate spacelike vector normalized by $m_\perp{}_\mu m_\perp{}^\mu=1$ that is orthogonal to $n$, i.e., $m_\perp{}^\mu n_\mu=0$ and thus it lies on $\Sigma$. This also imples that $[d/d\tau,d/d\lambda_\perp]=0$. Note that these conditions are compatible with $m_\pm{}_\mu m_\pm{}^\mu=0$. Both vectors $m_\pm$ are future-pointing and $m_{\pm}{}^\mu n_\mu=-1/2$. Finally, $m_+{}^\mu m_-{}_\mu=-1/2$. Furthermore,
\begin{align}\label{n:mplus:mminus:decomposition}
n=\frac{1}{2}(n+m_\perp)+\frac{1}{2}(n-m_\perp)=m_++m_-.
\end{align}
The construction given by \eqref{vector:plus:minus} and \eqref{n:mplus:mminus:decomposition} can be understood from standard vector addition rules: we add a timelike vector to a spacelike vector to obtain the vector of interest \cite{Carroll:2019}. For a pictorial understanding of the configuration see Figure~\ref{figure:paths:configuration}.

We recall the fact that $n(\phi_{\boldsymbol{k}}(x^\rho))=-i\Omega_{\boldsymbol{k}}\phi_{\boldsymbol{k}}(x^\rho)$. This informs us that the most general mode function $\phi_{\boldsymbol{k}}(x^\rho)$ can be decomposed as
\begin{align}\label{field:mode:decomposition}
\phi_{\boldsymbol{k}}(x^\rho)=\left(\alpha_+ e^{-i\Omega_{\boldsymbol{k}}(\xi+\lambda_\perp)}+\alpha_- e^{-i\Omega_{\boldsymbol{k}}(\xi-\lambda_\perp)} \right)\tilde{\phi}_{\boldsymbol{k}}(y^\rho_\perp),
\end{align}
where we can easily verify that if we call $E_\pm(\xi,\lambda_\perp):=\exp[-i\Omega_{\boldsymbol{k}}(\xi\pm\lambda_\perp)]$, we then have: first, $n(E_\sigma(\xi,\lambda_\perp))=-i\Omega_{\boldsymbol{k}}E_\sigma(\xi,\lambda_\perp)$; second, $m_\sigma(E_{\sigma'}(\xi,\lambda_\perp))=-i\delta_{\sigma\sigma'}\Omega_{\boldsymbol{k}}E_{\sigma'}(\xi,\lambda_\perp)$; third, we have $m_\perp(E_{\sigma}(\xi,\lambda_\perp))=-(1/2)\sigma i\Omega_{\boldsymbol{k}}E_{\sigma}(\xi,\lambda_\perp)$. Here the function $\tilde{\phi}_{\boldsymbol{k}}(y^\rho_\perp)$ collects the dependence on the remaining variables (that parametrize the surfaces $\Sigma$).
It should be clear that the components $E_+(\xi,\lambda_\perp)$ of the wave packet are propagating in the negative $\lambda_\perp$ direction, while the components $E_-(\xi,\lambda_\perp)$ of the wave packet are propagating in the positive $\lambda_\perp$ direction. 


We now observe that a photon can be engineered with a spatial profile that has many degrees of freedom. For example, it can have multiple peaks or no easily definable ``size'' \cite{Lundeen:Sutherland:2011,Chrapkiewicz:Radoslaw:2016}. Nevertheless, since the discussion of the redshift will effectively require considerations about measurements at a point (the users are pointlike), we need to be able to focus on the trajectory followed by a particular initial point at which the photon is located. In this sense, it is easier to assume that the photon is mostly localized along the direction of propagation, and therefore we assume that we can discard all effects due to the extension of the photon along directions that are orthogonal to that of propagation. These can be taken into account separately for general (weakly curved) spacetimes and some work has been dedicated to this task already in the literature \cite{Exirifard:Culf:2021,Exirifard:Karimi:2022}.

Our assumption naturally leads us to choose a function $F_{\boldsymbol{k}_0}(\boldsymbol{k})$ that localizes the photon on a particular null path. We choose the path parametrized by $\lambda_-=\frac{1}{2}(\xi-\lambda_\perp)$ without loss of generality, and we set $\lambda_-\equiv\tilde{\lambda}$ for convenience below. We then impose the constraint
\begin{align}\label{orthogonality:condition}
\int d^3k F_{\boldsymbol{k}_0}(\boldsymbol{k})\,\phi_{\boldsymbol{k}}(x^\rho)\approx&\int d^3k F_{\boldsymbol{k}_0}(\boldsymbol{k})\,e^{-i\Omega_{\boldsymbol{k}}(\xi-\lambda_\perp)} \tilde{\phi}_{\boldsymbol{k}}(y^\rho_\perp).
\end{align}
Here we have ignored the coefficient in the expansion \eqref{field:mode:decomposition} since it is irrelevant. Crucially, this implies that the photon moves along paths of constant $\lambda_-$.

We then proceed by recalling that vectors as elements of tangent spaces to manifolds naturally 
define directional derivatives.
In particular, we can compute the directional derivative of $F^{\text{s}}_{\boldsymbol{k}_0}(x^\mu)= \int d^3k F_{\boldsymbol{k}_0}(\boldsymbol{k})\phi_{\boldsymbol{k}}(x^\mu)$ along the vector $m_-$. We find
\begin{equation}\label{deviation:path:second}
\nabla_{m_-}\left(F^{\text{s}}_{\boldsymbol{k}_0}(x^\mu)\right)=\int d^3k F_{\boldsymbol{k}_0}(\boldsymbol{k})m_-{}^\mu\nabla_\mu\left(e^{-i\Omega_{\boldsymbol{k}}(\xi-\lambda_\perp)} \tilde{\phi}_{\boldsymbol{k}}(y^\rho_\perp)\right)=\int d^3k F_{\boldsymbol{k}_0}(\boldsymbol{k})n^\mu\partial_\mu\left(e^{-i\Omega_{\boldsymbol{k}}(\xi-\lambda_\perp)} \tilde{\phi}_{\boldsymbol{k}}(y^\rho_\perp)\right),
\end{equation}
where we have $\nabla_\mu f=\partial_\mu f$ when the covariant derivative $\nabla_\mu$ acts on functions. Here we have also written $m_-=n-m_+$ and we have used the fact that $m_+{}^\mu\partial_\mu(e^{-i\Omega_{\boldsymbol{k}}(\xi-\lambda_\perp)} \tilde{\phi}_{\boldsymbol{k}}(y^\rho_\perp))=m_+(e^{-i\Omega_{\boldsymbol{k}}(\xi-\lambda_\perp)}) \tilde{\phi}_{\boldsymbol{k}}(y^\rho_\perp)=0$.
  
We now recall that $n\equiv||K||^{-1}(d/d\xi)$ and obtain
\begin{align}\label{deviation:path:final:second}
\nabla_{m_-}(F^{\text{s}}_{\boldsymbol{k}_0}(x^\mu))
\approx&-i\int d^3k F_{\boldsymbol{k}_0}(\boldsymbol{k})\frac{\Omega_{\boldsymbol{k}}}{ \sqrt{-K^\rho  K_\rho}}e^{-i\Omega_{\boldsymbol{k}}(\xi-\lambda_\perp)} \tilde{\phi}_{\boldsymbol{k}}(y^\rho_\perp).
\end{align}
Finally, we use following reasoning: since we have $(d/d\xi)\phi_{\boldsymbol{k}}(x^\rho)=-i\Omega_{\boldsymbol{k}}\phi_{\boldsymbol{k}}(x^\rho)$ and $\tau=\sqrt{-K^\mu K_\mu}\xi$\footnote{Notice that $d\tau=\sqrt{-K^\mu K_\mu}d\xi$ implies $\tau=\sqrt{-K^\mu K_\mu}\xi$ since $\sqrt{-K^\mu K_\mu}$ is independent on $\xi$.} we can therefore see that it is natural to recast the eigenvalue equation for the modes $\phi_{\boldsymbol{k}}(x^\rho)$ as $(d/d\tau)\phi_{\boldsymbol{k}}(x^\rho)=-i\omega_{\boldsymbol{k}}\phi_{\boldsymbol{k}}(x^\rho)$, where $\omega_{\boldsymbol{k}}:=\Omega_{\boldsymbol{k}}/(\sqrt{-K^\rho  K_\rho})$. We also introduce $\tilde{\lambda}_\perp:=||K||\lambda_\perp$. Thus, using the identity $\hat{A}_{\boldsymbol{k}_0}(x^\rho)|_{\tilde{\lambda}}=\int_{\tilde{\lambda}_0}^{\tilde{\lambda}} d\tilde{\lambda}' \nabla_{m}\hat{A}_{\boldsymbol{k}_0}(x^\rho)+\hat{A}_{\boldsymbol{k}_0}(x^\rho)|_{\tilde{\lambda}_0}$ from calculus, we ultimately have
\begin{align}\label{first:useful:expression:fat:photon}
\hat{A}_{\boldsymbol{k}_0}(x^\rho)|_{\tilde{\lambda}}\approx -i\int_{\tilde{\lambda}_0}^{\tilde{\lambda}} d\tilde{\lambda}'\int d^3k \, \omega_{\boldsymbol{k}} e^{-i\omega_{\boldsymbol{k}}(\tau-\tilde{\lambda}_\perp)} F_{\boldsymbol{k}_0}(\boldsymbol{k}) \tilde{\phi}_{\boldsymbol{k}}(y^\rho_\perp) \hat{a}_{\boldsymbol{k}}+\hat{A}_{\boldsymbol{k}_0}(x^\rho)|_{\tilde{\lambda}_0}.
\end{align}
We now note that $e^{-i\omega(\tau-\tilde{\lambda}_\perp)}$ is an eigenfunction of $m_-=d/\lambda_-$ and these confined photons follow trajectories of constant $\lambda_-$, i.e., $\tilde{\lambda}_\perp=\tilde{\lambda}_\perp(0)+\tau-\tau_0$. We are therefore left with
\begin{align}\label{final:useful:expression:fat:photon}
\hat{A}_{\boldsymbol{k}_0}(x^\rho)|_{\tilde{\lambda}}\approx \int d^3k\, e^{-i\omega_{\boldsymbol{k}}(\tau-\tilde{\lambda}_\perp)} F_{\boldsymbol{k}_0}(\boldsymbol{k}) \tilde{\phi}_{\boldsymbol{k}}(y^\rho_\perp) \hat{a}_{\boldsymbol{k}},
\end{align}
with the understanding that const$=\tilde{\lambda}=\frac{1}{2}(\tau-\tilde{\lambda}_\perp)$ is evaluated at the hypersurface $\Sigma$ of interest. This tells us the respective values of $\tau$ and $\tilde{\lambda}_\perp$ at the point of interest. 

We here note that, in general, quantization of the Hilbert spaces using $\tau$ or $\tilde{\lambda}$ as external parameter at this stage in the Hamiltonian formulation can lead to inequivalent results. For this reason, we understand the expression \eqref{final:useful:expression:fat:photon} as follows: first, we perform the computations that lead here for a classical field, where $A_{\boldsymbol{k}_0}(x^\rho)|_{\tilde{\lambda}}$ are functions and not operators. This is obtained by demoting $\hat{a}_{\boldsymbol{k}}$ to scalar \textit{Fourier coefficients}\footnote{After French mathematician and physicist Jean-Baptiste Joseph Fourier (21 March 1768 -- 16 May 1830).}. Once the condition \eqref{final:useful:expression:fat:photon} is obtained for the classical computations, we promote the coefficients $\hat{A}_{\boldsymbol{k}_0}(x^\rho)|_{\tilde{\lambda}}$ to operators and quantize. This is the procedure that we assume is implicit in the process described above.  

The expression (\ref{final:useful:expression:fat:photon}) can be understood as follows: the evolution of a photon operator propagating between two static observers Alice and Bob, which is strongly confined along the direction of propagation, can be obtained as the evolution of the operator along the path of the observers once they have information of their relative positions. We can therefore identify $A_{\boldsymbol{k}_0}(x^\rho)|_{\tilde{\lambda}}\equiv A_{\boldsymbol{k}_0}(x^\rho)|_{\tau}$ with this understanding in mind. This is the key result of our work.

We now proceed and note that, since we have enforced the approximation that the photons are strongly confined along the direction of propagation, it is convenient to change from integration variables $\boldsymbol{k}\equiv(k_\text{x},k_\text{y},k_\text{z})$ to $\boldsymbol{k}_\text{n}\equiv(\omega,k_{\perp,1},k_{\perp,2})$. We can achieve this because $\omega_{\boldsymbol{k}}=\omega({\boldsymbol{k}})$, and therefore we can change variables accordingly, at least in principle. This, in turn, means that we will be able to write
\begin{align}\label{intermediate:operator:expression}
\hat{A}_{\boldsymbol{k}_0}(x^\rho)|_{\tau}=  \int d\omega\, e^{-i\omega(\tau-\tilde{\lambda}_\perp)}\int_{\perp,1,2} d^2k F_{\boldsymbol{k}_{\text{n},0}}(\boldsymbol{k}(\boldsymbol{k}_\text{n})) \, \tilde{\phi}_{\boldsymbol{k}(\boldsymbol{k}_\text{n})}(y^\rho)|_{\tau} \hat{a}_{\boldsymbol{k}(\boldsymbol{k}_\text{n})}.
\end{align}
We then focus on the operator $\hat{a}_{\boldsymbol{k}(\boldsymbol{k}_\text{n})}$. Our aim is to show that we can effectively write $\hat{a}_{\boldsymbol{k}}\approx f(\boldsymbol{k}_\text{n})\hat{a}_\omega$ for an appropriate function $f(\boldsymbol{k}_\text{n})$. The notation $\hat{a}_\omega$ used here stands to indicate that $\hat{a}_\omega$ is a function of $\omega$ and two other constants, which are effectively dropped. We would then like to absorb the function $f(\boldsymbol{k}_\text{n})$ into a new profile function $F_{\omega_0}(\omega)$. 
To arrive at this conclusion we start by recalling that $[\hat{a}_{\boldsymbol{k}}, \hat{a}_{\boldsymbol{k}'}]=\delta^3(\boldsymbol{k}-\boldsymbol{k}')$. 
We then write $\delta^3(\boldsymbol{k}-\boldsymbol{k}')=\delta^3(\boldsymbol{k}(\boldsymbol{k}_\text{n})-\boldsymbol{k}'(\boldsymbol{k}_\text{n}))$ and would have 
$\delta^3(\boldsymbol{k}-\boldsymbol{k}')=[\hat{a}_{\boldsymbol{k}},\hat{a}_{\boldsymbol{k}'}^\dag]=[\hat{a}_{\boldsymbol{k}(\boldsymbol{k}_\text{n})},\hat{a}_{\boldsymbol{k}'(\boldsymbol{k}_\text{n}')}^\dag]=\Theta(\boldsymbol{k}_\text{n})\Theta(\boldsymbol{k}_\text{n}')\delta^3(\boldsymbol{k}_\text{n}-\boldsymbol{k}_\text{n}')$.
The $\Theta$-functions do not need to be determined at this stage. We therefore have that $\hat{a}_{\boldsymbol{k}}=\hat{a}_{\boldsymbol{k}(\boldsymbol{k}_\text{n})}=\Theta(\boldsymbol{k}_\text{n})\hat{a}_{\boldsymbol{k}_\text{n}}$. 
We also know that $\bigl[\hat{A}_{\boldsymbol{k}_0}(x^\rho)|_{\tau},\hat{A}_{\boldsymbol{k}_0}^\dag(x^\rho)|_{\tau}\bigr]=1$. Using (\ref{intermediate:operator:expression}) and the relation $\hat{a}_{\boldsymbol{k}(\boldsymbol{k}_\text{n})}=\Theta(\boldsymbol{k}_\text{n})\hat{a}_{\boldsymbol{k}_\text{n}}$ we have
\begin{align}
1=\left[\hat{A}_{\boldsymbol{k}_0}(x^\rho)|_{\tau},\hat{A}_{\boldsymbol{k}_0}^\dag(x^\rho)|_{\tau}\right]=\int d\omega\int_{\perp,1,2} d^2k |F_{\boldsymbol{k}_{\text{n},0}}(\boldsymbol{k}(\boldsymbol{k}_\text{n})) \, \tilde{\phi}_{\boldsymbol{k}(\boldsymbol{k}_\text{n})}(y^\rho)|_{\tau}\Theta(\boldsymbol{k}_\text{n}) |^2.
\end{align}
We can also write this identity as $1=\int d\omega | F_{\omega_0}(\omega)\tilde{\phi}_{\omega}(y^\mu(\tau))|^2$, where we have introduced the functions $F_{\omega_0}(\omega)$ and $\tilde{\phi}_{\omega}(y^\mu(\tau))$ appropriately.

Recalling that we have assumed that the photon has most support along the direction of propagation, we come to the conclusion that we can approximate
\begin{align}\label{effective:frequency:operator}
\int_{\perp,1,2} d^2k F_{\boldsymbol{k}_{\text{n},0}}(\boldsymbol{k}(\boldsymbol{k}_\text{n})) \, \tilde{\phi}_{\boldsymbol{k}(\boldsymbol{k}_\text{n})}(y^\rho)|_{\tau} \Theta(\boldsymbol{k}_\text{n})\hat{a}_{\boldsymbol{k}_\text{n}}\approx F_{\omega_0}(\omega)\tilde{\phi}_{\omega}(y^\mu(\tau))\hat{a}_\omega,
\end{align}
where the operator $\hat{a}_\omega$ is effectively defined by this relation and satisfies $[\hat{a}_\omega,\hat{a}_\omega'^\dag]=\delta(\omega-\omega')$ while all other commutators vanish. Notice that $\int d\omega |F_{\omega_0}(\omega)\tilde{\phi}_{\omega}(y^\mu(\tau))|^2=1$ due to the constraint $\int d^3k |F_{\boldsymbol{k}_0,x|_\Sigma}(\boldsymbol{k})|^2=1$ discussed above, where $F_{\boldsymbol{k}_0,x|_\Sigma}(\boldsymbol{k}):=F_{\boldsymbol{k}_0}(\boldsymbol{k})\phi_{\boldsymbol{k}}(x^\mu)|_\Sigma$. This is all consistent with the fact that the photon operators (\ref{intermediate:operator:expression}) represent physical photons with a certain bandwidth profile.

The defining expression (\ref{effective:frequency:operator}) for the operators $\hat{a}_\omega$ is, stricto sensu, incorrect at face value because $\hat{a}_{\boldsymbol{k}_\text{n}}$ and $\hat{a}_\omega$ act on different Hilbert spaces --- the Hilbert spaces are determined by different numbers of degrees of freedom and they are not unitarily equivalent. Therefore, we will use the notation introduced here with the understanding that the Hilbert space determined by the $\hat{a}_\omega$ stands for, in a strict mathematical sense, a Hilbert space equivalent to that of the operators $\hat{a}_{\boldsymbol{k}_\text{n}}$. Furthermore, we restrict ourselves to the subspaces of constant $k_1,k_2$ given the effective strong confining of the photon to the direction perpendicular to $k_1$ and $k_2$. Therefore, the operators $\hat{a}_\omega$ should read $\hat{a}_{\omega,k_1,k_2}$. We will drop the dependence on $k_1,k_2$ and assume that it is from now on understood that we work in the subspace of the total Hilbert space defined by constant (vanishing) $k_1,k_2$.

Finally, combining all of the above means that we can express the field operator $\hat{A}_{\omega_0}$ at a particular (proper) time $\tau$ as
\begin{equation}\label{fat:operator:effective}
\hat{A}_{\omega_0}:=\int_0^{+\infty} d\omega \,e^{-i\omega(\tau-\tilde{\lambda}_\perp)}\,F_{\omega_0}(\omega)\tilde{\phi}_{\omega}(y^\mu(\tau)) \hat{a}_{\omega},
\end{equation}
where we have assumed that the frequency degree of freedom is always positive. Note that this final expression comes with the added constraint $\int_0^{+\infty} d\omega\,|F_{\omega_0}(\omega)\tilde{\phi}_{\omega}(y^\mu(\tau))|^2=1$ in order for the canonical commutation relations $[\hat{A}_{\omega_0},\hat{A}_{\omega_0}^\dag]=1$ to be satisfied.
We emphasize that this is an expression that is valid exactly \textit{only} if the observers are static. When the spacetime is dynamic, or there are nontrivial contribution from the directions perpendicular to that of the photon propagation, new work will be necessary to provide the correct expression.

\subsection{Gravitational redshift of sharp-momentum photon operators}
\begin{figure}[ht!]
\centering
\includegraphics[width=0.5\linewidth]{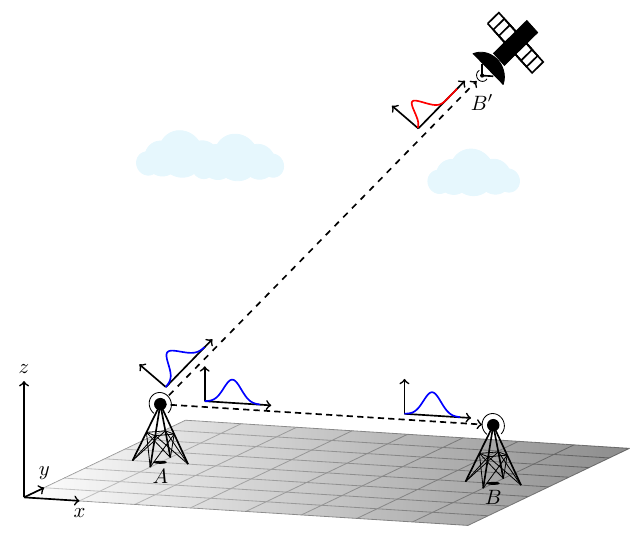}
\caption{Alice and Bob wish to perform the \textbf{Protocol} described above.}\label{miao:2}
\end{figure}

The effect of gravitational redshift on ideal free (quantum) photons, that is, on the excitations of the quantized free electromagnetic field, becomes manifest upon measurement. If our two observers Alice and Bob wish to determine the nature and magnitude of such effect they can na\"ively perform the following protocol which is borrowed from an empirical approach:

\newpage

\textit{\textbf{Protocol}---Alice and Bob are static with respect to each other and in flat spacetime. Each one of them is endowed with ideal photon sources and detectors, and we assume that there are no sources of decoherence, noise or other environmental disturbances. In this case, Alice prepares and sends a localized photon to Bob, who will detect the photon on arrival. He can compare the properties of such photon with those of the one that he keeps as local reference, and he will find that they match. This simple setup is depicted in the lower part of Figure~\ref{miao:2}. A second scenario sees Alice and Bob$^\prime$ located at a different position (e.g., on a satellite) in a curved spacetime. Alice sends a photon to Bob$^\prime$, who will potentially detect a (gravitational) redshift within the incoming photon, i.e., any frequency $\omega'$ as measured \textit{locally} by his clock will not coincide \textit{numerically} with the sharp frequency $\omega$ of the sent photon. The scheme is depicted in the upper part of Figure~\ref{miao:2}.} 

Now we can make a key observation. As far as Bob is concerned, i.e., from the perspective of his isolated laboratory, the photon sent by Alice might have changed with respect to the one he was expecting, and therefore he can study the properties of the transformation involved. He can do this \textit{irrespective} of where the incoming photon has originated or which specific physical process it has witnessed during its flight. Concretely, this means that Bob can assign a \textit{channel} to the process that affected the incoming photon, and seek for its properties. In particular, Bob wishes to implement the transformation $T$ that acts on the incoming operators $\hat{a}_\omega$ as a unitary operation that maps them to final operators $\hat{a}_{\omega'}$. 

Bob therefore assumes that there is a transformation $T(\alpha):\omega\rightarrow\omega'=\alpha\omega$ on \textit{each} sharp frequency $\omega$ of his \textit{local} spectrum. Note that here we have not used any particular form of the parameter $\alpha$, and indeed we do not need to specify any of its properties. For the moment we do not even assume that it is related to the gravitational redshift. Bob then looks for a \textit{unitary transformation $\hat{U}(\alpha)$} that implements $T(\alpha)$ through the following relation
\begin{equation}\label{unitary:transformation}
\hat{a}_{\omega'}=\hat{U}^\dag(\alpha)\,\hat{a}_{\omega}\,\hat{U}(\alpha)=\hat{a}_{\alpha\omega}
\end{equation}
for all $\alpha$, where $\hat{U}^\dag(\alpha)\hat{U}(\alpha)=\mathds{1}$. 

We now recall that the operators $\hat{a}_{\omega}$ are required to satisfy the canonical commutation relations $[\hat{a}_\omega,\hat{a}^\dag_{\omega'}]=\delta(\omega-\omega')$ while all others vanish. We also recall the fact that $[\hat{a}_{\alpha\omega},\hat{a}_{\alpha\omega'}^\dag]=\delta(\alpha\omega-\alpha\omega')$, and $\delta(f(x))=\sum_n\delta(x-x_{0,n})/|f'(x_{0,n})|$, where $x_{0,n}$ are the simple zeros of the function $f(x)$. This means that $\delta(\alpha\omega-\alpha\omega')=|\alpha|^{-1}\delta(\omega-\omega')$. Thus we can write
\begin{equation}\label{inconsistence:first}
\delta(\omega-\omega')=\hat{U}^\dag(\alpha)\delta(\omega-\omega')\hat{U}(\alpha)=\hat{U}^\dag(\alpha)\left[\hat{a}_{\omega},\hat{a}_{\omega'}^\dag\right]\hat{U}(\alpha)=\left[\hat{a}_{\alpha\omega},\hat{a}_{\alpha\omega'}^\dag\right]=\frac{1}{|\alpha|}\delta(\omega-\omega').
\end{equation}
It is clear that this series of equalities can be satisfied only if $\alpha=e^{i\theta}$ for some angle $\theta$. If we were to identify $\alpha$ with the redshift factor, i.e., $\alpha=\chi^2$, this would also imply $\chi=1$. Such identification would be motivated by the fact that the formally identical transformation $\omega'=\chi^2\omega$ is given by the main expression \eqref{general:redshift}. This means that only non-redshifted photons can be thought of as having witnessed the action of a unitary channel. We therefore conclude that, if the unitarity of the transformation is desired -- and we expect it to be since gravitational redshift can be ``undone'' by simply reflecting the photon back to the source -- the simple assumption that $T(\chi)$ act via $T(\chi):\omega\rightarrow\omega'=\chi^2\omega$ is incorrect.

One way to understand the inconsistency that has arisen in (\ref{inconsistence:first}) is to note that, in general, the \textit{magnitude} of the shift for each frequency is different. In fact, $|\alpha\omega'|>|\alpha\omega|$ for $\omega'>\omega$. Furthermore, the difference between two frequencies is ``stretched'' or ``compressed'' by this process, since also $|\alpha\omega'-\alpha\omega|>|\omega'-\omega|$ for $|\alpha|>1$, and $|\alpha\omega'-\alpha\omega|<|\omega'-\omega|$ for $|\alpha|<1$. However, the Dirac delta-function, which is not a function in the strict sense, does not have a ``width'' that can be stretched or compressed. One can say that the shape of the Dirac delta is ``rigid'' in this sense and does not change. We will solve the problem below by considering field operators that have compact support.

\subsection{Gravitational redshift of realistic photon operators}
We have considered the effects of gravitational redshift on ideal photons that have a sharp momentum. Here we move on to consider the effects of gravitational redshift on realistic photons.

The field operator $\hat{A}_{\omega_0}$ for the realistic photon is given by (\ref{fat:operator:effective}). In this section we would like to ask a question similar to the one posed above by Bob by replacing sharp momentum frequency field operators with realistic ones. The transformation $T(\alpha):\omega\rightarrow\omega'=\alpha\omega$ induces a unitary transformation of the operator $\hat{a}_{\omega}$ \textit{only}, and is poised to transform $\hat{A}_{\omega_0}\rightarrow\hat{A}_{\omega_0}'=\hat{U}^\dag(\alpha)\hat{A}_{\omega_0}\hat{U}(\alpha)$. This means that
{\small
\begin{equation}
\hat{A}_{\omega_0}=\int_0^{+\infty} d\omega F_{\omega_0}(\omega)e^{-i\omega(\tau-\tilde{\lambda}_\perp)}\tilde{\phi}_{\omega}(y^\mu(\tau)) \hat{a}_{\omega}\overset{T(\alpha)}{\rightarrow}\hat{A}_{\omega_0}'=\int_0^{+\infty} d\omega F_{\omega_0}(\omega)e^{-i\omega(\tau-\tilde{\lambda}_\perp)}\tilde{\phi}_{\omega}(y^\mu(\tau)) \hat{a}_{\alpha\omega}.
\end{equation}
}
Once more, assuming that the transformation (\ref{unitary:transformation}) holds, it is easy to compute the explicit expression for the commutator of the operators $\hat{A}_{\omega_0}',\hat{A}_{\omega_0}^{\prime\dag}$ after the redshift has applied. We have
\begin{equation}
1=\hat{U}^\dag(\alpha)\hat{U}(\alpha)=\hat{U}^\dag(\alpha)\left[\hat{A}_{\omega_0},\hat{A}_{\omega_0}^\dag\right]\hat{U}(\alpha)=\left[\hat{A}_{\omega_0}',\hat{A}_{\omega_0}^{\prime\dag}\right]=\frac{1}{|\alpha|}.
\end{equation}
As we found above, this equation can be satisfied only for the trivial case $|\alpha|=1$, which would be equivalent to the case of no redshift $\chi=1$ after identification $\alpha=\chi^2$. It is clear that we have not solved the problem by using the transformation $T(\alpha)$ on realistic operators, and therefore we conclude that \textit{it is not possible to obtain gravitational redshift in the form of a linear shift of the spectrum of sharp frequencies $\omega$ as a unitary operation acting on the set of field modes $\{\hat{a}_\omega\}$ alone}. This result corroborates the claim that the gravitational redshift  \textit{cannot} be interpreted simply as a shift in the sharp frequencies of the photons for all frequencies of the spectrum. Below we proceed to solve this conundrum.

\subsection{Quantum modelling of gravitational redshift}
We have now understood that we need to refine the question posed above by Bob. Specifically, Bob will ask the following: \textit{which transformation $T(\alpha)$ will be implemented by a unitary operator when acting on realistic photons?} To answer this question, we work backwards and  note that Bob will describe the received photon as $\hat{A}_{\omega_0'}=\int_0^\infty d\omega\,F'_{\omega_0'}(\omega)e^{-i\omega\tau}\phi_{\omega}'(y^\mu(\tau))\,\hat{a}_\omega$ as a function of the frequency $\omega$ as measured \textit{locally} in his laboratory with respect to his proper time $\tau$, while the expected photon has the expression $\hat{A}_{\omega_0}=\int_0^\infty d\omega\,F_{\omega_0}(\omega)e^{-i\omega(\tau-\tilde{\lambda}_\perp)}\phi_{\omega}(y^\mu(\tau))\,\hat{a}_\omega$. Bob then imposes a transformation where \textit{each} frequency variable $\omega$ that appears in the definition of $\hat{A}_{\omega_0}$ transforms according to $T(\alpha):\omega\rightarrow\omega'=\alpha\omega$, see \cite{Bruschi:Ralph:2014}. This means that the transformation for Bob reads
{\small
\begin{equation}\label{preliminary:transformation}
\hat{A}_{\omega_0}=\int_0^\infty d\omega F_{\omega_0}(\omega)e^{-i\omega(\tau-\tilde{\lambda}_\perp)}\tilde{\phi}_{\omega}(y^\mu(\tau)) \hat{a}_\omega \overset{T(\alpha)}{\rightarrow}\hat{A}_{\omega_0'}'=\alpha\int_0^\infty d\omega F_{\omega_0}(\alpha\omega)e^{-ig(\alpha)\omega(\tau-\tilde{\lambda}_\perp)}\tilde{\phi}_{\alpha\omega}(y^\mu(\tau)) \hat{a}_{\alpha\omega},
\end{equation}
}
where Bob needed to introduce a general function $g(\alpha)$ in front of $\omega\tau$ because of the fact that the photon sent is generated with respect to the proper time of Alice, which does not coincide with his own. Thus, according to \eqref{deviation:path:final:second}, \eqref{first:useful:expression:fat:photon} and the discussion in between, he needs to maintain some additional freedom in order to compensate for the difference in proper time between his clock and the clock at Alice's location. This function will be determined below.
Note that we now correctly have $[\hat{A}_{\omega_0'}',\hat{A}_{\omega_0'}^{'\dag}]=[\hat{A}_{\omega_0},\hat{A}_{\omega_0}^\dag]=1$ since $\int_0^{+\infty}|F_{\omega_0}(\omega)\tilde{\phi}_{\omega}(y^\mu(\tau))|^2=1$ as it is 
immediate to verify directly.

Bob is now tempted to make the following identification:
\begin{equation}\label{function:transformation}
F'_{\omega_0'}(\omega)\tilde{\phi}_{\omega}'(y^\mu(\tau))\equiv\sqrt{\alpha}\,F_{\omega_0}(\alpha\omega)\tilde{\phi}_{\alpha\omega}(y^\mu(\tau)),
\end{equation}
since it is easy to see that $\int_0^{+\infty}|F_{\omega_0}(\omega)\tilde{\phi}_{\omega}(y^\mu(\tau))|^2=1$ implies $\int_0^{+\infty}|F'_{\omega_0'}(\omega)\tilde{\phi}_{\omega}'(y^\mu(\tau))|^2=1$. 

If the identification \eqref{function:transformation} holds, it follows that Bob should be interested in then looking at the operator $\hat{a}_{\alpha\omega}$. Unfortunately, he already knows that $\bigl[\hat{a}_{\alpha\omega},\hat{a}_{\alpha\omega'}^\dag\bigr]=|\alpha|^{-1}\delta(\omega-\omega')$. Nevertheless, he is not discouraged and notices that he can introduce the operators
\begin{equation}\label{sharp:operator:transformation}
\hat{a}_{\omega}'\equiv \sqrt{|\alpha|}\,\hat{a}_{\alpha\omega},
\end{equation}
which now have the correct canonical commutation relations $[\hat{a}_{\omega}',\hat{a}_{\omega'}^{\prime\dag}]=\delta(\omega-\omega')$ as can be checked by direct inspection.

Bob is finally left with dealing with the arbitrary function $g(\alpha)$. He starts by noting that at Alice's location she will have prepared the photon using her proper time $\tau_{\text{A}}$ and frequencies $\omega_{\text{A}}$. In his own location, however, Bob will use instead his proper time $\tau_{\text{B}}$ and frequencies $\omega_{\text{B}}$.
Therefore, if Alice has sent a photon with phase $\exp[-i \omega_{\text{A}}(\tau_{\text{A}}-\tilde{\lambda}_{\perp,\text{A}})]$, Bob will define the expected photon in his laboratory with a phase $\exp[-i \omega_{\text{B}}(\tau_{\text{B}}-\tilde{\lambda}_{\perp,\text{B}})]$. Here we recall that $\tilde{\lambda}_\perp:=||K||\lambda_\perp$ and therefore $\tilde{\lambda}_{\perp,J}:=||K||_J\lambda_{\perp}$, where $J=$A,B. The crucial observation here is that, according to the transformations
{\small
\begin{align}
\frac{\omega_{\text{B}}}{\omega_{\text{A}}}=\frac{\sqrt{-K^\rho K_\rho}|_{x_{\text{A}}}}{\sqrt{-K^\rho K_\rho}|_{x_{\text{B}}}}=\chi^2=\frac{\tau_{\text{A}}}{\tau_{\text{B}}} \quad\Rightarrow\quad\omega_{\text{B}}\tau_{\text{B}}=\Omega \xi_\Sigma=\omega_{\text{A}}\tau_{\text{A}}\quad\text{and}\quad\tilde{\lambda}_{\perp,\text{A}}\omega_{\text{A}}=\Omega\lambda_\perp=\tilde{\lambda}_{\perp,\text{B}}\omega_{\text{B}}.
\end{align} 
}
These equations require us to make a few considerations. First it is clear that, if Alice has prepared the photon at time $\tau_{\text{A},0}$ it will be received at Bob's location at time $\tau_{\text{B}}=\Delta\tau_{\text{B}}+\tau_{\text{B},0}$, where $\tau_{\text{B},0}$ denotes the agreed upon local time at which he believes the photon has left Alice's lab, while $\Delta\tau_{\text{B}}$ is the lapse of time that he will associate to the travel. Therefore, $\tau_{\text{A}}$ is the time at which Alice believes the photon has arrived, and $\tau_{\text{A}}$ and $\tau_{\text{B}}$ are the proper times associated to the hypersurface $\Sigma$ of arrival. Both $\tau_{\text{A}}$ and $\tau_{\text{B}}$ can be obtained with respect to the Killing vector parameter $\xi$ using the relation $\tau_{\text{J}}=\sqrt{-K^\mu K_\mu}|_{\text{J}}\,\xi$ for J$=$A,B.  This is also true for the parameter $\tilde{\lambda}_\perp$ and therefore we can write $\tau_{\text{B}}-\tilde{\lambda}_{\perp,\text{B}}=\Delta\tau_{\text{B}}+\tau_{\text{B},0}-\Delta\tilde{\lambda}_{\perp,\text{B}}-\tilde{\lambda}_{\perp,\text{B},0}$. Since our photon moves along paths of constant $\tilde{\lambda}_-=\tau-\tilde{\lambda}_\perp$, where $\tilde{\lambda}_-=\lambda_-/||K||$, it follows that $\tilde{\lambda}_{-,0}=\tau_{\text{B}}-\tilde{\lambda}_{\perp,\text{B}}=\Delta\tau_{\text{B}}+\tau_{\text{B},0}-\Delta\tilde{\lambda}_{\perp,\text{B}}-\tilde{\lambda}_{\perp,\text{B},0}$ for an appropriate constant $\tilde{\lambda}_{-,0}$. Therefore, the photon initially located around $\tilde{\lambda}_{\perp,\text{B},0}=\tau_{\text{B},0}-\tilde{\lambda}_{-,0}$ at time $\tau_{\text{B},0}$, is now located around $\Delta\tilde{\lambda}_{\perp,\text{B}}+\tilde{\lambda}_{\perp,\text{B},0}=\Delta\tau_{\text{B}}+\tau_{\text{B},0}-\tilde{\lambda}_{-,0}$ after a laps of time $\Delta\tau_{\text{B}}$.
Since these phases are independent on the redshift, by the very definition of the proper time and the frequencies measured locally, it is natural for Bob to set $g(\alpha)=1$. Furthermore, $\omega\tilde{\lambda}_{-,0}=\omega(\tau-\tilde{\lambda}_\perp)=\omega(\Delta\tau+\tau_0-\Delta\tilde{\lambda}_\perp-\tilde{\lambda}_{\perp,0})$ locally according to Bob. The pictorial description is again in Figure~\ref{figure:paths:configuration}.

Bob now can fix the parameter $\alpha$ for the case of interest by noting that the transformation (\ref{preliminary:transformation}) leaves the phase $\omega(\tau-\tilde{\lambda}_\perp)$ invariant. The redshift (\ref{general:redshift:Killing:path:general}) is the physical transformation that also leaves the phase $\omega(\tau-\tilde{\lambda}_\perp)$ invariant.  
Therefore, Bob sets $\alpha=\chi^2$ which gives him the following transformation law for the bandwidth function of the physical photon:
\begin{equation}\label{function:transformation:final}
F_{\omega_0}(\omega)\tilde{\phi}_{\omega}(y^\mu(\tau))\rightarrow F'_{\omega_0'}(\omega)\tilde{\phi}_{\omega}'(y^\mu(\tau))\equiv\chi\,F_{\omega_0}(\chi^2\omega)\tilde{\phi}_{\chi^2\omega}(y^\mu(\tau)),
\end{equation}
as well as the transformation law for the physical photon operator:
\begin{equation}\label{physical:photon:operator:transformation:final}
\hat{A}_{\omega_0}\rightarrow \hat{A}'_{\omega_0'}=\hat{U}(\chi)^\dag\hat{A}_{\omega_0}\hat{U}(\chi)=\int_0^{+\infty}e^{-i\omega(\tau-\tilde{\lambda}_\perp)}F'_{\omega_0'}(\omega)\tilde{\phi}_{\omega}'(y^\mu(\tau))\hat{a}_\omega.
\end{equation}
These two equations are the main expressions of this work.

\subsection{Gravitational redshift as a mode-mixer}
We have shown that gravitational redshift can be viewed as a canonical transformation between a realistic photon operator and another one, thus preserving the canonical commutation relations. We also note that since $\hat{a}_\omega|0\rangle=0$ we also have that $\hat{A}'_{\omega_0'}|0\rangle=\hat{A}_{\omega_0}|0\rangle=0$. The functions  $F_{\omega_0}(\omega)e^{-i\omega(\tau-\tilde{\lambda}_\perp)}\tilde{\phi}_{\omega}(y^\mu(\tau))$ can be seen as an element of the orthonormal basis of the field expansion, which we can construct by seeking all functions $F_{\underline{\lambda}}(\omega)e^{-i\omega(\tau-\tilde{\lambda}_\perp)}\tilde{\phi}_{\omega}(y^\mu(\tau))$ that are orthogonal to $F_{\omega_0}(\omega)e^{-i\omega(\tau-\tilde{\lambda}_\perp)}\tilde{\phi}_{\omega}(y^\mu(\tau))$. Here $\underline{\lambda}$ is a collection of appropriate labels for the other mode functions and the exact nature of these labels is irrelevant. Orthonormality means that $\int_0^\infty F_{\underline{\lambda}}^*(\omega)F_{\omega_0}(\omega)|\tilde{\phi}_{\omega}(y^\mu(\tau))|^2=0$ for all $\underline{\lambda}$ and $\int_0^\infty F_{\underline{\lambda}}^*(\omega)F_{\underline{\lambda}'}(\omega)|\tilde{\phi}_{\omega}(y^\mu(\tau))|^2=\delta(\underline{\lambda}-\underline{\lambda}')$. Notice that the map $T$ can be seen as also inducing a transformation of the set $\{\phi_{\boldsymbol{k}}(x^\rho)\}$ of basis modes to the new set $\{F_{\omega_0}(\omega)e^{-i\omega(\tau-\tilde{\lambda}_\perp)}\tilde{\phi}_{\omega}(y^\mu(\tau)),F_{\underline{\lambda}}(\omega)e^{-i\omega(\tau-\tilde{\lambda}_\perp)}\tilde{\phi}_{\omega}(y^\mu(\tau))\}$ of basis modes, which is not uncommon in quantum field theory in flat or curved spacetime. For example, it appears in the derivation of the Unruh effect, where Unruh modes are obtained as particular linear combinations of Minkowski modes \cite{Unruh:1976,Bruschi:Louko:2010}.

We can construct the realistic photon operators $\hat{A}_{\underline{\lambda}}$ via
\begin{equation}
\hat{A}_{\underline{\lambda}}:=\int_0^{+\infty} d\omega e^{-i\omega(\tau-\tilde{\lambda}_\perp)}F_{\underline{\lambda}}(\omega)\tilde{\phi}_{\omega}(y^\mu(\tau)) \hat{a}_\omega,
\end{equation}
which obey the equal time canonical commutator relations $[\hat{A}_{\omega_0},\hat{A}_{\underline{\lambda}}]=0$ and $[\hat{A}_{\underline{\lambda}},\hat{A}_{\underline{\lambda}'}^\dag]=\delta(\underline{\lambda}-\underline{\lambda}')$. 
Furthermore, the transformations (\ref{function:transformation:final}) and (\ref{physical:photon:operator:transformation:final}) apply appropriately to \textit{all} of the photon operators $\{\hat{A}_{\omega_0},\hat{A}_{\underline{\lambda}}\}$, and thus between bases $\{F_{\omega_0},F_{\underline{\lambda}}\}$ and $\{F'_{\omega_0'},F'_{\underline{\lambda}'}\}$ of mode functions. 

We now collect all field operators $\{\hat{A}_{\omega_0},\hat{A}_{\underline{\lambda}}\}$ in the vector $\hat{\mathbb{X}}:=(\hat{A}_{\omega_0},\hat{A}_{\underline{\lambda}_1},\ldots,\hat{A}_{\omega_0}^\dag,\hat{A}_{\underline{\lambda}_1}^\dag,\ldots)^{\textrm{Tp}}$. Then, the transformation (\ref{physical:photon:operator:transformation:final}) implies that there exists a symplectic matrix $\boldsymbol{S}(\chi)$ such that
 \begin{equation}\label{arbitrary:transformation:representation}
\hat{\mathbb{X}}(\chi):=\hat{U}^\dag(\chi)\,\hat{\mathbb{X}}\,\hat{U}(\chi)\equiv\boldsymbol{S}(\chi)\,\hat{\mathbb{X}}.
\end{equation}
We have already presented the form of a generic symplectic matrix $\boldsymbol{S}$ in Subsection~\ref{section:liner:dynamics}. We note that, in the present case, there is no mixing between creation and annihilation operators and therefore the Bogoliubov beta-coefficients vanish, i.e., $\boldsymbol{B}\equiv0$. Thus, we see that $\boldsymbol{S}(\chi)=\boldsymbol{U}(\chi)\oplus\boldsymbol{U}^*(\chi)$, where $\boldsymbol{U}(\chi)$ is a unitary matrix \cite{Adesso:Ragy:2014}.

Let us pick $N-1$ particular modes $\hat{A}_{\omega_n}$ with $n\in \mathcal{I}:={1,...,N-1}$ and then build the operator $\hat{A}_\perp$ via the relation
\begin{equation}
\hat{A}_\perp:=\sum_{\underline{\lambda}'} c_{\underline{\lambda}'} \hat{A}_{\underline{\lambda}},
\end{equation}
where the sum is over all remaining operators labelled by $\underline{\lambda}'$ (i.e., except for those in the set $\{\hat{A}_{\omega_n}\}|_{n\in\mathcal{I}}$). Clearly, we must have: first, $\sum_{\underline{\lambda}'}|c_{\underline{\lambda}'}|^2=1$; second, $[\hat{A}_{\omega_n},\hat{A}_{\omega_m}^\dag]=\delta_{nm}$; and third, $[\hat{A}_\perp,\hat{A}_\perp^\dag]=1$, while all other commutators vanish.

We can then construct the vector $\hat{\mathbb{X}}_{\text{r}}:=(\hat{A}_{\omega_1},\hat{A}_{\omega_2},\ldots,\hat{A}_{\omega_{N}},\hat{A}_\perp)^{\textrm{Tp}}$ and therefore (\ref{arbitrary:transformation:representation}) simplifies to
 \begin{equation}\label{finite:transformation:representation}
\hat{\mathbb{X}}_{\text{r}}(\chi):=\hat{U}^\dag(\chi)\,\hat{\mathbb{X}}_{\text{r}}\,\hat{U}(\chi)\equiv\boldsymbol{U}(\chi)\,\hat{\mathbb{X}}_{\text{r}}.
\end{equation}
A unitary symplectic transformation is known in quantum optics as a \textit{mode-mixer} \cite{Scully:Zubairy:1997}. Let us select and employ $N-1$ modes of interest and 1 mode $F_\perp\equiv F_N$ that will constitute the ``orthogonal'' space, the totality of which form an orthonormal basis $\{F_n\}$. This basis will be mixed into a new one, which we divide accordingly into $N-1$ new modes and 1 orthogonal one $F_\perp'\equiv F_N'$, also forming an orthonormal basis $\{F_n'\}$. We can then see see that we can construct $(N-1)(N-1+1)=(N-1)N$ independent complex overlap functions of the form $\langle F_n',F_m\rangle$ for $n,m=1,...,N-1$ including both the modulus and the phase. The magnitude of the overlaps $\langle F_n',F_N\rangle$ for $n=1,...,N-1$ is fixed by the magnitudes $|\langle F_n',F_m\rangle|$ since the transformation between the two bases is canonical. We are left, however, with $N$ complex phases arg$(\langle F_n',F_N\rangle)$ that remain free parameters. The remaining overlaps $\langle F_N',F_n\rangle$ with $n=1,..,N-1$ are again fixed completely by the previous procedure. Thus, the total amount of free parameters is $N(N-1)+N=N^2=\textrm{dim}(U(N))$.\footnote{Note that this explanation clarifies the similar degree counting in \cite{Bruschi:Schell:2023}.} The group can be reduced to $SU(N)$ by extracting from the mixing matrix $\hat{U}(\chi)$ an overall phase that can be set to 1. The possibility of performing these identifications has already been discussed in the literature \cite{Menessen:Jones:2017}. 

\vspace{0.2cm}

\begin{tcolorbox}[colback=orange!3!white,colframe=orange!85!black,title= \textbf{Box 5:} Example: the Beamsplitter,label=example:beamsplitter]
We now apply the theoretical results discussed above. As an example we consider the case where $N=2$. In this case, we will have only the modes $\hat{A}_{\omega_0}$ and $\hat{A}_\perp$, where $\hat{A}_{\omega_0}$ is the mode of interest and $\hat{A}_\perp$ collects the remaining elements of the basis that we are not interested in.

Our vectors of operators read $\hat{\mathbb{X}}_{\text{r}}:=(\hat{A}_{\omega_0},\hat{A}_\perp)^{\textrm{Tp}}$ and $\hat{\mathbb{X}}_{\text{r}}(\chi):=(\hat{A}_{\omega_0}',\hat{A}_\perp')^{\textrm{Tp}}$, and we can write
\begin{align*}
\hat{\mathbb{X}}_{\text{r}}(\chi)=\boldsymbol{U}(\chi)\hat{\mathbb{X}}_{\text{r}},
\quad
\text{with}
\quad
\boldsymbol{U}(\chi)=
\begin{pmatrix}
\cos\theta & e^{i\phi}\sin\theta\\
-e^{-i\phi}\sin\theta & \cos\theta
\end{pmatrix},
\end{align*}
and we have $\theta=\theta(\chi)$ and $\phi=\phi(\chi)$. The angle $\theta$ and phase $\phi$ are obtained by employing the overlap of the two modes considered. In particular we have 
\begin{align*}
\cos\theta(\chi)\equiv|\langle1_{\omega_0}'|1_{\omega_0}\rangle|\quad\text{and}\quad\phi(\chi)\equiv\arg(\langle1_{\omega_0}'|1_\perp\rangle),
\end{align*}
where we have defined $|1_{\omega_0}\rangle:=\hat{A}_{\omega_0}^\dag|0\rangle$ and $|1_{\omega_0}'\rangle:=\hat{A}_{\omega_0}^{\prime\dag}|0\rangle$. It is possible to verify directly that these expressions give the desired result. Note that we have set the phase $\arg(\langle1_{\omega_0}'|1_{\omega_0}\rangle)$ to zero out of convenience and without loss of generality.
\end{tcolorbox}

\section{Applications}\label{section:Applications}
Here we report on a few applications of the predictions of this work that have been proposed in the literature. These are not exhaustive of all possibilities, which we leave open for study in future work.  

\subsection{Space-based quantum networks}
Quantum information tasks require the distribution of physical states that contain entanglement, which is the paramount resource for quantum information processing \cite{Nielsen:Chuang:2010}. Photons are a core system for the establishment of distant entangled nodes 
\cite{Duan:Lukin:2001,Vallone:Bacco:2015,Liorni:Kampermann:2021,Gundogan:Sidhu:2021,Mol:Esguerra:2023}, and they can be distributed either via optical fibre or free space links. Optical fibres are a convenient solution given the existing network which makes many options readily available. However, propagation in fibre is subject to significant distance limitations since photon loss limits the reachable distance to few hundreds of kilometres \cite{Okoshi:2012,Boaron:Boso:2018}. This, in turn, implies the need of a significant number of quantum repeaters in order to establish entangled states over useful distances \cite{Briegel:Duer:1998}. Space-based links are an obvious alternative in which photons are exchanged between nodes placed at different heights of the Earth's gravitational potential and propagate in free space. The \textit{Micius satellite}\footnote{After Chinese philosopher Mo Di -- latinized as Micius (c. 470 - c. 391 BC).} of the Chinese academy of Sciences (CAS) is an example of a dedicated (quantum) technology to study entanglement distribution between distant sources 
\cite{Ruihong:Ying:2019,Lu:Cao:2022}. 

Space-based science suffers from the obvious problem of being affected by gravity. In general, signals propagating between users located at different heights in the gravitational potential of a planet can lead to unwanted effects on the state of systems themselves. To date, many experiments have been performed to demonstrate single- and multi-photon exchange through free space links 
\cite{Vallone:Bacco:2015,Liao:Cai:2017,Calderaro:Agnesi:2018,Agnesi:Calderaro:2019}. Among the foreseen applications are quantum key distribution (QKD) 
\cite{Liao:Cai:2017}, Quantum Communication 
\cite{Sidu:Joshi:2021}, and Distributed Quantum Computing 
\cite{Cuomo:Caleffi:2020}. Regardless of the successes accrued so far, the level of precision of quantum systems is continuously increasing, leaving open the window to study the influence due to relativistic effects. 

\vspace{0.2cm}

\begin{tcolorbox}[colback=orange!3!white,colframe=orange!85!black,title=Quantum Key Distribution]
Quantum Key Distribution (QKD) is a protocol that enables two (or more) users to establish a common secret shared key 
\cite{Pirandola:Andersen:2020,Bennet:Brassard:1984,Bennet:Brassard:1992}. 
The basic idea is that Alice and Bob will attempt to share a secret key to be used in an encryption scheme, and they will try to detect any eavesdropper.
The two users require a random number generator and a scheme to encode classical bits into polarization states of photons, and to detect them.
One of the most studied protocols that contains the core of the idea is known as BB84 \cite{Bennet:Brassard:1984}, and reads as follows:
\begin{itemize}
	\item Alice uses the random number generator to pick either a $HV$ or a $45^\circ$ polarization basis. She encodes the first bit of information in the chosen basis, records both bit and basis, and sends the photon to Bob.
	\item Bob uses his random number generator to pick either a $HV$ or a $45^\circ$ polarization basis. He detects the incoming photon and notes the outcome and the basis chosen.
	\item Alice and Bob reveal publicly \textit{the choice of basis} made for each transmission. They keep only the bits that correspond to the cases where they chose the same basis.
	\item Alice and Bob ``sacrifice'', or reveal publicly, a significant fraction of the bits that they now share and perform some statistical analysis on them. They know that if the \textit{quantum bit error rate} (QBER), defined as the fraction of bits that differ, rises above a certain value provided by security proofs, then the communication was hacked and they have to abort the protocol 
 \cite{Shor:Preskill:2000,Renner:2005}; 
	\item If the protocol was securely executed the unrevealed shared bits are used to create a secret key.
\end{itemize}
This simple scheme relies on the \textit{no cloning theorem} \cite{Wootters:Zurek:1982,Dieks:1982,Ghirardi:2013}, which states that it is not possible to perfectly clone a quantum state in quantum mechanics. This in turn implies that an eavesdropper, when trying to tamper with the communication, will inevitably leave a mark that can be traced and detected. Thus, QKD in principle provides a scheme that is secured by the laws of physics, rather than by the complexity of inverting complicated functions 
\cite{Ekert:1991,Gisin:Ribordy:Tittel:2002}.
\end{tcolorbox}

\vspace{0.2cm}

A recent trend of work has initiated the study of the effects of gravity on the propagation of photons between different users located at different heights in the gravitational potential of the Earth 
\cite{Bruschi:Ralph:2014,Bruschi:Datta:2014,Kohlrus:Bruschi:2017,Kohlrus:Bruschi:Fuentes:2019,Bruschi:Chatzinotas:2021,Barzel:Bruschi:2022,Bruschi:Schell:2023}. The core idea is to model photons and excitations of a quantum field in (weakly) curved spacetime. Gravitational redshift is one of the key effects that modify the quantum state of the photons. Since standard quantum information protocols require stability of the state during free space propagation (ignoring, say, sources of noise or decoherence), it becomes evident that quantification of gravitational effects can inform on the need, or lack thereof, of keeping track of their impact. 

Among the growing literature on this topic there are studies that investigate the effects of gravity on quantum communication 
\cite{Bruschi:Ralph:2014,Exirifard:Culf:2021}, quantum key distribution 
\cite{Vilasini:Portmann:2019,Xu:Ma:2020}, quantum illumination 
\cite{Tan:Erkmen:2008,Liu:Wen:2022} to name a few. In general, it is clear that nontrivial modifications of standard protocols are predicted and it has even been argued that the effects need not be small 
\cite{Bruschi:Ralph:2014}. As an example, one can consider a simple QKD protocol between Alice and Bob and attempt to evaluate the overall effect of gravity on a key figure of merit known as the \textit{quantum bit error rate} (QBER). For a brief overview of the key idea behind QKD see the Panel `Quantum Key Distribution' below.

Protocols for securely sharing a key are almost always studied using quantum mechanics alone. Nevertheless, it is both of fundamental interest and of practical concern to assess the influence of relativistic features on quantum communication protocols, including QKD. Preliminary work has been performed in this direction providing a positive answer: the QBER is predicted to be affected, in ideal conditions, even in the case of weak gravity as that found on Earth or in neighbouring space.
This conclusion has been obtained in the literature by studying a specific simple protocol 
\cite{Bruschi:Ralph:2014}. There, a set of steps devised to establish a maximally entangled state between two memories located at different heights is executed. The key step is the interference of two identical photons at a $50:50$ beamsplitter, where one photon has been propagating on Earth while the other arrives from the source located somewhere above. If the two photons are not identical, a standard entanglement swapping scheme will not ultimately provide a maximally entangled state shared by the two users, but instead a mixed state where the mixedness depends on the gravitational redshift \cite{Zukowski:Zeilinger:1993,Sen:Sen:2005}. While the effects can be made extremely small by appropriately choosing the free parameters, photons that are defined by a frequency distribution that is extremely peaked can suffer effects that lead to a non-negligible effect on the QBER. It has been shown that for Gaussian-shaped photon frequency distributions with few hundred KHz bandwidths one can expect a QBER$\sim1\%$ purely as a result of gravitational effects.  
Note that, while in general the magnitude of the effect will strongly depend on optimization of all parameters, including the specific choice of protocol, the fact that an effect exists in the first place is due solely to photon mismatch. Therefore, any protocol that requires quantum interference between photons that are produced at different heights in a gravitational field will in general witness an effect. Particular setups, such as those including systems travelling with relative velocity  with respect to each other and thus introducing additional Doppler shifts, can experience a compensation and therefore cancellation of the total effect.

\subsection{Sensing}
In relativistic quantum metrology, we are interested in estimating parameters encoded in the evolution of quantum states of a quantum field as they propagate through curved spacetime. Here, the process in which the measurement of a physical parameter is performed by means of quantum phenomena is referred to as sensing \cite{Degen2017}. The interest in the topic has promoted a growing literature in the field and the development of novel tools such as those briefly mentioned here. The specific aim is to employ the information encoded in the deformation of wave packets of light as the propagate in curved spacetime to perform sensing protocols for the measurement of relevant physical parameters \cite{Bruschi:Datta:2014,Kohlrus:Bruschi:Fuentes:2019}. 

Among the possible schemes and implementations, an interesting case is that consisting of two-mode entangled states that are exchanged between users placed at different locations in curved spacetime with the ambition of estimating relevant parameters, such as the Schwarzschild radius of the Earth or the distance between source and receiver \cite{Bruschi:Datta:2014}. The sender, Alice, prepares and sends a pulse of light to the receiver, Bob, who will compare it with the expected one. Propagation in curved spacetime can then be modelled in first approximation as two beamsplitting operations acting independently on each mode, which is mixed with an orthogonal one that cannot be measured. The problem is thus consisting of four modes $(b_1b_2c_1c_2)$, of which Alice can manipulate only $(b_1b_2)$, i.e., wavepackets that are centred around the frequencies $\omega_1$ and $\omega_2$, for which she prepares a two-mode squeezed state with initial reduced covariance matrix $\boldsymbol{\sigma}^{b_1b_2}(r)=\boldsymbol{S}_{\textrm{TMS}}(r)\boldsymbol{S}^\dag_{\textrm{TMS}}(r)$ defined in \hyperref[exampleCovarianceMatrix]{Box 4}. As presented in the previous section and analogously to \hyperref[example:beamsplitter]{Box 5}, the orthogonal modes $c_1$ and $c_2$ are inaccessible and the effects of propagation between Alice and Bob on the modes $b_1$ and $b_2$ can be modelled as two mode-mixing operations $\boldsymbol{S}_{\textrm{BS}}(\theta)=\boldsymbol{R}(\theta)\oplus\boldsymbol{R}(\theta)$ acting on each pair $b_1,c_1$ and $b_2,c_2$ independently. The angle $\theta$ can (and, in general, will) be different as explained before.

We consider that the orthogonal modes are initially in the vacuum state for simplicity of presentation, which means that their initial reduced covariance matrix $\boldsymbol{\sigma}^{c_1c_2}_0=\mathds{1}_4$ where $\mathbb{1}_4$ is the $4\times4$ identity matrix.  We define our operator vector as $\mathbb{X}:=(\hat{b}_1,\hat{b}_2,\hat{c}_1,\hat{c}_2,\hat{b}^\dag_1,\hat{b}^\dag_2,\hat{c}^\dag_1,\hat{c}^\dag_2)$. Then, the full channel $\boldsymbol{S}^{\textrm{full}}(\theta_1,\theta_2)$ reads in matrix form
\begin{align}
\boldsymbol{S}^{\textrm{full}}(\theta_1,\theta_2) = 
   \begin{pmatrix}    
\cos\theta_1 &0 & \sin\theta_1 & 0 & 0 & 0 & 0 & 0\\
0 &\cos\theta_2 & 0 &\sin\theta_2 & 0 & 0 & 0 & 0\\
-\sin\theta_1 &0 & \cos\theta_1  & 0 & 0 & 0 & 0 & 0\\
0 &-\sin\theta_2 & 0 & \cos\theta_2 & 0 & 0 & 0 & 0\\
0 &0 & 0 & 0 & \cos\theta_1 & 0 & \sin\theta_1 & 0\\
0 &0 & 0 & 0 & 0 & \cos\theta_2 & 0 & \sin\theta_2\\
0 &0 & 0 & 0 & -\sin\theta_1 & 0 & \cos\theta_1 & 0\\
0 &0 & 0 & 0 & 0 & -\sin\theta_2 & 0 & \cos\theta_2
\end{pmatrix},
\end{align}
while the initial and final full covariance matrices $\boldsymbol{\sigma}^{\textrm{full}}(r)$ and $\boldsymbol{\sigma}^{\textrm{full}}_{\theta_1,\theta_2}(r)$ of the system respectively read
\begin{align}
\boldsymbol{\sigma}^{\textrm{full}}(r) =
\begin{pmatrix}    
\cosh(2r)\,\mathds{1}_2 &0 & \sinh(2r)\,\boldsymbol{\sigma}_{\textrm{y}} & 0 \\
0 & \mathds{1}_2 & 0 & 0 \\
\sinh(2r)\,\boldsymbol{\sigma}_{\textrm{y}} &0 & \cosh(2r)\,\mathds{1}_2 & 0 \\
0 &0 & 0 & \mathds{1}_2 
\end{pmatrix},
\quad
\boldsymbol{\sigma}^{\textrm{full}}_{\theta_1,\theta_2}(r)=\boldsymbol{S}^{\textrm{full}}(\theta_1,\theta_2)\boldsymbol{\sigma}^{\textrm{full}}(r)\boldsymbol{S}^{\textrm{full}}{}^\dag(\theta_1,\theta_2).
\end{align}
Here, $\boldsymbol{\sigma}_{\textrm{y}}$ is one of the Pauli matrices. Once the full final state $\boldsymbol{\sigma}^{\textrm{full}}_{\theta_1,\theta_2}(r)$ is computed following this procedure, one traces over the ancillary modes $c_1$ and $c_2$ by simply removing the corresponding rows and columns obtaining the final reduced state $\boldsymbol{\sigma}^{b_1,b_2}_{\theta_1,\theta_2}(r)$ of modes $b_1,b_2$. Finally, we can exploit the fidelity $\mathcal{F}$ in the covariance matrix formalism as defined in \eqref{eq:Fidelity} to find the QFI. This can be done for single-parameter estimation when it occurs that $\theta_1=\theta_2\equiv\theta$. In turn, this provides a bound on the estimation error.

\subsection{Testing fundamental theories of Nature}
As a last broad application of the tools discussed here we mention the possibility of testing our fundamental theories of Nature. General relativity and quantum mechanics in particular are well established within their respective domains of validity. Nevertheless, it is commonly believed that at very high energies, or very small length scales, quantum gravitational effects should appear 
\cite{Bassi:Grossart:2017}. In our work we do not attempt to study such effects but we note that interesting information can still be extracted at the energy and length scales where quantum field theory in curved spacetime is believed to apply. Instead, we offer a brief overview on a few potential applications for testing new physics.

\textit{Testing via interferometry}---In the past few decades it has become evident that interferometric setups have the potential for 
revolutionizing experimental investigation of many phenomena due to the very high precisions that can be reached 
\cite{Peters:Chung:2001,Yang:Zhang:2018}. Interferometers, whether of light or matter, have now been proposed as key instruments in a myriad of experiments, such as 
MAcroscopic Quantum ResOnator (MAQRO) \cite{Kaltenbaek:Aspelmeyer:2016}, Laser Interferometer Space Antenna (LISA) \cite{Amaro-Seoane:Audley:2017}, and Atomic Experiment for Dark Matter and Gravity Exploration in Space (AEDGE) \cite{El-Neaj:Alpigiani:2020}. We believe that interferometric detection of the modifications of the quantum state of light discussed in this work will be the core approach to the different avenues addressed below.

\textit{Quantum field theory in curved spacetime}---Our results depend on the validity of quantum field theory in curved spacetime. Therefore, probing the validity of the transformation \eqref{finite:transformation:representation} for a few modes of light and for different redshifts $\chi$ (i.e., different configurations of the Alice-Bob positioning) can be used to test the theory. Multimode mixing can induce quantum interference 
\cite{Pan:Chen:2012}, and it has even been shown that it can induce Hong-Ou-Mandel-like quantum interference \cite{Bruschi:Schell:2023}. The conditions necessary for such phenomena to manifest can be obtained, for example, by engineering the input modes $\hat{A}_{\omega_n}$ to have multiple peaks that alternate. In principle, such photons could be generated using optical parametric oscillators \cite{Scholz:Koch:2009,Wahl:Roerhlicke:2013}. From the formalism presented here it should also become evident that single bell-shaped modes that do not overlap before the transformation is induced lead to the destruction of the interference effect sought after \cite{Bruschi:Schell:2023}. Successful experimental detection of this quantum interference effect would  support the validity of quantum field theory in curved spacetime, at least in the weak curvature regime (such as that of the Earth). This would improve our current 
understanding of the theory which lacks experimental corroboration regardless of the many signature theoretical predictions put forward \cite{Birrell:Davies:1982,Wald:1995}. 

\textit{Testing Equivalence Principles}---Equivalence Principles are foundational guidelines for the development of theories of gravity \cite{Misner:Thorne:1973,Carroll:2019}. The Einstein Equivalence Principle (EEP) in particular states that the laws of physics in curved spacetime reduce to those of special relativity locally (i.e., in regions of spacetime that are ``small enough'') \cite{Misner:Thorne:1973,Carroll:2019}. 
The EEP is a fundamental statement about the most basic workings of Nature and it is therefore a matter of fundamental interest to know if and in which regimes this principle holds. To date, many experiments have been performed to test the EEP in a classical setup, and new experiments are planned \cite{Altschul:Bailey:2015,Tino:Cacciapuoti:2020,Bassi:Cacciapuoti:2022}. An even more compelling problem is the validity of the EEP in the quantum domain. More concretely, it is of great interest to verify this principle within a framework where both gravitational and quantum mechanical features of physical systems play a role. It is usually implicitly assumed that the EEP does apply. Nevertheless, there are different arguments in favour of testing it for free falling quantum systems, which would be greatly beneficial for our current understanding of quantum systems in gravitational fields \cite{Altschul:Bailey:2015}.

The effects described here can be used to develop new ways of testing the EEP. However, we want to emphasize that they do not solve the problem of the EEP for gravitating quantum matter \cite{Zych:Brukner:2015}. The physical principles described here set themselves apart from many experiments that have been proposed and already performed, since we would not use massive particles \cite{Tino:Cacciapuoti:2020}, but photons (which are modelled as massless excitations of a quantum field). Photons can be engineered to propagate (i.e., ``free fall'') between two users at different heights in the gravitational potential of the Earth or another planet, and the shift that is induced can be measured using interferometric setups \cite{Scully:Zubairy:1997,Barzel:Bruschi:2022}. The degree of control over photons and the high precisions that modern photonics have reached very high levels and this would enable the possiblity to test the universality of the gravitational redshift as a function, for example, of the initial (quantum) state of the photon, of the motion of photons (i.e., of the parameters of the trajectory), and of the polarization. Gravitational redshift can be derived from first principles as a direct consequence of the EEP applied to two accelerated objects that exchange electromagnetic pulses \cite{Carroll:2019}. Therefore, we conclude that our tools can provide yet another way to explore and verify the EEP. 

\textit{Modified and Novel Theories of Gravity}---Advanced and new theories of Nature can predict deviations from those of general relativity to be expected to manifest within specific (usually high-energy or small-scale) regimes. Many proposals have now been put forward to test different aspects of novel physics within space-based experiments 
\cite{Gasbarri:Belenchia:2021,Belenchia:Carlesso:2022}. A key aspect of the space-based setup is that photons that propagate through spacetime can witness deviations from predicted kinematics. For example, these effects might arise as a consequence of asymmetries due to anisotropies of the background spacetimes, of the presence of ultraviolet cutoffs in the field theory or of the coarse graining of spacetime \cite{Bassi:Cacciapuoti:2022}. Regardless of the particular effect that might be of interest, propagation through a long baseline can lead to the cumulation of effects and therefore an overall detectable signal. In general the effects might be extremely small, which is why one expects that large distances must be traversed by photons before a successful measurement can be made. Luckily, as is evident from the fact that the light of distant stars can reach us, achieving (very) long distances for photons for these purposes is possible at least in principle.
 
\textit{Neutrino Physics}---Mode mixing is an ubiquitous operation in quantum optics \cite{Scully:Zubairy:1997}, as well as a key physical phenomenon in many 
areas of physics, such as nanomechanics \cite{Yamaguchi:Hajime:2012}, surface acoustic waves \cite{Tam:Hu:1989} or plasmonics \cite{Lou:Daoxin:2012} to name a few. Neutrino physics is also an area where mode mixing has played a revolutionary role that has changed our understanding of high-energy physical processes. Neutrinos were originally postulated to be massless, a property that was successively found to be incompatible with experiments. In fact, neutrinos were found to ``mix flavours'' \cite{Bellini:Ludhova:2014,Salas:Forero:2020}, and this can be explained through quantum field theory only if they are massive. This phenomenon, known as \textit{neutrino oscillations}\footnote{First postulated theoretically by Italian physicist Bruno Pontecorvo (22 August 1913 -- 24 September 1993).}, is a form of mode mixing during which three distinct operators (one for each \textit{flavour}) are combined unitarily into three new ones \cite{Bellini:Ludhova:2014}. Three-mode mixing, in particular, can be thought of as a \textit{tritter} \cite{Menessen:Jones:2017}.  

\vspace{0.2cm}

\begin{tcolorbox}[colback=orange!3!white,colframe=orange!85!black,title=Neutrino oscillations]
Neutrino physics is a branch of high energy physics \cite{Bellini:Ludhova:2014}. Here we are interested only in the phenomenon called \textit{neutrino oscillations}, which occurs when the neutrinos are massive. In this case, there are three neutrino field operators $\hat{\nu}_k$ that mix via a $3\times3$ unitary matrix $\boldsymbol{U}_{\text{mix}}$, which is a function of three \textit{mixing angles} $\theta_{12}$, $\theta_{23}$ and $\theta_{13}$, as well as a Dirac-phase $\delta$.  
Let $\hat{\mathbb{X}}_{\text{r}}:=(\hat{\nu}_1,\hat{\nu}_2,\hat{\nu}_3)^{\textrm{Tp}}$ and $\hat{\mathbb{X}}_{\text{r}}':=(\hat{\nu}_1',\hat{\nu}_2',\hat{\nu}_3')^{\textrm{Tp}}$, where the unprimed operators $\hat{\nu}_k$ determine the left handed components of the neutrino fields, while the primed operators $\hat{\nu}_k'$ are determined by the \textit{flavour} degrees of freedom \cite{Bellini:Ludhova:2014}. We have 
\begin{align*}
\hat{\mathbb{X}}_{\text{r}}'=\boldsymbol{U}_{\text{mix}}\hat{\mathbb{X}}_{\text{r}},
\quad
\text{where}
\quad
\boldsymbol{U}_{\text{mix}}=
\begin{pmatrix}
c_{12}c_{13} & s_{12}c_{13} & s_{13}e^{-i\delta}\\
-s_{12}c_{23}-c_{12}s_{23}s_{13}e^{i\delta} & c_{12}c_{23}-s_{12}s_{23}s_{13}e^{i\delta} & s_{23}c_{13}\\
s_{12}s_{23}-c_{12}c_{23}s_{13}e^{i\delta} & -c_{12}s_{23}-s_{12}c_{23}s_{13}e^{i\delta} & c_{23}c_{13}
\end{pmatrix}.
\end{align*}
It is immediate to verify that $\boldsymbol{U}_{\text{mix}}\boldsymbol{U}_{\text{mix}}^\dag=\mathds{1}$. This matrix is known as \textit{tritter} in quantum optics \cite{Menessen:Jones:2017}, and is another special case of the general matrix $\boldsymbol{U}$ that appears in (\ref{finite:transformation:representation}).
\end{tcolorbox}

\vspace{0.2cm}

Physical neutrinos, like every other realistic particle, will not have infinitely sharp momentum but will be best characterized by a wavepacket of ideal sharp-momentum field excitations. This will require  updating the mathematical technology developed for ideal sharp-momentum particles in order to take care of all realistic features. Our results can help in addressing some of the issues, including adding the effects of weak gravitational backgrounds on the propagation of the neutrinos as wave-packets.

\textit{Astrophysics}---Finally, we note here that the electromagnetic field is ubiquitous since light and other forms of electromagnetic radiation are constantly propagating in all directions. The sources can be very mundane objects, such as 
man-made devices on Earth, as well as the Sun. Light emitted by impossibly distant objects can also reach the globe that we live on, thus being detected by our telescopes and observatories. Indeed astrophysics and astronomy are largely based on the ability to detect (excitations of) the electromagnetic field in a variety of frequency ranges, from which information is extracted.

The redshift of astrophysical and astronomical entities is well established theoretically. Recently, attention has been turned toward the verification of the predictions by direct analysis of the data \cite{Do:Hees:2019,Gutierrez:Ramos-Chernenko:2022}. Detected light carries crucial information on a wide variety of properties of sources, from mass-to-radius ratio 
\cite{vonHippel:1996,Chandra:Hwang:2020}, to magnetic fields \cite{Mosquera:Salim:2004}. Redshift in this context is usually considered from a classical perspective in the sense of ignoring the quantum properties of light. Nevertheless, as we have argued here, gravitational redshift of realistic (quantum) photons can add significant new possibilities to the measurements performed based on interferometric setups as compared to classical light.
We envision that applying the techniques presented in this work could provide additional channels from which information regarding sources can be extracted.

\section{Considerations and Outlook}\label{section:ConsiderationsnOutlook}
The material presented here is developed as a comprehensive review of the previous work on this topic. To date, the studies reviewed have been purely theoretical in nature, and more work is necessary to fully characterize the overall effect, as well as to include it in a practical scheme where it can be gauged against competing effects. If one does not wish to measure and employ the effect specifically, but cannot fully compensate for it because of external constraints, it would be desirable to embed the effects of the gravitational redshift in an overall expression that includes all relevant sources of noise, loss, etc... One such expression is called the \textit{radar equation}, which has been developed for this purpose 
\cite{Degnan:1993,Vallone:Bacco:2015}. The aim of this quantity is to provide an effective rate of incident photons on the target given a variety of impediments: scattering of photons, turbulence, weather conditions, beam widening, pointing errors to name a few. Effects due to gravitational redshift, in particular when a quantum protocol is employed, could be included in such an expression.

It is also of great interest to actively attempt to measure the effects of gravitational redshift for purposes of foundational experiments or sensing. A promising avenue to implement the tests discussed above is the use of Cubesats and other similar crafts \cite{Saeed:Elzanaty:2020,Araniti:Iera:2022,Bloser:Murphy:2022} that are now being considered for use in space-based classical \cite{Moore:Woods:2016,Shkolnik:2018,Bowman:Vandenbussche:2022} and quantum experiments \cite{Oi:Ling:2017,Joshi:Pneaar:2018,Mazzarella:Lowe:2020}. In this case, small and relatively inexpensive satellites can be deployed at a fraction of the cost of conventional missions, where the craft itself can be potentially loaded with all necessary equipment to perform (reasonable) long-range experiments. One idea is to use a collection of such satellites as sources of photons to be detected on Earth \cite{Belenchia:Carlesso:2022}. These satellites have short lives compared to their traditional counterparts but can allow for proof-of-principle experiments to be performed, which can in turn be used to support planning and development of larger scale missions.

Finally, states that exhibit quantum coherence can be used as resources for quantum computing \cite{Nielsen:Chuang:2010}. It remains an open question how this final aspect can be used constructively in concrete applications.

\section{Conclusions}\label{section:Conclusions}
We have reviewed an approach to study photons that propagate in curved spacetime between a sender and a receiver. Our main objective was to provide the formalism that allows to model the transformation of the photon spectrum distribution, as well as its quantum state, when the two users witness differing local gravitational effects. We have reviewed the existing work that initiated this approach and we extended it to include $3+1$-dimensions. We have focused on photons that are strongly confined along the direction of propagation, for which the evolution of the wavepackets can be significantly simplified. It is possible to effectively consider a one dimensional problem for all purposes, which allows to model the photon spectrum using the frequency variable alone. In particular, it has been found that the gravitational redshift present deforms the wavepacket in a non-trivial way, meaning that a rigid shift of the whole spectrum occurs together with a deformation.
It has also been shown that the quantum state of the photon is effectively subject to a mode-mixing operation that exchanges excitations between different modes of the electromagnetic field. The intensity of this mixing depends directly on the strength of the gravitational redshift. Particle creation due to dynamical backgrounds or other effects has been neglected, although it can be included if required by the specific scenario. Such addition is expected to significantly increase the complexity of the analysis. Furthermore, we have discussed potential applications of the predictions of this work, such as sensing and changes in the QBER of a QKD protocol, as well as potential novel insights to be gained in advanced modern theories, such as modified theories of gravity and neutrino physics. We leave it to future work to explore such avenues.
Finally, we note that testing the existence of this effect can contribute in demonstrating quantum field theory in (weakly) curved spacetime, and therefore add new insight to our understanding of physics at the overlap of quantum mechanics and general relativity.

\section*{Contribution statement}
DEB conceived the work and elaborated the bulk of the material. AWS oversaw the experimental aspects of the claims. LAA produced the figures and contributed towards elaboration of the scientific material. All authors contributed to writing the manuscript.

\acknowledgments
We thank Timothy C. Ralph, Leila Khouri, Valente Pranubon, Thomas Mieling and Claus Kiefer for useful comments and discussions. We extend special thanks to Mohsen Esmail-Zadeh for questions on the nature of the phases and mode functions in the frequency distribution, and to Jorma Louko for  questions on the correct approach on how to decompose the propagation of the photon in curved spacetime.


\bibliographystyle{apsrev4-2}
\bibliography{ProceedQuaGraRedBib}

\begin{thebibliography}{176}%
\makeatletter
\providecommand \@ifxundefined [1]{%
 \@ifx{#1\undefined}
}%
\providecommand \@ifnum [1]{%
 \ifnum #1\expandafter \@firstoftwo
 \else \expandafter \@secondoftwo
 \fi
}%
\providecommand \@ifx [1]{%
 \ifx #1\expandafter \@firstoftwo
 \else \expandafter \@secondoftwo
 \fi
}%
\providecommand \natexlab [1]{#1}%
\providecommand \enquote  [1]{``#1''}%
\providecommand \bibnamefont  [1]{#1}%
\providecommand \bibfnamefont [1]{#1}%
\providecommand \citenamefont [1]{#1}%
\providecommand \href@noop [0]{\@secondoftwo}%
\providecommand \href [0]{\begingroup \@sanitize@url \@href}%
\providecommand \@href[1]{\@@startlink{#1}\@@href}%
\providecommand \@@href[1]{\endgroup#1\@@endlink}%
\providecommand \@sanitize@url [0]{\catcode `\\12\catcode `\$12\catcode
  `\&12\catcode `\#12\catcode `\^12\catcode `\_12\catcode `\%12\relax}%
\providecommand \@@startlink[1]{}%
\providecommand \@@endlink[0]{}%
\providecommand \url  [0]{\begingroup\@sanitize@url \@url }%
\providecommand \@url [1]{\endgroup\@href {#1}{\urlprefix }}%
\providecommand \urlprefix  [0]{URL }%
\providecommand \Eprint [0]{\href }%
\providecommand \doibase [0]{https://doi.org/}%
\providecommand \selectlanguage [0]{\@gobble}%
\providecommand \bibinfo  [0]{\@secondoftwo}%
\providecommand \bibfield  [0]{\@secondoftwo}%
\providecommand \translation [1]{[#1]}%
\providecommand \BibitemOpen [0]{}%
\providecommand \bibitemStop [0]{}%
\providecommand \bibitemNoStop [0]{.\EOS\space}%
\providecommand \EOS [0]{\spacefactor3000\relax}%
\providecommand \BibitemShut  [1]{\csname bibitem#1\endcsname}%
\let\auto@bib@innerbib\@empty
\bibitem [{\citenamefont {Einstein}(1916)}]{Einstein:1916}%
  \BibitemOpen
  \bibfield  {author} {\bibinfo {author} {\bibfnamefont {A.}~\bibnamefont
  {Einstein}},\ }\href
  {https://doi.org/https://doi.org/10.1002/andp.19163540702} {\bibfield
  {journal} {\bibinfo  {journal} {Annalen der Physik}\ }\textbf {\bibinfo
  {volume} {354}},\ \bibinfo {pages} {769} (\bibinfo {year} {1916})},\ \Eprint
  {https://arxiv.org/abs/https://onlinelibrary.wiley.com/doi/pdf/10.1002/andp.19163540702}
  {https://onlinelibrary.wiley.com/doi/pdf/10.1002/andp.19163540702}
  \BibitemShut {NoStop}%
\bibitem [{\citenamefont {Earman}\ and\ \citenamefont
  {Glymour}(1980)}]{Earman:Glymour:1980}%
  \BibitemOpen
  \bibfield  {author} {\bibinfo {author} {\bibfnamefont {J.}~\bibnamefont
  {Earman}}\ and\ \bibinfo {author} {\bibfnamefont {C.}~\bibnamefont
  {Glymour}},\ }\href
  {https://doi.org/https://doi.org/10.1016/0039-3681(80)90025-4} {\bibfield
  {journal} {\bibinfo  {journal} {Studies in History and Philosophy of Science
  Part A}\ }\textbf {\bibinfo {volume} {11}},\ \bibinfo {pages} {175} (\bibinfo
  {year} {1980})}\BibitemShut {NoStop}%
\bibitem [{\citenamefont {Wald}(1984)}]{Wald:1984}%
  \BibitemOpen
  \bibfield  {author} {\bibinfo {author} {\bibfnamefont {R.~M.}\ \bibnamefont
  {Wald}},\ }\href@noop {} {\emph {\bibinfo {title} {{General relativity}}}}\
  (\bibinfo  {publisher} {The University of Chicago Press},\ \bibinfo {address}
  {Chicago, IL},\ \bibinfo {year} {1984})\BibitemShut {NoStop}%
\bibitem [{\citenamefont {Carroll}(2019)}]{Carroll:2019}%
  \BibitemOpen
  \bibfield  {author} {\bibinfo {author} {\bibfnamefont {S.~M.}\ \bibnamefont
  {Carroll}},\ }\href {https://doi.org/10.1017/9781108770385} {\emph {\bibinfo
  {title} {{Spacetime and Geometry: An Introduction to General Relativity}}}}\
  (\bibinfo  {publisher} {Cambridge University Press},\ \bibinfo {year}
  {2019})\BibitemShut {NoStop}%
\bibitem [{\citenamefont {{Einstein}}(1915)}]{Einstein:1915}%
  \BibitemOpen
  \bibfield  {author} {\bibinfo {author} {\bibfnamefont {A.}~\bibnamefont
  {{Einstein}}},\ }\href@noop {} {\bibfield  {journal} {\bibinfo  {journal}
  {Sitzungsberichte der K\&ouml;niglich Preussischen Akademie der
  Wissenschaften}\ ,\ \bibinfo {pages} {844}} (\bibinfo {year}
  {1915})}\BibitemShut {NoStop}%
\bibitem [{\citenamefont {Pound}\ and\ \citenamefont
  {Rebka}(1959)}]{Pound:Rebka:1959}%
  \BibitemOpen
  \bibfield  {author} {\bibinfo {author} {\bibfnamefont {R.~V.}\ \bibnamefont
  {Pound}}\ and\ \bibinfo {author} {\bibfnamefont {G.~A.}\ \bibnamefont
  {Rebka}},\ }\href {https://doi.org/10.1103/PhysRevLett.3.439} {\bibfield
  {journal} {\bibinfo  {journal} {Phys. Rev. Lett.}\ }\textbf {\bibinfo
  {volume} {3}},\ \bibinfo {pages} {439} (\bibinfo {year} {1959})}\BibitemShut
  {NoStop}%
\bibitem [{\citenamefont {Pound}\ and\ \citenamefont
  {Snider}(1964)}]{Pound:Snider:1964}%
  \BibitemOpen
  \bibfield  {author} {\bibinfo {author} {\bibfnamefont {R.~V.}\ \bibnamefont
  {Pound}}\ and\ \bibinfo {author} {\bibfnamefont {J.~L.}\ \bibnamefont
  {Snider}},\ }\href {https://doi.org/10.1103/PhysRevLett.13.539} {\bibfield
  {journal} {\bibinfo  {journal} {Phys. Rev. Lett.}\ }\textbf {\bibinfo
  {volume} {13}},\ \bibinfo {pages} {539} (\bibinfo {year} {1964})}\BibitemShut
  {NoStop}%
\bibitem [{\citenamefont {{M\"uller, A.}}\ and\ \citenamefont {{Wold,
  M.}}(2006)}]{Mueller:Wold:2006}%
  \BibitemOpen
  \bibfield  {author} {\bibinfo {author} {\bibnamefont {{M\"uller, A.}}}\ and\
  \bibinfo {author} {\bibnamefont {{Wold, M.}}},\ }\href
  {https://doi.org/10.1051/0004-6361:20065615} {\bibfield  {journal} {\bibinfo
  {journal} {A\&A}\ }\textbf {\bibinfo {volume} {457}},\ \bibinfo {pages} {485}
  (\bibinfo {year} {2006})}\BibitemShut {NoStop}%
\bibitem [{\citenamefont {Chou}\ \emph {et~al.}(2010)\citenamefont {Chou},
  \citenamefont {Hume}, \citenamefont {Rosenband},\ and\ \citenamefont
  {Wineland}}]{Chou:Hume:2010}%
  \BibitemOpen
  \bibfield  {author} {\bibinfo {author} {\bibfnamefont {C.~W.}\ \bibnamefont
  {Chou}}, \bibinfo {author} {\bibfnamefont {D.~B.}\ \bibnamefont {Hume}},
  \bibinfo {author} {\bibfnamefont {T.}~\bibnamefont {Rosenband}},\ and\
  \bibinfo {author} {\bibfnamefont {D.~J.}\ \bibnamefont {Wineland}},\ }\href
  {https://doi.org/10.1126/science.1192720} {\bibfield  {journal} {\bibinfo
  {journal} {Science}\ }\textbf {\bibinfo {volume} {329}},\ \bibinfo {pages}
  {1630} (\bibinfo {year} {2010})}\BibitemShut {NoStop}%
\bibitem [{\citenamefont {M{\"u}ller}\ \emph {et~al.}(2010)\citenamefont
  {M{\"u}ller}, \citenamefont {Peters},\ and\ \citenamefont
  {Chu}}]{Mueller:Peters:2010}%
  \BibitemOpen
  \bibfield  {author} {\bibinfo {author} {\bibfnamefont {H.}~\bibnamefont
  {M{\"u}ller}}, \bibinfo {author} {\bibfnamefont {A.}~\bibnamefont {Peters}},\
  and\ \bibinfo {author} {\bibfnamefont {S.}~\bibnamefont {Chu}},\ }\href
  {https://doi.org/10.1038/nature08776} {\bibfield  {journal} {\bibinfo
  {journal} {Nature}\ }\textbf {\bibinfo {volume} {463}},\ \bibinfo {pages}
  {926} (\bibinfo {year} {2010})}\BibitemShut {NoStop}%
\bibitem [{\citenamefont {Will}(2014)}]{Clifford:2014}%
  \BibitemOpen
  \bibfield  {author} {\bibinfo {author} {\bibfnamefont {C.~M.}\ \bibnamefont
  {Will}},\ }\href {https://doi.org/10.12942/lrr-2014-4} {\bibfield  {journal}
  {\bibinfo  {journal} {Living Reviews in Relativity}\ }\textbf {\bibinfo
  {volume} {17}},\ \bibinfo {pages} {4} (\bibinfo {year} {2014})}\BibitemShut
  {NoStop}%
\bibitem [{\citenamefont {Herrmann}\ \emph {et~al.}(2018)\citenamefont
  {Herrmann}, \citenamefont {Finke}, \citenamefont {L\"ulf}, \citenamefont
  {Kichakova}, \citenamefont {Puetzfeld}, \citenamefont {Knickmann},
  \citenamefont {List}, \citenamefont {Rievers}, \citenamefont {Giorgi},
  \citenamefont {G\"unther}, \citenamefont {Dittus}, \citenamefont
  {Prieto-Cerdeira}, \citenamefont {Dilssner}, \citenamefont {Gonzalez},
  \citenamefont {Sch\"onemann}, \citenamefont {Ventura-Traveset},\ and\
  \citenamefont {L\"ammerzahl}}]{Herrmann:Finke:2018}%
  \BibitemOpen
  \bibfield  {author} {\bibinfo {author} {\bibfnamefont {S.}~\bibnamefont
  {Herrmann}}, \bibinfo {author} {\bibfnamefont {F.}~\bibnamefont {Finke}},
  \bibinfo {author} {\bibfnamefont {M.}~\bibnamefont {L\"ulf}}, \bibinfo
  {author} {\bibfnamefont {O.}~\bibnamefont {Kichakova}}, \bibinfo {author}
  {\bibfnamefont {D.}~\bibnamefont {Puetzfeld}}, \bibinfo {author}
  {\bibfnamefont {D.}~\bibnamefont {Knickmann}}, \bibinfo {author}
  {\bibfnamefont {M.}~\bibnamefont {List}}, \bibinfo {author} {\bibfnamefont
  {B.}~\bibnamefont {Rievers}}, \bibinfo {author} {\bibfnamefont
  {G.}~\bibnamefont {Giorgi}}, \bibinfo {author} {\bibfnamefont
  {C.}~\bibnamefont {G\"unther}}, \bibinfo {author} {\bibfnamefont
  {H.}~\bibnamefont {Dittus}}, \bibinfo {author} {\bibfnamefont
  {R.}~\bibnamefont {Prieto-Cerdeira}}, \bibinfo {author} {\bibfnamefont
  {F.}~\bibnamefont {Dilssner}}, \bibinfo {author} {\bibfnamefont
  {F.}~\bibnamefont {Gonzalez}}, \bibinfo {author} {\bibfnamefont
  {E.}~\bibnamefont {Sch\"onemann}}, \bibinfo {author} {\bibfnamefont
  {J.}~\bibnamefont {Ventura-Traveset}},\ and\ \bibinfo {author} {\bibfnamefont
  {C.}~\bibnamefont {L\"ammerzahl}},\ }\href
  {https://doi.org/10.1103/PhysRevLett.121.231102} {\bibfield  {journal}
  {\bibinfo  {journal} {Phys. Rev. Lett.}\ }\textbf {\bibinfo {volume} {121}},\
  \bibinfo {pages} {231102} (\bibinfo {year} {2018})}\BibitemShut {NoStop}%
\bibitem [{\citenamefont {Litvinov}\ \emph {et~al.}(2018)\citenamefont
  {Litvinov}, \citenamefont {Rudenko}, \citenamefont {Alakoz}, \citenamefont
  {Bach}, \citenamefont {Bartel}, \citenamefont {Belonenko}, \citenamefont
  {Belousov}, \citenamefont {Bietenholz}, \citenamefont {Biriukov},
  \citenamefont {Carman} \emph {et~al.}}]{Litvinov:Rudenko:2018}%
  \BibitemOpen
  \bibfield  {author} {\bibinfo {author} {\bibfnamefont {D.}~\bibnamefont
  {Litvinov}}, \bibinfo {author} {\bibfnamefont {V.}~\bibnamefont {Rudenko}},
  \bibinfo {author} {\bibfnamefont {A.}~\bibnamefont {Alakoz}}, \bibinfo
  {author} {\bibfnamefont {U.}~\bibnamefont {Bach}}, \bibinfo {author}
  {\bibfnamefont {N.}~\bibnamefont {Bartel}}, \bibinfo {author} {\bibfnamefont
  {A.}~\bibnamefont {Belonenko}}, \bibinfo {author} {\bibfnamefont
  {K.}~\bibnamefont {Belousov}}, \bibinfo {author} {\bibfnamefont
  {M.}~\bibnamefont {Bietenholz}}, \bibinfo {author} {\bibfnamefont
  {A.}~\bibnamefont {Biriukov}}, \bibinfo {author} {\bibfnamefont
  {R.}~\bibnamefont {Carman}}, \emph {et~al.},\ }\href
  {https://doi.org/https://doi.org/10.1016/j.physleta.2017.09.014} {\bibfield
  {journal} {\bibinfo  {journal} {Physics Letters A}\ }\textbf {\bibinfo
  {volume} {382}},\ \bibinfo {pages} {2192} (\bibinfo {year}
  {2018})}\BibitemShut {NoStop}%
\bibitem [{\citenamefont {Delva}\ \emph {et~al.}(2018)\citenamefont {Delva},
  \citenamefont {Puchades}, \citenamefont {Sch\"onemann}, \citenamefont
  {Dilssner}, \citenamefont {Courde}, \citenamefont {Bertone}, \citenamefont
  {Gonzalez}, \citenamefont {Hees}, \citenamefont {Le~Poncin-Lafitte},
  \citenamefont {Meynadier}, \citenamefont {Prieto-Cerdeira}, \citenamefont
  {Sohet}, \citenamefont {Ventura-Traveset},\ and\ \citenamefont
  {Wolf}}]{Delva:Puchades:2018}%
  \BibitemOpen
  \bibfield  {author} {\bibinfo {author} {\bibfnamefont {P.}~\bibnamefont
  {Delva}}, \bibinfo {author} {\bibfnamefont {N.}~\bibnamefont {Puchades}},
  \bibinfo {author} {\bibfnamefont {E.}~\bibnamefont {Sch\"onemann}}, \bibinfo
  {author} {\bibfnamefont {F.}~\bibnamefont {Dilssner}}, \bibinfo {author}
  {\bibfnamefont {C.}~\bibnamefont {Courde}}, \bibinfo {author} {\bibfnamefont
  {S.}~\bibnamefont {Bertone}}, \bibinfo {author} {\bibfnamefont
  {F.}~\bibnamefont {Gonzalez}}, \bibinfo {author} {\bibfnamefont
  {A.}~\bibnamefont {Hees}}, \bibinfo {author} {\bibfnamefont {C.}~\bibnamefont
  {Le~Poncin-Lafitte}}, \bibinfo {author} {\bibfnamefont {F.}~\bibnamefont
  {Meynadier}}, \bibinfo {author} {\bibfnamefont {R.}~\bibnamefont
  {Prieto-Cerdeira}}, \bibinfo {author} {\bibfnamefont {B.}~\bibnamefont
  {Sohet}}, \bibinfo {author} {\bibfnamefont {J.}~\bibnamefont
  {Ventura-Traveset}},\ and\ \bibinfo {author} {\bibfnamefont {P.}~\bibnamefont
  {Wolf}},\ }\href {https://doi.org/10.1103/PhysRevLett.121.231101} {\bibfield
  {journal} {\bibinfo  {journal} {Phys. Rev. Lett.}\ }\textbf {\bibinfo
  {volume} {121}},\ \bibinfo {pages} {231101} (\bibinfo {year}
  {2018})}\BibitemShut {NoStop}%
\bibitem [{\citenamefont {Di~Pumpo}\ \emph {et~al.}(2021)\citenamefont
  {Di~Pumpo}, \citenamefont {Ufrecht}, \citenamefont {Friedrich}, \citenamefont
  {Giese}, \citenamefont {Schleich},\ and\ \citenamefont
  {Unruh}}]{DiPumpo:Ufrecht:2021}%
  \BibitemOpen
  \bibfield  {author} {\bibinfo {author} {\bibfnamefont {F.}~\bibnamefont
  {Di~Pumpo}}, \bibinfo {author} {\bibfnamefont {C.}~\bibnamefont {Ufrecht}},
  \bibinfo {author} {\bibfnamefont {A.}~\bibnamefont {Friedrich}}, \bibinfo
  {author} {\bibfnamefont {E.}~\bibnamefont {Giese}}, \bibinfo {author}
  {\bibfnamefont {W.~P.}\ \bibnamefont {Schleich}},\ and\ \bibinfo {author}
  {\bibfnamefont {W.~G.}\ \bibnamefont {Unruh}},\ }\href
  {https://doi.org/10.1103/PRXQuantum.2.040333} {\bibfield  {journal} {\bibinfo
   {journal} {PRX Quantum}\ }\textbf {\bibinfo {volume} {2}},\ \bibinfo {pages}
  {040333} (\bibinfo {year} {2021})}\BibitemShut {NoStop}%
\bibitem [{\citenamefont {Bothwell}\ \emph {et~al.}(2022)\citenamefont
  {Bothwell}, \citenamefont {Kennedy}, \citenamefont {Aeppli}, \citenamefont
  {Kedar}, \citenamefont {Robinson}, \citenamefont {Oelker}, \citenamefont
  {Staron},\ and\ \citenamefont {Ye}}]{Bothwell:Kennedy:2022}%
  \BibitemOpen
  \bibfield  {author} {\bibinfo {author} {\bibfnamefont {T.}~\bibnamefont
  {Bothwell}}, \bibinfo {author} {\bibfnamefont {C.~J.}\ \bibnamefont
  {Kennedy}}, \bibinfo {author} {\bibfnamefont {A.}~\bibnamefont {Aeppli}},
  \bibinfo {author} {\bibfnamefont {D.}~\bibnamefont {Kedar}}, \bibinfo
  {author} {\bibfnamefont {J.~M.}\ \bibnamefont {Robinson}}, \bibinfo {author}
  {\bibfnamefont {E.}~\bibnamefont {Oelker}}, \bibinfo {author} {\bibfnamefont
  {A.}~\bibnamefont {Staron}},\ and\ \bibinfo {author} {\bibfnamefont
  {J.}~\bibnamefont {Ye}},\ }\href {https://doi.org/10.1038/s41586-021-04349-7}
  {\bibfield  {journal} {\bibinfo  {journal} {Nature}\ }\textbf {\bibinfo
  {volume} {602}},\ \bibinfo {pages} {420} (\bibinfo {year}
  {2022})}\BibitemShut {NoStop}%
\bibitem [{\citenamefont {Adams}(1925)}]{Adams:1925}%
  \BibitemOpen
  \bibfield  {author} {\bibinfo {author} {\bibfnamefont {W.~S.}\ \bibnamefont
  {Adams}},\ }\href {https://doi.org/10.1073/pnas.11.7.382} {\bibfield
  {journal} {\bibinfo  {journal} {Proceedings of the National Academy of
  Sciences}\ }\textbf {\bibinfo {volume} {11}},\ \bibinfo {pages} {382}
  (\bibinfo {year} {1925})},\ \Eprint
  {https://arxiv.org/abs/https://www.pnas.org/doi/pdf/10.1073/pnas.11.7.382}
  {https://www.pnas.org/doi/pdf/10.1073/pnas.11.7.382} \BibitemShut {NoStop}%
\bibitem [{\citenamefont {von Hippel}(1996)}]{vonHippel:1996}%
  \BibitemOpen
  \bibfield  {author} {\bibinfo {author} {\bibfnamefont {T.}~\bibnamefont {von
  Hippel}},\ }\href {https://doi.org/10.1086/309911} {\bibfield  {journal}
  {\bibinfo  {journal} {The Astrophysical Journal}\ }\textbf {\bibinfo {volume}
  {458}},\ \bibinfo {pages} {L37} (\bibinfo {year} {1996})}\BibitemShut
  {NoStop}%
\bibitem [{\citenamefont {Schunck}\ and\ \citenamefont
  {Liddle}(1997)}]{Schunck:Liddle:1997}%
  \BibitemOpen
  \bibfield  {author} {\bibinfo {author} {\bibfnamefont {F.~E.}\ \bibnamefont
  {Schunck}}\ and\ \bibinfo {author} {\bibfnamefont {A.~R.}\ \bibnamefont
  {Liddle}},\ }\href
  {https://doi.org/https://doi.org/10.1016/S0370-2693(97)00559-5} {\bibfield
  {journal} {\bibinfo  {journal} {Physics Letters B}\ }\textbf {\bibinfo
  {volume} {404}},\ \bibinfo {pages} {25} (\bibinfo {year} {1997})}\BibitemShut
  {NoStop}%
\bibitem [{\citenamefont {Cottam}\ \emph {et~al.}(2002)\citenamefont {Cottam},
  \citenamefont {Paerels},\ and\ \citenamefont {Mendez}}]{Cottam:Pearels:2002}%
  \BibitemOpen
  \bibfield  {author} {\bibinfo {author} {\bibfnamefont {J.}~\bibnamefont
  {Cottam}}, \bibinfo {author} {\bibfnamefont {F.}~\bibnamefont {Paerels}},\
  and\ \bibinfo {author} {\bibfnamefont {M.}~\bibnamefont {Mendez}},\ }\href
  {https://doi.org/10.1038/nature01159} {\bibfield  {journal} {\bibinfo
  {journal} {Nature}\ }\textbf {\bibinfo {volume} {420}},\ \bibinfo {pages}
  {51} (\bibinfo {year} {2002})}\BibitemShut {NoStop}%
\bibitem [{\citenamefont {Falcon}\ \emph {et~al.}(2010)\citenamefont {Falcon},
  \citenamefont {Winget}, \citenamefont {Montgomery},\ and\ \citenamefont
  {Williams}}]{Falcon:Winget:2010}%
  \BibitemOpen
  \bibfield  {author} {\bibinfo {author} {\bibfnamefont {R.~E.}\ \bibnamefont
  {Falcon}}, \bibinfo {author} {\bibfnamefont {D.~E.}\ \bibnamefont {Winget}},
  \bibinfo {author} {\bibfnamefont {M.~H.}\ \bibnamefont {Montgomery}},\ and\
  \bibinfo {author} {\bibfnamefont {K.~A.}\ \bibnamefont {Williams}},\ }\href
  {https://doi.org/10.1088/0004-637X/712/1/585} {\bibfield  {journal} {\bibinfo
   {journal} {The Astrophysical Journal}\ }\textbf {\bibinfo {volume} {712}},\
  \bibinfo {pages} {585} (\bibinfo {year} {2010})}\BibitemShut {NoStop}%
\bibitem [{\citenamefont {{Pasquini, L.}}\ \emph {et~al.}(2011)\citenamefont
  {{Pasquini, L.}}, \citenamefont {{Melo, C.}}, \citenamefont {{Chavero, C.}},
  \citenamefont {{Dravins, D.}}, \citenamefont {{Ludwig, H.-G.}}, \citenamefont
  {{Bonifacio, P.}},\ and\ \citenamefont {{De La Reza,
  R.}}}]{Pasquini:Melo:2011}%
  \BibitemOpen
  \bibfield  {author} {\bibinfo {author} {\bibnamefont {{Pasquini, L.}}},
  \bibinfo {author} {\bibnamefont {{Melo, C.}}}, \bibinfo {author}
  {\bibnamefont {{Chavero, C.}}}, \bibinfo {author} {\bibnamefont {{Dravins,
  D.}}}, \bibinfo {author} {\bibnamefont {{Ludwig, H.-G.}}}, \bibinfo {author}
  {\bibnamefont {{Bonifacio, P.}}},\ and\ \bibinfo {author} {\bibnamefont {{De
  La Reza, R.}}},\ }\href {https://doi.org/10.1051/0004-6361/201015337}
  {\bibfield  {journal} {\bibinfo  {journal} {A\&A}\ }\textbf {\bibinfo
  {volume} {526}},\ \bibinfo {pages} {A127} (\bibinfo {year}
  {2011})}\BibitemShut {NoStop}%
\bibitem [{\citenamefont {{GRAVITY Collaboration}}\ \emph
  {et~al.}(2018)\citenamefont {{GRAVITY Collaboration}}, \citenamefont
  {{Abuter, R.}}, \citenamefont {{Amorim, A.}}, \citenamefont {{Anugu, N.}},
  \citenamefont {{Baub\"ock, M.}}, \citenamefont {{Benisty, M.}}, \citenamefont
  {{Berger, J. P.}}, \citenamefont {{Blind, N.}}, \citenamefont {{Bonnet, H.}},
  \citenamefont {{Brandner, W.}}, \citenamefont {{Buron, A.}}, \citenamefont
  {{Collin, C.}} \emph {et~al.}}]{Abuter:Amorim:2018}%
  \BibitemOpen
  \bibfield  {author} {\bibinfo {author} {\bibnamefont {{GRAVITY
  Collaboration}}}, \bibinfo {author} {\bibnamefont {{Abuter, R.}}}, \bibinfo
  {author} {\bibnamefont {{Amorim, A.}}}, \bibinfo {author} {\bibnamefont
  {{Anugu, N.}}}, \bibinfo {author} {\bibnamefont {{Baub\"ock, M.}}}, \bibinfo
  {author} {\bibnamefont {{Benisty, M.}}}, \bibinfo {author} {\bibnamefont
  {{Berger, J. P.}}}, \bibinfo {author} {\bibnamefont {{Blind, N.}}}, \bibinfo
  {author} {\bibnamefont {{Bonnet, H.}}}, \bibinfo {author} {\bibnamefont
  {{Brandner, W.}}}, \bibinfo {author} {\bibnamefont {{Buron, A.}}}, \bibinfo
  {author} {\bibnamefont {{Collin, C.}}}, \emph {et~al.},\ }\href
  {https://doi.org/10.1051/0004-6361/201833718} {\bibfield  {journal} {\bibinfo
   {journal} {A\&A}\ }\textbf {\bibinfo {volume} {615}},\ \bibinfo {pages}
  {L15} (\bibinfo {year} {2018})}\BibitemShut {NoStop}%
\bibitem [{\citenamefont {Chandra}\ \emph {et~al.}(2020)\citenamefont
  {Chandra}, \citenamefont {Hwang}, \citenamefont {Zakamska},\ and\
  \citenamefont {Cheng}}]{Chandra:Hwang:2020}%
  \BibitemOpen
  \bibfield  {author} {\bibinfo {author} {\bibfnamefont {V.}~\bibnamefont
  {Chandra}}, \bibinfo {author} {\bibfnamefont {H.-C.}\ \bibnamefont {Hwang}},
  \bibinfo {author} {\bibfnamefont {N.~L.}\ \bibnamefont {Zakamska}},\ and\
  \bibinfo {author} {\bibfnamefont {S.}~\bibnamefont {Cheng}},\ }\href
  {https://doi.org/10.3847/1538-4357/aba8a2} {\bibfield  {journal} {\bibinfo
  {journal} {The Astrophysical Journal}\ }\textbf {\bibinfo {volume} {899}},\
  \bibinfo {pages} {146} (\bibinfo {year} {2020})}\BibitemShut {NoStop}%
\bibitem [{\citenamefont {El-Badry}(2022)}]{El-Badry:2022}%
  \BibitemOpen
  \bibfield  {author} {\bibinfo {author} {\bibfnamefont {K.}~\bibnamefont
  {El-Badry}},\ }\href {https://doi.org/10.3847/2515-5172/ac7c16} {\bibfield
  {journal} {\bibinfo  {journal} {Research Notes of the AAS}\ }\textbf
  {\bibinfo {volume} {6}},\ \bibinfo {pages} {137} (\bibinfo {year}
  {2022})}\BibitemShut {NoStop}%
\bibitem [{\citenamefont {Nu{\~n}ez}\ and\ \citenamefont
  {Nowakowski}(2010)}]{Nunez:Nowakowski:2010}%
  \BibitemOpen
  \bibfield  {author} {\bibinfo {author} {\bibfnamefont {P.~D.}\ \bibnamefont
  {Nu{\~n}ez}}\ and\ \bibinfo {author} {\bibfnamefont {M.}~\bibnamefont
  {Nowakowski}},\ }\href {https://doi.org/10.1007/s12036-010-0006-9} {\bibfield
   {journal} {\bibinfo  {journal} {Journal of Astrophysics and Astronomy}\
  }\textbf {\bibinfo {volume} {31}},\ \bibinfo {pages} {105} (\bibinfo {year}
  {2010})}\BibitemShut {NoStop}%
\bibitem [{\citenamefont {Peres}\ and\ \citenamefont
  {Terno}(2004)}]{Peres:Terno:2004}%
  \BibitemOpen
  \bibfield  {author} {\bibinfo {author} {\bibfnamefont {A.}~\bibnamefont
  {Peres}}\ and\ \bibinfo {author} {\bibfnamefont {D.~R.}\ \bibnamefont
  {Terno}},\ }\href {https://doi.org/10.1103/RevModPhys.76.93} {\bibfield
  {journal} {\bibinfo  {journal} {Reviews of Modern Physics}\ }\textbf
  {\bibinfo {volume} {76}},\ \bibinfo {pages} {93} (\bibinfo {year} {2004})},\
  \bibinfo {note} {publisher: American Physical Society}\BibitemShut {NoStop}%
\bibitem [{\citenamefont {Hohensee}\ \emph {et~al.}(2011)\citenamefont
  {Hohensee}, \citenamefont {Chu}, \citenamefont {Peters},\ and\ \citenamefont
  {M\"uller}}]{Hohensee:Chu:2011}%
  \BibitemOpen
  \bibfield  {author} {\bibinfo {author} {\bibfnamefont {M.~A.}\ \bibnamefont
  {Hohensee}}, \bibinfo {author} {\bibfnamefont {S.}~\bibnamefont {Chu}},
  \bibinfo {author} {\bibfnamefont {A.}~\bibnamefont {Peters}},\ and\ \bibinfo
  {author} {\bibfnamefont {H.}~\bibnamefont {M\"uller}},\ }\href
  {https://doi.org/10.1103/PhysRevLett.106.151102} {\bibfield  {journal}
  {\bibinfo  {journal} {Phys. Rev. Lett.}\ }\textbf {\bibinfo {volume} {106}},\
  \bibinfo {pages} {151102} (\bibinfo {year} {2011})}\BibitemShut {NoStop}%
\bibitem [{\citenamefont {FLORIDES}(2002)}]{Florides:2002}%
  \BibitemOpen
  \bibfield  {author} {\bibinfo {author} {\bibfnamefont {P.~S.}\ \bibnamefont
  {FLORIDES}},\ }\href {https://doi.org/10.1142/S0217751X02011862} {\bibfield
  {journal} {\bibinfo  {journal} {International Journal of Modern Physics A}\
  }\textbf {\bibinfo {volume} {17}},\ \bibinfo {pages} {2759} (\bibinfo {year}
  {2002})},\ \Eprint
  {https://arxiv.org/abs/https://doi.org/10.1142/S0217751X02011862}
  {https://doi.org/10.1142/S0217751X02011862} \BibitemShut {NoStop}%
\bibitem [{\citenamefont {Okun}(2000)}]{Okun:2000}%
  \BibitemOpen
  \bibfield  {author} {\bibinfo {author} {\bibfnamefont {L.~B.}\ \bibnamefont
  {Okun}},\ }\href {https://doi.org/10.1142/S0217732300002358} {\bibfield
  {journal} {\bibinfo  {journal} {Modern Physics Letters A}\ }\textbf {\bibinfo
  {volume} {15}},\ \bibinfo {pages} {1941} (\bibinfo {year}
  {2000})}\BibitemShut {NoStop}%
\bibitem [{\citenamefont {Okun}\ \emph {et~al.}(2000)\citenamefont {Okun},
  \citenamefont {Selivanov},\ and\ \citenamefont
  {Telegdi}}]{Okun:Selivanov:2000}%
  \BibitemOpen
  \bibfield  {author} {\bibinfo {author} {\bibfnamefont {L.~B.}\ \bibnamefont
  {Okun}}, \bibinfo {author} {\bibfnamefont {K.~G.}\ \bibnamefont
  {Selivanov}},\ and\ \bibinfo {author} {\bibfnamefont {V.~L.}\ \bibnamefont
  {Telegdi}},\ }\href {https://doi.org/10.1119/1.19382} {\bibfield  {journal}
  {\bibinfo  {journal} {American Journal of Physics}\ }\textbf {\bibinfo
  {volume} {68}},\ \bibinfo {pages} {115} (\bibinfo {year} {2000})},\ \Eprint
  {https://arxiv.org/abs/https://doi.org/10.1119/1.19382}
  {https://doi.org/10.1119/1.19382} \BibitemShut {NoStop}%
\bibitem [{\citenamefont {Kiefer}(2012)}]{Kiefer:2012}%
  \BibitemOpen
  \bibfield  {author} {\bibinfo {author} {\bibfnamefont {C.}~\bibnamefont
  {Kiefer}},\ }\href
  {https://doi.org/10.1093/acprof:oso/9780199585205.001.0001} {\emph {\bibinfo
  {title} {{Quantum Gravity}}}}\ (\bibinfo  {publisher} {Oxford University
  Press},\ \bibinfo {year} {2012})\BibitemShut {NoStop}%
\bibitem [{\citenamefont {Green}\ \emph {et~al.}(2012)\citenamefont {Green},
  \citenamefont {Schwarz},\ and\ \citenamefont
  {Witten}}]{Green:Schwarz:Witten:2012}%
  \BibitemOpen
  \bibfield  {author} {\bibinfo {author} {\bibfnamefont {M.~B.}\ \bibnamefont
  {Green}}, \bibinfo {author} {\bibfnamefont {J.~H.}\ \bibnamefont {Schwarz}},\
  and\ \bibinfo {author} {\bibfnamefont {E.}~\bibnamefont {Witten}},\ }\href
  {https://doi.org/10.1017/CBO9781139248563} {\emph {\bibinfo {title}
  {Superstring Theory: 25th Anniversary Edition}}},\ \bibinfo {series}
  {Cambridge Monographs on Mathematical Physics}, Vol.~\bibinfo {volume} {1}\
  (\bibinfo  {publisher} {Cambridge University Press},\ \bibinfo {year}
  {2012})\BibitemShut {NoStop}%
\bibitem [{\citenamefont {Polchinski}(1998)}]{Polchinski:1998}%
  \BibitemOpen
  \bibfield  {author} {\bibinfo {author} {\bibfnamefont {J.}~\bibnamefont
  {Polchinski}},\ }\href {https://doi.org/10.1017/CBO9780511816079} {\emph
  {\bibinfo {title} {String Theory}}},\ \bibinfo {series} {Cambridge Monographs
  on Mathematical Physics}, Vol.~\bibinfo {volume} {1}\ (\bibinfo  {publisher}
  {Cambridge University Press},\ \bibinfo {year} {1998})\BibitemShut {NoStop}%
\bibitem [{\citenamefont {Rovelli}(1998)}]{Rovelli:1998}%
  \BibitemOpen
  \bibfield  {author} {\bibinfo {author} {\bibfnamefont {C.}~\bibnamefont
  {Rovelli}},\ }\href {https://doi.org/10.12942/lrr-1998-1} {\bibfield
  {journal} {\bibinfo  {journal} {Living Reviews in Relativity}\ }\textbf
  {\bibinfo {volume} {1}},\ \bibinfo {pages} {1} (\bibinfo {year}
  {1998})}\BibitemShut {NoStop}%
\bibitem [{\citenamefont {Nielsen}\ and\ \citenamefont
  {Chuang}(2010)}]{Nielsen:Chuang:2010}%
  \BibitemOpen
  \bibfield  {author} {\bibinfo {author} {\bibfnamefont {M.~A.}\ \bibnamefont
  {Nielsen}}\ and\ \bibinfo {author} {\bibfnamefont {I.~L.}\ \bibnamefont
  {Chuang}},\ }\href {https://doi.org/10.1017/CBO9780511976667} {\emph
  {\bibinfo {title} {Quantum Computation and Quantum Information: 10th
  Anniversary Edition}}}\ (\bibinfo  {publisher} {Cambridge University Press},\
  \bibinfo {year} {2010})\BibitemShut {NoStop}%
\bibitem [{\citenamefont {Hu}\ \emph {et~al.}(2012)\citenamefont {Hu},
  \citenamefont {Lin},\ and\ \citenamefont {Louko}}]{Hu:Lin:Louko:2012}%
  \BibitemOpen
  \bibfield  {author} {\bibinfo {author} {\bibfnamefont {B.~L.}\ \bibnamefont
  {Hu}}, \bibinfo {author} {\bibfnamefont {S.-Y.}\ \bibnamefont {Lin}},\ and\
  \bibinfo {author} {\bibfnamefont {J.}~\bibnamefont {Louko}},\ }\href
  {https://doi.org/10.1088/0264-9381/29/22/224005} {\bibfield  {journal}
  {\bibinfo  {journal} {Classical and Quantum Gravity}\ }\textbf {\bibinfo
  {volume} {29}},\ \bibinfo {pages} {224005} (\bibinfo {year}
  {2012})}\BibitemShut {NoStop}%
\bibitem [{\citenamefont {Bassi}\ \emph {et~al.}(2013)\citenamefont {Bassi},
  \citenamefont {Lochan}, \citenamefont {Satin}, \citenamefont {Singh},\ and\
  \citenamefont {Ulbricht}}]{Bassi:Lochan:2013}%
  \BibitemOpen
  \bibfield  {author} {\bibinfo {author} {\bibfnamefont {A.}~\bibnamefont
  {Bassi}}, \bibinfo {author} {\bibfnamefont {K.}~\bibnamefont {Lochan}},
  \bibinfo {author} {\bibfnamefont {S.}~\bibnamefont {Satin}}, \bibinfo
  {author} {\bibfnamefont {T.~P.}\ \bibnamefont {Singh}},\ and\ \bibinfo
  {author} {\bibfnamefont {H.}~\bibnamefont {Ulbricht}},\ }\href
  {https://doi.org/10.1103/RevModPhys.85.471} {\bibfield  {journal} {\bibinfo
  {journal} {Rev. Mod. Phys.}\ }\textbf {\bibinfo {volume} {85}},\ \bibinfo
  {pages} {471} (\bibinfo {year} {2013})}\BibitemShut {NoStop}%
\bibitem [{\citenamefont {Bassi}\ \emph {et~al.}(2017)\citenamefont {Bassi},
  \citenamefont {Gro{\ss}ardt},\ and\ \citenamefont
  {Ulbricht}}]{Bassi:Grossart:2017}%
  \BibitemOpen
  \bibfield  {author} {\bibinfo {author} {\bibfnamefont {A.}~\bibnamefont
  {Bassi}}, \bibinfo {author} {\bibfnamefont {A.}~\bibnamefont
  {Gro{\ss}ardt}},\ and\ \bibinfo {author} {\bibfnamefont {H.}~\bibnamefont
  {Ulbricht}},\ }\href {https://doi.org/10.1088/1361-6382/aa864f} {\bibfield
  {journal} {\bibinfo  {journal} {Classical and Quantum Gravity}\ }\textbf
  {\bibinfo {volume} {34}},\ \bibinfo {pages} {193002} (\bibinfo {year}
  {2017})}\BibitemShut {NoStop}%
\bibitem [{\citenamefont {Howl}\ \emph {et~al.}(2019)\citenamefont {Howl},
  \citenamefont {Penrose},\ and\ \citenamefont {Fuentes}}]{Howl:Penrose:2019}%
  \BibitemOpen
  \bibfield  {author} {\bibinfo {author} {\bibfnamefont {R.}~\bibnamefont
  {Howl}}, \bibinfo {author} {\bibfnamefont {R.}~\bibnamefont {Penrose}},\ and\
  \bibinfo {author} {\bibfnamefont {I.}~\bibnamefont {Fuentes}},\ }\href
  {https://doi.org/10.1088/1367-2630/ab104a} {\bibfield  {journal} {\bibinfo
  {journal} {New Journal of Physics}\ }\textbf {\bibinfo {volume} {21}},\
  \bibinfo {pages} {043047} (\bibinfo {year} {2019})}\BibitemShut {NoStop}%
\bibitem [{\citenamefont {Carney}\ \emph {et~al.}(2019)\citenamefont {Carney},
  \citenamefont {Stamp},\ and\ \citenamefont {Taylor}}]{Carney:Stamp:2019}%
  \BibitemOpen
  \bibfield  {author} {\bibinfo {author} {\bibfnamefont {D.}~\bibnamefont
  {Carney}}, \bibinfo {author} {\bibfnamefont {P.~C.~E.}\ \bibnamefont
  {Stamp}},\ and\ \bibinfo {author} {\bibfnamefont {J.~M.}\ \bibnamefont
  {Taylor}},\ }\href {https://doi.org/10.1088/1361-6382/aaf9ca} {\bibfield
  {journal} {\bibinfo  {journal} {Classical and Quantum Gravity}\ }\textbf
  {\bibinfo {volume} {36}},\ \bibinfo {pages} {034001} (\bibinfo {year}
  {2019})}\BibitemShut {NoStop}%
\bibitem [{\citenamefont {Downes}\ \emph {et~al.}(2013)\citenamefont {Downes},
  \citenamefont {Ralph},\ and\ \citenamefont {Walk}}]{Downes:Ralph:2013}%
  \BibitemOpen
  \bibfield  {author} {\bibinfo {author} {\bibfnamefont {T.~G.}\ \bibnamefont
  {Downes}}, \bibinfo {author} {\bibfnamefont {T.~C.}\ \bibnamefont {Ralph}},\
  and\ \bibinfo {author} {\bibfnamefont {N.}~\bibnamefont {Walk}},\ }\href
  {https://doi.org/10.1103/PhysRevA.87.012327} {\bibfield  {journal} {\bibinfo
  {journal} {Phys. Rev. A}\ }\textbf {\bibinfo {volume} {87}},\ \bibinfo
  {pages} {012327} (\bibinfo {year} {2013})}\BibitemShut {NoStop}%
\bibitem [{\citenamefont {Bruschi}\ \emph
  {et~al.}(2014{\natexlab{a}})\citenamefont {Bruschi}, \citenamefont {Ralph},
  \citenamefont {Fuentes}, \citenamefont {Jennewein},\ and\ \citenamefont
  {Razavi}}]{Bruschi:Ralph:2014}%
  \BibitemOpen
  \bibfield  {author} {\bibinfo {author} {\bibfnamefont {D.~E.}\ \bibnamefont
  {Bruschi}}, \bibinfo {author} {\bibfnamefont {T.~C.}\ \bibnamefont {Ralph}},
  \bibinfo {author} {\bibfnamefont {I.}~\bibnamefont {Fuentes}}, \bibinfo
  {author} {\bibfnamefont {T.}~\bibnamefont {Jennewein}},\ and\ \bibinfo
  {author} {\bibfnamefont {M.}~\bibnamefont {Razavi}},\ }\href
  {https://doi.org/10.1103/PhysRevD.90.045041} {\bibfield  {journal} {\bibinfo
  {journal} {Phys. Rev. D}\ }\textbf {\bibinfo {volume} {90}},\ \bibinfo
  {pages} {045041} (\bibinfo {year} {2014}{\natexlab{a}})}\BibitemShut
  {NoStop}%
\bibitem [{\citenamefont {Bruschi}\ \emph
  {et~al.}(2014{\natexlab{b}})\citenamefont {Bruschi}, \citenamefont {Datta},
  \citenamefont {Ursin}, \citenamefont {Ralph},\ and\ \citenamefont
  {Fuentes}}]{Bruschi:Datta:2014}%
  \BibitemOpen
  \bibfield  {author} {\bibinfo {author} {\bibfnamefont {D.~E.}\ \bibnamefont
  {Bruschi}}, \bibinfo {author} {\bibfnamefont {A.}~\bibnamefont {Datta}},
  \bibinfo {author} {\bibfnamefont {R.}~\bibnamefont {Ursin}}, \bibinfo
  {author} {\bibfnamefont {T.~C.}\ \bibnamefont {Ralph}},\ and\ \bibinfo
  {author} {\bibfnamefont {I.}~\bibnamefont {Fuentes}},\ }\href
  {https://doi.org/10.1103/PhysRevD.90.124001} {\bibfield  {journal} {\bibinfo
  {journal} {Phys. Rev. D}\ }\textbf {\bibinfo {volume} {90}},\ \bibinfo
  {pages} {124001} (\bibinfo {year} {2014}{\natexlab{b}})}\BibitemShut
  {NoStop}%
\bibitem [{\citenamefont {Berera}(2020)}]{Berera:2020}%
  \BibitemOpen
  \bibfield  {author} {\bibinfo {author} {\bibfnamefont {A.}~\bibnamefont
  {Berera}},\ }\href {https://doi.org/10.1103/PhysRevD.102.063005} {\bibfield
  {journal} {\bibinfo  {journal} {Phys. Rev. D}\ }\textbf {\bibinfo {volume}
  {102}},\ \bibinfo {pages} {063005} (\bibinfo {year} {2020})}\BibitemShut
  {NoStop}%
\bibitem [{\citenamefont {Berera}\ \emph {et~al.}(2021)\citenamefont {Berera},
  \citenamefont {Brahma}, \citenamefont {Brandenberger}, \citenamefont
  {Calder\'on-Figueroa},\ and\ \citenamefont {Heavens}}]{Berera:Brahma:2021}%
  \BibitemOpen
  \bibfield  {author} {\bibinfo {author} {\bibfnamefont {A.}~\bibnamefont
  {Berera}}, \bibinfo {author} {\bibfnamefont {S.}~\bibnamefont {Brahma}},
  \bibinfo {author} {\bibfnamefont {R.}~\bibnamefont {Brandenberger}}, \bibinfo
  {author} {\bibfnamefont {J.}~\bibnamefont {Calder\'on-Figueroa}},\ and\
  \bibinfo {author} {\bibfnamefont {A.}~\bibnamefont {Heavens}},\ }\href
  {https://doi.org/10.1103/PhysRevD.104.063519} {\bibfield  {journal} {\bibinfo
   {journal} {Phys. Rev. D}\ }\textbf {\bibinfo {volume} {104}},\ \bibinfo
  {pages} {063519} (\bibinfo {year} {2021})}\BibitemShut {NoStop}%
\bibitem [{\citenamefont {Berera}\ and\ \citenamefont
  {Calder\'on-Figueroa}(2022)}]{Berera:Calderon:2022}%
  \BibitemOpen
  \bibfield  {author} {\bibinfo {author} {\bibfnamefont {A.}~\bibnamefont
  {Berera}}\ and\ \bibinfo {author} {\bibfnamefont {J.}~\bibnamefont
  {Calder\'on-Figueroa}},\ }\href {https://doi.org/10.1103/PhysRevD.105.123033}
  {\bibfield  {journal} {\bibinfo  {journal} {Phys. Rev. D}\ }\textbf {\bibinfo
  {volume} {105}},\ \bibinfo {pages} {123033} (\bibinfo {year}
  {2022})}\BibitemShut {NoStop}%
\bibitem [{\citenamefont {Hu}\ and\ \citenamefont
  {Verdaguer}(2008)}]{Hu:Verdaguer:2008}%
  \BibitemOpen
  \bibfield  {author} {\bibinfo {author} {\bibfnamefont {B.~L.}\ \bibnamefont
  {Hu}}\ and\ \bibinfo {author} {\bibfnamefont {E.}~\bibnamefont {Verdaguer}},\
  }\href {https://doi.org/10.12942/lrr-2008-3} {\bibfield  {journal} {\bibinfo
  {journal} {Living Reviews in Relativity}\ }\textbf {\bibinfo {volume} {11}},\
  \bibinfo {pages} {3} (\bibinfo {year} {2008})}\BibitemShut {NoStop}%
\bibitem [{\citenamefont {Bruschi}\ and\ \citenamefont
  {Wilhelm}(2020)}]{Bruschi:Wilhelm:2020}%
  \BibitemOpen
  \bibfield  {author} {\bibinfo {author} {\bibfnamefont {D.~E.}\ \bibnamefont
  {Bruschi}}\ and\ \bibinfo {author} {\bibfnamefont {F.~K.}\ \bibnamefont
  {Wilhelm}},\ }\Eprint {https://arxiv.org/abs/2006.11768} {arXiv:2006.11768
  [quant-ph]}  (\bibinfo {year} {2020})\BibitemShut {NoStop}%
\bibitem [{\citenamefont {Smith}\ and\ \citenamefont
  {Ahmadi}(2019)}]{Smith:Ahmadi:2019}%
  \BibitemOpen
  \bibfield  {author} {\bibinfo {author} {\bibfnamefont {A.~R.~H.}\
  \bibnamefont {Smith}}\ and\ \bibinfo {author} {\bibfnamefont
  {M.}~\bibnamefont {Ahmadi}},\ }\href
  {https://doi.org/10.22331/q-2019-07-08-160} {\bibfield  {journal} {\bibinfo
  {journal} {{Quantum}}\ }\textbf {\bibinfo {volume} {3}},\ \bibinfo {pages}
  {160} (\bibinfo {year} {2019})}\BibitemShut {NoStop}%
\bibitem [{\citenamefont {Smith}\ and\ \citenamefont
  {Ahmadi}(2020)}]{Smith:Ahmadi:2020}%
  \BibitemOpen
  \bibfield  {author} {\bibinfo {author} {\bibfnamefont {A.~R.~H.}\
  \bibnamefont {Smith}}\ and\ \bibinfo {author} {\bibfnamefont
  {M.}~\bibnamefont {Ahmadi}},\ }\href
  {https://doi.org/10.1038/s41467-020-18264-4} {\bibfield  {journal} {\bibinfo
  {journal} {Nature Communications}\ }\textbf {\bibinfo {volume} {11}},\
  \bibinfo {pages} {5360} (\bibinfo {year} {2020})}\BibitemShut {NoStop}%
\bibitem [{\citenamefont {Bose}\ \emph {et~al.}(2017)\citenamefont {Bose},
  \citenamefont {Mazumdar}, \citenamefont {Morley}, \citenamefont {Ulbricht},
  \citenamefont {Toro\ifmmode~\check{s}\else \v{s}\fi{}}, \citenamefont
  {Paternostro}, \citenamefont {Geraci}, \citenamefont {Barker}, \citenamefont
  {Kim},\ and\ \citenamefont {Milburn}}]{Bose:Mazumdar:2017}%
  \BibitemOpen
  \bibfield  {author} {\bibinfo {author} {\bibfnamefont {S.}~\bibnamefont
  {Bose}}, \bibinfo {author} {\bibfnamefont {A.}~\bibnamefont {Mazumdar}},
  \bibinfo {author} {\bibfnamefont {G.~W.}\ \bibnamefont {Morley}}, \bibinfo
  {author} {\bibfnamefont {H.}~\bibnamefont {Ulbricht}}, \bibinfo {author}
  {\bibfnamefont {M.}~\bibnamefont {Toro\ifmmode~\check{s}\else \v{s}\fi{}}},
  \bibinfo {author} {\bibfnamefont {M.}~\bibnamefont {Paternostro}}, \bibinfo
  {author} {\bibfnamefont {A.~A.}\ \bibnamefont {Geraci}}, \bibinfo {author}
  {\bibfnamefont {P.~F.}\ \bibnamefont {Barker}}, \bibinfo {author}
  {\bibfnamefont {M.~S.}\ \bibnamefont {Kim}},\ and\ \bibinfo {author}
  {\bibfnamefont {G.}~\bibnamefont {Milburn}},\ }\href
  {https://doi.org/10.1103/PhysRevLett.119.240401} {\bibfield  {journal}
  {\bibinfo  {journal} {Phys. Rev. Lett.}\ }\textbf {\bibinfo {volume} {119}},\
  \bibinfo {pages} {240401} (\bibinfo {year} {2017})}\BibitemShut {NoStop}%
\bibitem [{\citenamefont {Marletto}\ and\ \citenamefont
  {Vedral}(2017)}]{Marletto:Vedral:2017}%
  \BibitemOpen
  \bibfield  {author} {\bibinfo {author} {\bibfnamefont {C.}~\bibnamefont
  {Marletto}}\ and\ \bibinfo {author} {\bibfnamefont {V.}~\bibnamefont
  {Vedral}},\ }\href {https://doi.org/10.1103/PhysRevLett.119.240402}
  {\bibfield  {journal} {\bibinfo  {journal} {Phys. Rev. Lett.}\ }\textbf
  {\bibinfo {volume} {119}},\ \bibinfo {pages} {240402} (\bibinfo {year}
  {2017})}\BibitemShut {NoStop}%
\bibitem [{\citenamefont {Bose}\ \emph {et~al.}(2022)\citenamefont {Bose},
  \citenamefont {Mazumdar}, \citenamefont {Schut},\ and\ \citenamefont
  {Toro\ifmmode~\check{s}\else \v{s}\fi{}}}]{Bose:Mazumdar:2022}%
  \BibitemOpen
  \bibfield  {author} {\bibinfo {author} {\bibfnamefont {S.}~\bibnamefont
  {Bose}}, \bibinfo {author} {\bibfnamefont {A.}~\bibnamefont {Mazumdar}},
  \bibinfo {author} {\bibfnamefont {M.}~\bibnamefont {Schut}},\ and\ \bibinfo
  {author} {\bibfnamefont {M.}~\bibnamefont {Toro\ifmmode~\check{s}\else
  \v{s}\fi{}}},\ }\href {https://doi.org/10.1103/PhysRevD.105.106028}
  {\bibfield  {journal} {\bibinfo  {journal} {Phys. Rev. D}\ }\textbf {\bibinfo
  {volume} {105}},\ \bibinfo {pages} {106028} (\bibinfo {year}
  {2022})}\BibitemShut {NoStop}%
\bibitem [{\citenamefont {Williams}\ \emph {et~al.}(2016)\citenamefont
  {Williams}, \citenamefont {wey Chiow}, \citenamefont {Yu},\ and\
  \citenamefont {M{\"u}ller}}]{Williams:Chiow:2016}%
  \BibitemOpen
  \bibfield  {author} {\bibinfo {author} {\bibfnamefont {J.}~\bibnamefont
  {Williams}}, \bibinfo {author} {\bibfnamefont {S.}~\bibnamefont {wey Chiow}},
  \bibinfo {author} {\bibfnamefont {N.}~\bibnamefont {Yu}},\ and\ \bibinfo
  {author} {\bibfnamefont {H.}~\bibnamefont {M{\"u}ller}},\ }\href
  {https://doi.org/10.1088/1367-2630/18/2/025018} {\bibfield  {journal}
  {\bibinfo  {journal} {New Journal of Physics}\ }\textbf {\bibinfo {volume}
  {18}},\ \bibinfo {pages} {025018} (\bibinfo {year} {2016})}\BibitemShut
  {NoStop}%
\bibitem [{\citenamefont {Tino}(2021)}]{Tino:2021}%
  \BibitemOpen
  \bibfield  {author} {\bibinfo {author} {\bibfnamefont {G.~M.}\ \bibnamefont
  {Tino}},\ }\href {https://doi.org/10.1088/2058-9565/abd83e} {\bibfield
  {journal} {\bibinfo  {journal} {Quantum Science and Technology}\ }\textbf
  {\bibinfo {volume} {6}},\ \bibinfo {pages} {024014} (\bibinfo {year}
  {2021})}\BibitemShut {NoStop}%
\bibitem [{\citenamefont {Kohlrus}\ \emph {et~al.}(2017)\citenamefont
  {Kohlrus}, \citenamefont {Bruschi}, \citenamefont {Louko},\ and\
  \citenamefont {Fuentes}}]{Kohlrus:Bruschi:2017}%
  \BibitemOpen
  \bibfield  {author} {\bibinfo {author} {\bibfnamefont {J.}~\bibnamefont
  {Kohlrus}}, \bibinfo {author} {\bibfnamefont {D.~E.}\ \bibnamefont
  {Bruschi}}, \bibinfo {author} {\bibfnamefont {J.}~\bibnamefont {Louko}},\
  and\ \bibinfo {author} {\bibfnamefont {I.}~\bibnamefont {Fuentes}},\ }\href
  {https://doi.org/10.1140/epjqt/s40507-017-0061-0} {\bibfield  {journal}
  {\bibinfo  {journal} {EPJ Quantum Technology}\ }\textbf {\bibinfo {volume}
  {4}},\ \bibinfo {pages} {7} (\bibinfo {year} {2017})}\BibitemShut {NoStop}%
\bibitem [{\citenamefont {Bruschi}\ \emph {et~al.}(2021)\citenamefont
  {Bruschi}, \citenamefont {Chatzinotas}, \citenamefont {Wilhelm},\ and\
  \citenamefont {Schell}}]{Bruschi:Chatzinotas:2021}%
  \BibitemOpen
  \bibfield  {author} {\bibinfo {author} {\bibfnamefont {D.~E.}\ \bibnamefont
  {Bruschi}}, \bibinfo {author} {\bibfnamefont {S.}~\bibnamefont
  {Chatzinotas}}, \bibinfo {author} {\bibfnamefont {F.~K.}\ \bibnamefont
  {Wilhelm}},\ and\ \bibinfo {author} {\bibfnamefont {A.~W.}\ \bibnamefont
  {Schell}},\ }\bibfield  {journal} {\bibinfo  {journal} {Physical Review D}\
  }\textbf {\bibinfo {volume} {104}},\ \href
  {https://doi.org/10.1103/physrevd.104.085015} {10.1103/physrevd.104.085015}
  (\bibinfo {year} {2021})\BibitemShut {NoStop}%
\bibitem [{\citenamefont {Bruschi}\ and\ \citenamefont
  {Schell}(2023)}]{Bruschi:Schell:2023}%
  \BibitemOpen
  \bibfield  {author} {\bibinfo {author} {\bibfnamefont {D.~E.}\ \bibnamefont
  {Bruschi}}\ and\ \bibinfo {author} {\bibfnamefont {A.~W.}\ \bibnamefont
  {Schell}},\ }\href {https://doi.org/https://doi.org/10.1002/andp.202200468}
  {\bibfield  {journal} {\bibinfo  {journal} {Annalen der Physik}\ }\textbf
  {\bibinfo {volume} {535}},\ \bibinfo {pages} {2200468} (\bibinfo {year}
  {2023})},\ \Eprint
  {https://arxiv.org/abs/https://onlinelibrary.wiley.com/doi/pdf/10.1002/andp.202200468}
  {https://onlinelibrary.wiley.com/doi/pdf/10.1002/andp.202200468} \BibitemShut
  {NoStop}%
\bibitem [{\citenamefont {Pan}\ \emph {et~al.}(2012)\citenamefont {Pan},
  \citenamefont {Chen}, \citenamefont {Lu}, \citenamefont {Weinfurter},
  \citenamefont {Zeilinger},\ and\ \citenamefont {\ifmmode~\dot{Z}\else
  \.{Z}\fi{}ukowski}}]{Pan:Chen:2012}%
  \BibitemOpen
  \bibfield  {author} {\bibinfo {author} {\bibfnamefont {J.-W.}\ \bibnamefont
  {Pan}}, \bibinfo {author} {\bibfnamefont {Z.-B.}\ \bibnamefont {Chen}},
  \bibinfo {author} {\bibfnamefont {C.-Y.}\ \bibnamefont {Lu}}, \bibinfo
  {author} {\bibfnamefont {H.}~\bibnamefont {Weinfurter}}, \bibinfo {author}
  {\bibfnamefont {A.}~\bibnamefont {Zeilinger}},\ and\ \bibinfo {author}
  {\bibfnamefont {M.}~\bibnamefont {\ifmmode~\dot{Z}\else \.{Z}\fi{}ukowski}},\
  }\href {https://doi.org/10.1103/RevModPhys.84.777} {\bibfield  {journal}
  {\bibinfo  {journal} {Rev. Mod. Phys.}\ }\textbf {\bibinfo {volume} {84}},\
  \bibinfo {pages} {777} (\bibinfo {year} {2012})}\BibitemShut {NoStop}%
\bibitem [{\citenamefont {Scully}\ and\ \citenamefont
  {Zubairy}(1997)}]{Scully:Zubairy:1997}%
  \BibitemOpen
  \bibfield  {author} {\bibinfo {author} {\bibfnamefont {M.~O.}\ \bibnamefont
  {Scully}}\ and\ \bibinfo {author} {\bibfnamefont {M.~S.}\ \bibnamefont
  {Zubairy}},\ }\href {https://doi.org/10.1017/CBO9780511813993} {\emph
  {\bibinfo {title} {Quantum Optics}}}\ (\bibinfo  {publisher} {Cambridge
  University Press},\ \bibinfo {year} {1997})\BibitemShut {NoStop}%
\bibitem [{\citenamefont {Exirifard}\ \emph {et~al.}(2021)\citenamefont
  {Exirifard}, \citenamefont {Culf},\ and\ \citenamefont
  {Karimi}}]{Exirifard:Culf:2021}%
  \BibitemOpen
  \bibfield  {author} {\bibinfo {author} {\bibfnamefont {Q.}~\bibnamefont
  {Exirifard}}, \bibinfo {author} {\bibfnamefont {E.}~\bibnamefont {Culf}},\
  and\ \bibinfo {author} {\bibfnamefont {E.}~\bibnamefont {Karimi}},\
  }\bibfield  {journal} {\bibinfo  {journal} {Communications Physics}\ }\textbf
  {\bibinfo {volume} {4}},\ \href {https://doi.org/10.1038/s42005-021-00671-8}
  {10.1038/s42005-021-00671-8} (\bibinfo {year} {2021})\BibitemShut {NoStop}%
\bibitem [{\citenamefont {Exirifard}\ and\ \citenamefont
  {Karimi}(2022)}]{Exirifard:Karimi:2022}%
  \BibitemOpen
  \bibfield  {author} {\bibinfo {author} {\bibfnamefont {Q.}~\bibnamefont
  {Exirifard}}\ and\ \bibinfo {author} {\bibfnamefont {E.}~\bibnamefont
  {Karimi}},\ }\bibfield  {journal} {\bibinfo  {journal} {Physical Review D}\
  }\textbf {\bibinfo {volume} {105}},\ \href
  {https://doi.org/10.1103/physrevd.105.084016} {10.1103/physrevd.105.084016}
  (\bibinfo {year} {2022})\BibitemShut {NoStop}%
\bibitem [{\citenamefont {{Misner}}\ \emph {et~al.}(1973)\citenamefont
  {{Misner}}, \citenamefont {{Thorne}},\ and\ \citenamefont
  {{Wheeler}}}]{Misner:Thorne:1973}%
  \BibitemOpen
  \bibfield  {author} {\bibinfo {author} {\bibfnamefont {C.~W.}\ \bibnamefont
  {{Misner}}}, \bibinfo {author} {\bibfnamefont {K.~S.}\ \bibnamefont
  {{Thorne}}},\ and\ \bibinfo {author} {\bibfnamefont {J.~A.}\ \bibnamefont
  {{Wheeler}}},\ }\href@noop {} {\emph {\bibinfo {title} {San Francisco:
  W.H.~Freeman and Co., 1973}}}\ (\bibinfo  {publisher} {W. H. Freeman and
  Company},\ \bibinfo {year} {1973})\BibitemShut {NoStop}%
\bibitem [{\citenamefont {Birrell}\ and\ \citenamefont
  {Davies}(1982)}]{Birrell:Davies:1982}%
  \BibitemOpen
  \bibfield  {author} {\bibinfo {author} {\bibfnamefont {N.~D.}\ \bibnamefont
  {Birrell}}\ and\ \bibinfo {author} {\bibfnamefont {P.~C.~W.}\ \bibnamefont
  {Davies}},\ }\href {https://doi.org/10.1017/CBO9780511622632} {\emph
  {\bibinfo {title} {Quantum Fields in Curved Space}}},\ Cambridge Monographs
  on Mathematical Physics\ (\bibinfo  {publisher} {Cambridge University
  Press},\ \bibinfo {year} {1982})\BibitemShut {NoStop}%
\bibitem [{\citenamefont {Wald}(1995)}]{Wald:1995}%
  \BibitemOpen
  \bibfield  {author} {\bibinfo {author} {\bibfnamefont {R.~M.}\ \bibnamefont
  {Wald}},\ }\href@noop {} {\emph {\bibinfo {title} {{Quantum Field Theory in
  Curved Space-Time and Black Hole Thermodynamics}}}},\ Chicago Lectures in
  Physics\ (\bibinfo  {publisher} {University of Chicago Press},\ \bibinfo
  {address} {Chicago, IL},\ \bibinfo {year} {1995})\BibitemShut {NoStop}%
\bibitem [{\citenamefont {Srednicki}(2007)}]{Srednicki:2007}%
  \BibitemOpen
  \bibfield  {author} {\bibinfo {author} {\bibfnamefont {M.}~\bibnamefont
  {Srednicki}},\ }\href {https://doi.org/10.1017/CBO9780511813917} {\emph
  {\bibinfo {title} {Quantum Field Theory}}}\ (\bibinfo  {publisher} {Cambridge
  University Press},\ \bibinfo {year} {2007})\BibitemShut {NoStop}%
\bibitem [{\citenamefont {Barbado}\ \emph {et~al.}(2020)\citenamefont
  {Barbado}, \citenamefont {B{\'{a}}ez-Camargo},\ and\ \citenamefont
  {Fuentes}}]{Barbado:Baez-Camargo:2020}%
  \BibitemOpen
  \bibfield  {author} {\bibinfo {author} {\bibfnamefont {L.~C.}\ \bibnamefont
  {Barbado}}, \bibinfo {author} {\bibfnamefont {A.~L.}\ \bibnamefont
  {B{\'{a}}ez-Camargo}},\ and\ \bibinfo {author} {\bibfnamefont
  {I.}~\bibnamefont {Fuentes}},\ }\bibfield  {journal} {\bibinfo  {journal}
  {The European Physical Journal C}\ }\textbf {\bibinfo {volume} {80}},\ \href
  {https://doi.org/10.1140/epjc/s10052-020-8369-9}
  {10.1140/epjc/s10052-020-8369-9} (\bibinfo {year} {2020})\BibitemShut
  {NoStop}%
\bibitem [{\citenamefont {Barbado}\ \emph {et~al.}(2021)\citenamefont
  {Barbado}, \citenamefont {B{\'{a}}ez-Camargo},\ and\ \citenamefont
  {Fuentes}}]{Barbado:Baez-Camargo:2021}%
  \BibitemOpen
  \bibfield  {author} {\bibinfo {author} {\bibfnamefont {L.~C.}\ \bibnamefont
  {Barbado}}, \bibinfo {author} {\bibfnamefont {A.~L.}\ \bibnamefont
  {B{\'{a}}ez-Camargo}},\ and\ \bibinfo {author} {\bibfnamefont
  {I.}~\bibnamefont {Fuentes}},\ }\bibfield  {journal} {\bibinfo  {journal}
  {The European Physical Journal C}\ }\textbf {\bibinfo {volume} {81}},\ \href
  {https://doi.org/10.1140/epjc/s10052-021-09737-x}
  {10.1140/epjc/s10052-021-09737-x} (\bibinfo {year} {2021})\BibitemShut
  {NoStop}%
\bibitem [{\citenamefont {Maybee}\ \emph {et~al.}(2019)\citenamefont {Maybee},
  \citenamefont {Hodgson}, \citenamefont {Beige},\ and\ \citenamefont
  {Purdy}}]{Maybee:Hodgson:2019}%
  \BibitemOpen
  \bibfield  {author} {\bibinfo {author} {\bibfnamefont {B.}~\bibnamefont
  {Maybee}}, \bibinfo {author} {\bibfnamefont {D.}~\bibnamefont {Hodgson}},
  \bibinfo {author} {\bibfnamefont {A.}~\bibnamefont {Beige}},\ and\ \bibinfo
  {author} {\bibfnamefont {R.}~\bibnamefont {Purdy}},\ }\bibfield  {journal}
  {\bibinfo  {journal} {Entropy}\ }\textbf {\bibinfo {volume} {21}},\ \href
  {https://doi.org/10.3390/e21090844} {10.3390/e21090844} (\bibinfo {year}
  {2019})\BibitemShut {NoStop}%
\bibitem [{\citenamefont {Hodgson}\ \emph {et~al.}(2022)\citenamefont
  {Hodgson}, \citenamefont {Southall}, \citenamefont {Purdy},\ and\
  \citenamefont {Beige}}]{Hodgson:Southall:2022}%
  \BibitemOpen
  \bibfield  {author} {\bibinfo {author} {\bibfnamefont {D.}~\bibnamefont
  {Hodgson}}, \bibinfo {author} {\bibfnamefont {J.}~\bibnamefont {Southall}},
  \bibinfo {author} {\bibfnamefont {R.}~\bibnamefont {Purdy}},\ and\ \bibinfo
  {author} {\bibfnamefont {A.}~\bibnamefont {Beige}},\ }\bibfield  {journal}
  {\bibinfo  {journal} {Frontiers in Photonics}\ }\textbf {\bibinfo {volume}
  {3}},\ \href {https://doi.org/10.3389/fphot.2022.978855}
  {10.3389/fphot.2022.978855} (\bibinfo {year} {2022})\BibitemShut {NoStop}%
\bibitem [{\citenamefont {Keller}\ \emph {et~al.}(2004)\citenamefont {Keller},
  \citenamefont {Lange}, \citenamefont {Hayasaka}, \citenamefont {Lange},\ and\
  \citenamefont {Walther}}]{Keller:Lange:2004}%
  \BibitemOpen
  \bibfield  {author} {\bibinfo {author} {\bibfnamefont {M.}~\bibnamefont
  {Keller}}, \bibinfo {author} {\bibfnamefont {B.}~\bibnamefont {Lange}},
  \bibinfo {author} {\bibfnamefont {K.}~\bibnamefont {Hayasaka}}, \bibinfo
  {author} {\bibfnamefont {W.}~\bibnamefont {Lange}},\ and\ \bibinfo {author}
  {\bibfnamefont {H.}~\bibnamefont {Walther}},\ }\href
  {https://doi.org/10.1038/nature02961} {\bibfield  {journal} {\bibinfo
  {journal} {Nature}\ }\textbf {\bibinfo {volume} {431}},\ \bibinfo {pages}
  {1075} (\bibinfo {year} {2004})}\BibitemShut {NoStop}%
\bibitem [{\citenamefont {Nisbet-Jones}\ \emph {et~al.}(2011)\citenamefont
  {Nisbet-Jones}, \citenamefont {Dilley}, \citenamefont {Ljunggren},\ and\
  \citenamefont {Kuhn}}]{Nisbet-Jones:Dilley:2011}%
  \BibitemOpen
  \bibfield  {author} {\bibinfo {author} {\bibfnamefont {P.~B.~R.}\
  \bibnamefont {Nisbet-Jones}}, \bibinfo {author} {\bibfnamefont
  {J.}~\bibnamefont {Dilley}}, \bibinfo {author} {\bibfnamefont
  {D.}~\bibnamefont {Ljunggren}},\ and\ \bibinfo {author} {\bibfnamefont
  {A.}~\bibnamefont {Kuhn}},\ }\href
  {https://doi.org/10.1088/1367-2630/13/10/103036} {\bibfield  {journal}
  {\bibinfo  {journal} {New Journal of Physics}\ }\textbf {\bibinfo {volume}
  {13}},\ \bibinfo {pages} {103036} (\bibinfo {year} {2011})}\BibitemShut
  {NoStop}%
\bibitem [{\citenamefont {Chi}\ \emph {et~al.}(2021)\citenamefont {Chi},
  \citenamefont {Wang},\ and\ \citenamefont {Yao}}]{Chi:Wang:2021}%
  \BibitemOpen
  \bibfield  {author} {\bibinfo {author} {\bibfnamefont {H.}~\bibnamefont
  {Chi}}, \bibinfo {author} {\bibfnamefont {C.}~\bibnamefont {Wang}},\ and\
  \bibinfo {author} {\bibfnamefont {J.}~\bibnamefont {Yao}},\ }\href
  {https://doi.org/10.1109/JMW.2021.3085868} {\bibfield  {journal} {\bibinfo
  {journal} {IEEE Journal of Microwaves}\ }\textbf {\bibinfo {volume} {1}},\
  \bibinfo {pages} {787} (\bibinfo {year} {2021})}\BibitemShut {NoStop}%
\bibitem [{\citenamefont {Unruh}(1976)}]{Unruh:1976}%
  \BibitemOpen
  \bibfield  {author} {\bibinfo {author} {\bibfnamefont {W.~G.}\ \bibnamefont
  {Unruh}},\ }\href {https://doi.org/10.1103/PhysRevD.14.870} {\bibfield
  {journal} {\bibinfo  {journal} {Phys. Rev. D}\ }\textbf {\bibinfo {volume}
  {14}},\ \bibinfo {pages} {870} (\bibinfo {year} {1976})}\BibitemShut
  {NoStop}%
\bibitem [{\citenamefont {Bruschi}\ \emph {et~al.}(2010)\citenamefont
  {Bruschi}, \citenamefont {Louko}, \citenamefont
  {Mart{\'{\i}}n-Mart{\'{\i}}nez}, \citenamefont {Dragan},\ and\ \citenamefont
  {Fuentes}}]{Bruschi:Louko:2010}%
  \BibitemOpen
  \bibfield  {author} {\bibinfo {author} {\bibfnamefont {D.~E.}\ \bibnamefont
  {Bruschi}}, \bibinfo {author} {\bibfnamefont {J.}~\bibnamefont {Louko}},
  \bibinfo {author} {\bibfnamefont {E.}~\bibnamefont
  {Mart{\'{\i}}n-Mart{\'{\i}}nez}}, \bibinfo {author} {\bibfnamefont
  {A.}~\bibnamefont {Dragan}},\ and\ \bibinfo {author} {\bibfnamefont
  {I.}~\bibnamefont {Fuentes}},\ }\bibfield  {journal} {\bibinfo  {journal}
  {Physical Review A}\ }\textbf {\bibinfo {volume} {82}},\ \href
  {https://doi.org/10.1103/physreva.82.042332} {10.1103/physreva.82.042332}
  (\bibinfo {year} {2010})\BibitemShut {NoStop}%
\bibitem [{\citenamefont {Ashby}(2003)}]{Ashby:2003}%
  \BibitemOpen
  \bibfield  {author} {\bibinfo {author} {\bibfnamefont {N.}~\bibnamefont
  {Ashby}},\ }\href {https://doi.org/10.12942/lrr-2003-1} {\bibfield  {journal}
  {\bibinfo  {journal} {Living Reviews in Relativity}\ }\textbf {\bibinfo
  {volume} {6}},\ \bibinfo {pages} {1} (\bibinfo {year} {2003})}\BibitemShut
  {NoStop}%
\bibitem [{\citenamefont {Wilhelm}\ and\ \citenamefont
  {Dwivedi}(2014)}]{Wilhelm:Bhola:2014}%
  \BibitemOpen
  \bibfield  {author} {\bibinfo {author} {\bibfnamefont {K.}~\bibnamefont
  {Wilhelm}}\ and\ \bibinfo {author} {\bibfnamefont {B.~N.}\ \bibnamefont
  {Dwivedi}},\ }\href
  {https://doi.org/https://doi.org/10.1016/j.newast.2014.01.012} {\bibfield
  {journal} {\bibinfo  {journal} {New Astronomy}\ }\textbf {\bibinfo {volume}
  {31}},\ \bibinfo {pages} {8} (\bibinfo {year} {2014})}\BibitemShut {NoStop}%
\bibitem [{\citenamefont {Mieling}(2021)}]{Mieling:2021}%
  \BibitemOpen
  \bibfield  {author} {\bibinfo {author} {\bibfnamefont {T.~B.}\ \bibnamefont
  {Mieling}},\ }\href {https://doi.org/10.1088/1361-6382/ac15db} {\bibfield
  {journal} {\bibinfo  {journal} {Classical and Quantum Gravity}\ }\textbf
  {\bibinfo {volume} {38}},\ \bibinfo {pages} {175007} (\bibinfo {year}
  {2021})}\BibitemShut {NoStop}%
\bibitem [{\citenamefont {Parker}(1969)}]{Parker:1969}%
  \BibitemOpen
  \bibfield  {author} {\bibinfo {author} {\bibfnamefont {L.}~\bibnamefont
  {Parker}},\ }\href {https://doi.org/10.1103/PhysRev.183.1057} {\bibfield
  {journal} {\bibinfo  {journal} {Phys. Rev.}\ }\textbf {\bibinfo {volume}
  {183}},\ \bibinfo {pages} {1057} (\bibinfo {year} {1969})}\BibitemShut
  {NoStop}%
\bibitem [{\citenamefont {Parker}(1971)}]{Parker:1971}%
  \BibitemOpen
  \bibfield  {author} {\bibinfo {author} {\bibfnamefont {L.}~\bibnamefont
  {Parker}},\ }\href {https://doi.org/10.1103/PhysRevD.3.346} {\bibfield
  {journal} {\bibinfo  {journal} {Phys. Rev. D}\ }\textbf {\bibinfo {volume}
  {3}},\ \bibinfo {pages} {346} (\bibinfo {year} {1971})}\BibitemShut {NoStop}%
\bibitem [{\citenamefont {Christ}\ \emph {et~al.}(2013)\citenamefont {Christ},
  \citenamefont {Brecht}, \citenamefont {Mauerer},\ and\ \citenamefont
  {Silberhorn}}]{Christ:Brecht:2013}%
  \BibitemOpen
  \bibfield  {author} {\bibinfo {author} {\bibfnamefont {A.}~\bibnamefont
  {Christ}}, \bibinfo {author} {\bibfnamefont {B.}~\bibnamefont {Brecht}},
  \bibinfo {author} {\bibfnamefont {W.}~\bibnamefont {Mauerer}},\ and\ \bibinfo
  {author} {\bibfnamefont {C.}~\bibnamefont {Silberhorn}},\ }\href
  {https://doi.org/10.1088/1367-2630/15/5/053038} {\bibfield  {journal}
  {\bibinfo  {journal} {New Journal of Physics}\ }\textbf {\bibinfo {volume}
  {15}},\ \bibinfo {pages} {053038} (\bibinfo {year} {2013})}\BibitemShut
  {NoStop}%
\bibitem [{\citenamefont {Couteau}(2018)}]{Couteau:2018}%
  \BibitemOpen
  \bibfield  {author} {\bibinfo {author} {\bibfnamefont {C.}~\bibnamefont
  {Couteau}},\ }\href {https://doi.org/10.1080/00107514.2018.1488463}
  {\bibfield  {journal} {\bibinfo  {journal} {Contemporary Physics}\ }\textbf
  {\bibinfo {volume} {59}},\ \bibinfo {pages} {291} (\bibinfo {year} {2018})},\
  \Eprint {https://arxiv.org/abs/https://doi.org/10.1080/00107514.2018.1488463}
  {https://doi.org/10.1080/00107514.2018.1488463} \BibitemShut {NoStop}%
\bibitem [{\citenamefont {Hong}\ \emph {et~al.}(1987)\citenamefont {Hong},
  \citenamefont {Ou},\ and\ \citenamefont {Mandel}}]{Hong:Ou:1987}%
  \BibitemOpen
  \bibfield  {author} {\bibinfo {author} {\bibfnamefont {C.~K.}\ \bibnamefont
  {Hong}}, \bibinfo {author} {\bibfnamefont {Z.~Y.}\ \bibnamefont {Ou}},\ and\
  \bibinfo {author} {\bibfnamefont {L.}~\bibnamefont {Mandel}},\ }\href
  {https://doi.org/10.1103/PhysRevLett.59.2044} {\bibfield  {journal} {\bibinfo
   {journal} {Phys. Rev. Lett.}\ }\textbf {\bibinfo {volume} {59}},\ \bibinfo
  {pages} {2044} (\bibinfo {year} {1987})}\BibitemShut {NoStop}%
\bibitem [{\citenamefont {Bouchard}\ \emph {et~al.}(2020)\citenamefont
  {Bouchard}, \citenamefont {Sit}, \citenamefont {Zhang}, \citenamefont
  {Fickler}, \citenamefont {Miatto}, \citenamefont {Yao}, \citenamefont
  {Sciarrino},\ and\ \citenamefont {Karimi}}]{Bouchard:Sit:2021}%
  \BibitemOpen
  \bibfield  {author} {\bibinfo {author} {\bibfnamefont {F.}~\bibnamefont
  {Bouchard}}, \bibinfo {author} {\bibfnamefont {A.}~\bibnamefont {Sit}},
  \bibinfo {author} {\bibfnamefont {Y.}~\bibnamefont {Zhang}}, \bibinfo
  {author} {\bibfnamefont {R.}~\bibnamefont {Fickler}}, \bibinfo {author}
  {\bibfnamefont {F.~M.}\ \bibnamefont {Miatto}}, \bibinfo {author}
  {\bibfnamefont {Y.}~\bibnamefont {Yao}}, \bibinfo {author} {\bibfnamefont
  {F.}~\bibnamefont {Sciarrino}},\ and\ \bibinfo {author} {\bibfnamefont
  {E.}~\bibnamefont {Karimi}},\ }\href
  {https://doi.org/10.1088/1361-6633/abcd7a} {\bibfield  {journal} {\bibinfo
  {journal} {Reports on Progress in Physics}\ }\textbf {\bibinfo {volume}
  {84}},\ \bibinfo {pages} {012402} (\bibinfo {year} {2020})}\BibitemShut
  {NoStop}%
\bibitem [{\citenamefont {Frisk~Kockum}(2021)}]{Kockum:2021}%
  \BibitemOpen
  \bibfield  {author} {\bibinfo {author} {\bibfnamefont {A.}~\bibnamefont
  {Frisk~Kockum}},\ }in\ \href@noop {} {\emph {\bibinfo {booktitle}
  {International Symposium on Mathematics, Quantum Theory, and
  Cryptography}}},\ \bibinfo {editor} {edited by\ \bibinfo {editor}
  {\bibfnamefont {T.}~\bibnamefont {Takagi}}, \bibinfo {editor} {\bibfnamefont
  {M.}~\bibnamefont {Wakayama}}, \bibinfo {editor} {\bibfnamefont
  {K.}~\bibnamefont {Tanaka}}, \bibinfo {editor} {\bibfnamefont
  {N.}~\bibnamefont {Kunihiro}}, \bibinfo {editor} {\bibfnamefont
  {K.}~\bibnamefont {Kimoto}},\ and\ \bibinfo {editor} {\bibfnamefont
  {Y.}~\bibnamefont {Ikematsu}}}\ (\bibinfo  {publisher} {Springer Singapore},\
  \bibinfo {address} {Singapore},\ \bibinfo {year} {2021})\ pp.\ \bibinfo
  {pages} {125--146}\BibitemShut {NoStop}%
\bibitem [{\citenamefont {Soro}\ and\ \citenamefont
  {Kockum}(2022)}]{Soro:Kockum:2022}%
  \BibitemOpen
  \bibfield  {author} {\bibinfo {author} {\bibfnamefont {A.}~\bibnamefont
  {Soro}}\ and\ \bibinfo {author} {\bibfnamefont {A.~F.}\ \bibnamefont
  {Kockum}},\ }\href {https://doi.org/10.1103/PhysRevA.105.023712} {\bibfield
  {journal} {\bibinfo  {journal} {Phys. Rev. A}\ }\textbf {\bibinfo {volume}
  {105}},\ \bibinfo {pages} {023712} (\bibinfo {year} {2022})}\BibitemShut
  {NoStop}%
\bibitem [{\citenamefont {Drummond}\ and\ \citenamefont
  {Corney}(2006)}]{Drummond:Corney:2006}%
  \BibitemOpen
  \bibfield  {author} {\bibinfo {author} {\bibfnamefont {P.~D.}\ \bibnamefont
  {Drummond}}\ and\ \bibinfo {author} {\bibfnamefont {J.}~\bibnamefont
  {Corney}},\ }in\ \href {https://doi.org/10.1364/FIO.2006.FME1} {\emph
  {\bibinfo {booktitle} {Frontiers in Optics}}}\ (\bibinfo  {publisher} {Optica
  Publishing Group},\ \bibinfo {year} {2006})\ p.\ \bibinfo {pages}
  {FME1}\BibitemShut {NoStop}%
\bibitem [{\citenamefont {Lukishova}(2014)}]{Lukishova:2014}%
  \BibitemOpen
  \bibfield  {author} {\bibinfo {author} {\bibfnamefont {S.~G.}\ \bibnamefont
  {Lukishova}},\ }\href {https://doi.org/10.1088/1742-6596/497/1/012008}
  {\bibfield  {journal} {\bibinfo  {journal} {Journal of Physics: Conference
  Series}\ }\textbf {\bibinfo {volume} {497}},\ \bibinfo {pages} {012008}
  (\bibinfo {year} {2014})}\BibitemShut {NoStop}%
\bibitem [{\citenamefont {Hawking}(1974)}]{Hawking:1974}%
  \BibitemOpen
  \bibfield  {author} {\bibinfo {author} {\bibfnamefont {S.~W.}\ \bibnamefont
  {Hawking}},\ }\href {https://doi.org/10.1038/248030a0} {\bibfield  {journal}
  {\bibinfo  {journal} {Nature}\ }\textbf {\bibinfo {volume} {248}},\ \bibinfo
  {pages} {30} (\bibinfo {year} {1974})}\BibitemShut {NoStop}%
\bibitem [{\citenamefont {Grishchuk}\ and\ \citenamefont
  {Sidorov}(1990)}]{Grishchuk:Sidorov:1990}%
  \BibitemOpen
  \bibfield  {author} {\bibinfo {author} {\bibfnamefont {L.~P.}\ \bibnamefont
  {Grishchuk}}\ and\ \bibinfo {author} {\bibfnamefont {Y.~V.}\ \bibnamefont
  {Sidorov}},\ }\href {https://doi.org/10.1103/PhysRevD.42.3413} {\bibfield
  {journal} {\bibinfo  {journal} {Phys. Rev. D}\ }\textbf {\bibinfo {volume}
  {42}},\ \bibinfo {pages} {3413} (\bibinfo {year} {1990})}\BibitemShut
  {NoStop}%
\bibitem [{\citenamefont {Hu}\ \emph {et~al.}(1994)\citenamefont {Hu},
  \citenamefont {Kang},\ and\ \citenamefont {Matacz}}]{Hu:Kang:Matacz:1994}%
  \BibitemOpen
  \bibfield  {author} {\bibinfo {author} {\bibfnamefont {B.~L.}\ \bibnamefont
  {Hu}}, \bibinfo {author} {\bibfnamefont {G.}~\bibnamefont {Kang}},\ and\
  \bibinfo {author} {\bibfnamefont {A.}~\bibnamefont {Matacz}},\ }\href
  {https://doi.org/10.1142/S0217751X94000455} {\bibfield  {journal} {\bibinfo
  {journal} {International Journal of Modern Physics A}\ }\textbf {\bibinfo
  {volume} {09}},\ \bibinfo {pages} {991} (\bibinfo {year} {1994})},\ \bibinfo
  {note} {publisher: World Scientific Publishing Co.}\BibitemShut {Stop}%
\bibitem [{\citenamefont {Dodonov}(2010)}]{Dodonov:2010}%
  \BibitemOpen
  \bibfield  {author} {\bibinfo {author} {\bibfnamefont {V.~V.}\ \bibnamefont
  {Dodonov}},\ }\href {https://doi.org/10.1088/0031-8949/82/03/038105}
  {\bibfield  {journal} {\bibinfo  {journal} {Physica Scripta}\ }\textbf
  {\bibinfo {volume} {82}},\ \bibinfo {pages} {038105} (\bibinfo {year}
  {2010})}\BibitemShut {NoStop}%
\bibitem [{\citenamefont {Bruschi}\ \emph {et~al.}(2012)\citenamefont
  {Bruschi}, \citenamefont {Fuentes},\ and\ \citenamefont
  {Louko}}]{Bruschi:Fuentes:2012}%
  \BibitemOpen
  \bibfield  {author} {\bibinfo {author} {\bibfnamefont {D.~E.}\ \bibnamefont
  {Bruschi}}, \bibinfo {author} {\bibfnamefont {I.}~\bibnamefont {Fuentes}},\
  and\ \bibinfo {author} {\bibfnamefont {J.}~\bibnamefont {Louko}},\ }\href
  {https://doi.org/10.1103/PhysRevD.85.061701} {\bibfield  {journal} {\bibinfo
  {journal} {Phys. Rev. D}\ }\textbf {\bibinfo {volume} {85}},\ \bibinfo
  {pages} {061701} (\bibinfo {year} {2012})}\BibitemShut {NoStop}%
\bibitem [{\citenamefont {Bruschi}\ \emph {et~al.}(2013)\citenamefont
  {Bruschi}, \citenamefont {Dragan}, \citenamefont {Lee}, \citenamefont
  {Fuentes},\ and\ \citenamefont {Louko}}]{Bruschi:Lee:2013}%
  \BibitemOpen
  \bibfield  {author} {\bibinfo {author} {\bibfnamefont {D.~E.}\ \bibnamefont
  {Bruschi}}, \bibinfo {author} {\bibfnamefont {A.}~\bibnamefont {Dragan}},
  \bibinfo {author} {\bibfnamefont {A.~R.}\ \bibnamefont {Lee}}, \bibinfo
  {author} {\bibfnamefont {I.}~\bibnamefont {Fuentes}},\ and\ \bibinfo {author}
  {\bibfnamefont {J.}~\bibnamefont {Louko}},\ }\href
  {https://doi.org/10.1103/PhysRevLett.111.090504} {\bibfield  {journal}
  {\bibinfo  {journal} {Phys. Rev. Lett.}\ }\textbf {\bibinfo {volume} {111}},\
  \bibinfo {pages} {090504} (\bibinfo {year} {2013})}\BibitemShut {NoStop}%
\bibitem [{\citenamefont {Adesso}\ \emph {et~al.}(2014)\citenamefont {Adesso},
  \citenamefont {Ragy},\ and\ \citenamefont {Lee}}]{Adesso:Ragy:2014}%
  \BibitemOpen
  \bibfield  {author} {\bibinfo {author} {\bibfnamefont {G.}~\bibnamefont
  {Adesso}}, \bibinfo {author} {\bibfnamefont {S.}~\bibnamefont {Ragy}},\ and\
  \bibinfo {author} {\bibfnamefont {A.~R.}\ \bibnamefont {Lee}},\ }\href@noop
  {} {\bibfield  {journal} {\bibinfo  {journal} {Open Syst. Inf. Dyn.}\
  }\textbf {\bibinfo {volume} {21}},\ \bibinfo {pages} {1440001} (\bibinfo
  {year} {2014})}\BibitemShut {NoStop}%
\bibitem [{\citenamefont {Mart\'{\i}n-Mart\'{\i}nez}\ \emph
  {et~al.}(2013)\citenamefont {Mart\'{\i}n-Mart\'{\i}nez}, \citenamefont
  {Brown}, \citenamefont {Donnelly},\ and\ \citenamefont
  {Kempf}}]{Brown:Martin-Martinez:2013}%
  \BibitemOpen
  \bibfield  {author} {\bibinfo {author} {\bibfnamefont {E.}~\bibnamefont
  {Mart\'{\i}n-Mart\'{\i}nez}}, \bibinfo {author} {\bibfnamefont {E.~G.}\
  \bibnamefont {Brown}}, \bibinfo {author} {\bibfnamefont {W.}~\bibnamefont
  {Donnelly}},\ and\ \bibinfo {author} {\bibfnamefont {A.}~\bibnamefont
  {Kempf}},\ }\href {https://doi.org/10.1103/PhysRevA.88.052310} {\bibfield
  {journal} {\bibinfo  {journal} {Phys. Rev. A}\ }\textbf {\bibinfo {volume}
  {88}},\ \bibinfo {pages} {052310} (\bibinfo {year} {2013})}\BibitemShut
  {NoStop}%
\bibitem [{\citenamefont {Bruschi}\ and\ \citenamefont
  {Xuereb}(2018)}]{Bruschi:Xuereb:2018}%
  \BibitemOpen
  \bibfield  {author} {\bibinfo {author} {\bibfnamefont {D.~E.}\ \bibnamefont
  {Bruschi}}\ and\ \bibinfo {author} {\bibfnamefont {A.}~\bibnamefont
  {Xuereb}},\ }\href {https://doi.org/10.1088/1367-2630/aaca27} {\bibfield
  {journal} {\bibinfo  {journal} {New Journal of Physics}\ }\textbf {\bibinfo
  {volume} {20}},\ \bibinfo {pages} {065004} (\bibinfo {year} {2018})},\
  \bibinfo {note} {publisher: IOP Publishing}\BibitemShut {NoStop}%
\bibitem [{\citenamefont {Williamson}(1936)}]{Williamson:1923}%
  \BibitemOpen
  \bibfield  {author} {\bibinfo {author} {\bibfnamefont {J.}~\bibnamefont
  {Williamson}},\ }\href@noop {} {\bibfield  {journal} {\bibinfo  {journal}
  {American Journal of Mathematics}\ }\textbf {\bibinfo {volume} {58}},\
  \bibinfo {pages} {141} (\bibinfo {year} {1936})}\BibitemShut {NoStop}%
\bibitem [{\citenamefont {Giovannetti}\ \emph {et~al.}(2011)\citenamefont
  {Giovannetti}, \citenamefont {Lloyd},\ and\ \citenamefont
  {Maccone}}]{Giovannetti:Lloyd:2011}%
  \BibitemOpen
  \bibfield  {author} {\bibinfo {author} {\bibfnamefont {V.}~\bibnamefont
  {Giovannetti}}, \bibinfo {author} {\bibfnamefont {S.}~\bibnamefont {Lloyd}},\
  and\ \bibinfo {author} {\bibfnamefont {L.}~\bibnamefont {Maccone}},\ }\href
  {https://doi.org/10.1038/nphoton.2011.35} {\bibfield  {journal} {\bibinfo
  {journal} {Nature Photonics}\ }\textbf {\bibinfo {volume} {5}},\ \bibinfo
  {pages} {222} (\bibinfo {year} {2011})}\BibitemShut {NoStop}%
\bibitem [{\citenamefont {Hovhannisyan}\ \emph {et~al.}(2021)\citenamefont
  {Hovhannisyan}, \citenamefont {J\o{}rgensen}, \citenamefont {Landi},
  \citenamefont {Alhambra}, \citenamefont {Brask},\ and\ \citenamefont
  {Perarnau-Llobet}}]{Hovhannisyan:Jorgensen:2021}%
  \BibitemOpen
  \bibfield  {author} {\bibinfo {author} {\bibfnamefont {K.~V.}\ \bibnamefont
  {Hovhannisyan}}, \bibinfo {author} {\bibfnamefont {M.~R.}\ \bibnamefont
  {J\o{}rgensen}}, \bibinfo {author} {\bibfnamefont {G.~T.}\ \bibnamefont
  {Landi}}, \bibinfo {author} {\bibfnamefont {A.~M.}\ \bibnamefont {Alhambra}},
  \bibinfo {author} {\bibfnamefont {J.~B.}\ \bibnamefont {Brask}},\ and\
  \bibinfo {author} {\bibfnamefont {M.}~\bibnamefont {Perarnau-Llobet}},\
  }\href {https://doi.org/10.1103/PRXQuantum.2.020322} {\bibfield  {journal}
  {\bibinfo  {journal} {PRX Quantum}\ }\textbf {\bibinfo {volume} {2}},\
  \bibinfo {pages} {020322} (\bibinfo {year} {2021})}\BibitemShut {NoStop}%
\bibitem [{\citenamefont {Wang}\ \emph {et~al.}(2019)\citenamefont {Wang},
  \citenamefont {Wu}, \citenamefont {Ma}, \citenamefont {Cai}, \citenamefont
  {Hu}, \citenamefont {Mu}, \citenamefont {Xu}, \citenamefont {Chen},
  \citenamefont {Wang}, \citenamefont {Song}, \citenamefont {Yuan},
  \citenamefont {Zou}, \citenamefont {Duan},\ and\ \citenamefont
  {Sun}}]{Wang:Wu:2019}%
  \BibitemOpen
  \bibfield  {author} {\bibinfo {author} {\bibfnamefont {W.}~\bibnamefont
  {Wang}}, \bibinfo {author} {\bibfnamefont {Y.}~\bibnamefont {Wu}}, \bibinfo
  {author} {\bibfnamefont {Y.}~\bibnamefont {Ma}}, \bibinfo {author}
  {\bibfnamefont {W.}~\bibnamefont {Cai}}, \bibinfo {author} {\bibfnamefont
  {L.}~\bibnamefont {Hu}}, \bibinfo {author} {\bibfnamefont {X.}~\bibnamefont
  {Mu}}, \bibinfo {author} {\bibfnamefont {Y.}~\bibnamefont {Xu}}, \bibinfo
  {author} {\bibfnamefont {Z.-J.}\ \bibnamefont {Chen}}, \bibinfo {author}
  {\bibfnamefont {H.}~\bibnamefont {Wang}}, \bibinfo {author} {\bibfnamefont
  {Y.~P.}\ \bibnamefont {Song}}, \bibinfo {author} {\bibfnamefont
  {H.}~\bibnamefont {Yuan}}, \bibinfo {author} {\bibfnamefont {C.~L.}\
  \bibnamefont {Zou}}, \bibinfo {author} {\bibfnamefont {L.~M.}\ \bibnamefont
  {Duan}},\ and\ \bibinfo {author} {\bibfnamefont {L.}~\bibnamefont {Sun}},\
  }\href {https://doi.org/10.1038/s41467-019-12290-7} {\bibfield  {journal}
  {\bibinfo  {journal} {Nature Communications}\ }\textbf {\bibinfo {volume}
  {10}},\ \bibinfo {pages} {4382} (\bibinfo {year} {2019})}\BibitemShut
  {NoStop}%
\bibitem [{\citenamefont {Ahmadi}\ \emph {et~al.}(2014)\citenamefont {Ahmadi},
  \citenamefont {Bruschi},\ and\ \citenamefont
  {Fuentes}}]{Ahmadi:Bruschi:2014}%
  \BibitemOpen
  \bibfield  {author} {\bibinfo {author} {\bibfnamefont {M.}~\bibnamefont
  {Ahmadi}}, \bibinfo {author} {\bibfnamefont {D.~E.}\ \bibnamefont
  {Bruschi}},\ and\ \bibinfo {author} {\bibfnamefont {I.}~\bibnamefont
  {Fuentes}},\ }\href {https://doi.org/10.1103/PhysRevD.89.065028} {\bibfield
  {journal} {\bibinfo  {journal} {Phys. Rev. D}\ }\textbf {\bibinfo {volume}
  {89}},\ \bibinfo {pages} {065028} (\bibinfo {year} {2014})}\BibitemShut
  {NoStop}%
\bibitem [{\citenamefont {Polino}\ \emph {et~al.}(2020)\citenamefont {Polino},
  \citenamefont {Valeri}, \citenamefont {Spagnolo},\ and\ \citenamefont
  {Sciarrino}}]{Polino:Valeri:2020}%
  \BibitemOpen
  \bibfield  {author} {\bibinfo {author} {\bibfnamefont {E.}~\bibnamefont
  {Polino}}, \bibinfo {author} {\bibfnamefont {M.}~\bibnamefont {Valeri}},
  \bibinfo {author} {\bibfnamefont {N.}~\bibnamefont {Spagnolo}},\ and\
  \bibinfo {author} {\bibfnamefont {F.}~\bibnamefont {Sciarrino}},\ }\href
  {https://doi.org/10.1116/5.0007577} {\bibfield  {journal} {\bibinfo
  {journal} {AVS Quantum Science}\ }\textbf {\bibinfo {volume} {2}},\ \bibinfo
  {pages} {024703} (\bibinfo {year} {2020})},\ \bibinfo {note} {publisher:
  American Vacuum Society}\BibitemShut {NoStop}%
\bibitem [{\citenamefont {Kohlrus}\ \emph {et~al.}(2019)\citenamefont
  {Kohlrus}, \citenamefont {Bruschi},\ and\ \citenamefont
  {Fuentes}}]{Kohlrus:Bruschi:Fuentes:2019}%
  \BibitemOpen
  \bibfield  {author} {\bibinfo {author} {\bibfnamefont {J.}~\bibnamefont
  {Kohlrus}}, \bibinfo {author} {\bibfnamefont {D.~E.}\ \bibnamefont
  {Bruschi}},\ and\ \bibinfo {author} {\bibfnamefont {I.}~\bibnamefont
  {Fuentes}},\ }\href {https://doi.org/10.1103/PhysRevA.99.032350} {\bibfield
  {journal} {\bibinfo  {journal} {Phys. Rev. A}\ }\textbf {\bibinfo {volume}
  {99}},\ \bibinfo {pages} {032350} (\bibinfo {year} {2019})}\BibitemShut
  {NoStop}%
\bibitem [{\citenamefont {Braunstein}\ and\ \citenamefont
  {Caves}(1994)}]{Braunstein:Caves:1994}%
  \BibitemOpen
  \bibfield  {author} {\bibinfo {author} {\bibfnamefont {S.~L.}\ \bibnamefont
  {Braunstein}}\ and\ \bibinfo {author} {\bibfnamefont {C.~M.}\ \bibnamefont
  {Caves}},\ }\href {https://doi.org/10.1103/PhysRevLett.72.3439} {\bibfield
  {journal} {\bibinfo  {journal} {Phys. Rev. Lett.}\ }\textbf {\bibinfo
  {volume} {72}},\ \bibinfo {pages} {3439} (\bibinfo {year}
  {1994})}\BibitemShut {NoStop}%
\bibitem [{\citenamefont {Vidrighin}\ \emph {et~al.}(2014)\citenamefont
  {Vidrighin}, \citenamefont {Donati}, \citenamefont {Genoni}, \citenamefont
  {Jin}, \citenamefont {Kolthammer}, \citenamefont {Kim}, \citenamefont
  {Datta}, \citenamefont {Barbieri},\ and\ \citenamefont
  {Walmsley}}]{Vidrighin:Donati:2014}%
  \BibitemOpen
  \bibfield  {author} {\bibinfo {author} {\bibfnamefont {M.~D.}\ \bibnamefont
  {Vidrighin}}, \bibinfo {author} {\bibfnamefont {G.}~\bibnamefont {Donati}},
  \bibinfo {author} {\bibfnamefont {M.~G.}\ \bibnamefont {Genoni}}, \bibinfo
  {author} {\bibfnamefont {X.-M.}\ \bibnamefont {Jin}}, \bibinfo {author}
  {\bibfnamefont {W.~S.}\ \bibnamefont {Kolthammer}}, \bibinfo {author}
  {\bibfnamefont {M.}~\bibnamefont {Kim}}, \bibinfo {author} {\bibfnamefont
  {A.}~\bibnamefont {Datta}}, \bibinfo {author} {\bibfnamefont
  {M.}~\bibnamefont {Barbieri}},\ and\ \bibinfo {author} {\bibfnamefont
  {I.~A.}\ \bibnamefont {Walmsley}},\ }\href
  {https://doi.org/10.1038/ncomms4532} {\bibfield  {journal} {\bibinfo
  {journal} {Nature Communications}\ }\textbf {\bibinfo {volume} {5}},\
  \bibinfo {pages} {3532} (\bibinfo {year} {2014})}\BibitemShut {NoStop}%
\bibitem [{\citenamefont {Marian}\ and\ \citenamefont
  {Marian}(2008)}]{Marian:Marian:2008}%
  \BibitemOpen
  \bibfield  {author} {\bibinfo {author} {\bibfnamefont {P.}~\bibnamefont
  {Marian}}\ and\ \bibinfo {author} {\bibfnamefont {T.~A.}\ \bibnamefont
  {Marian}},\ }\href {https://doi.org/10.1103/PhysRevA.77.062319} {\bibfield
  {journal} {\bibinfo  {journal} {Phys. Rev. A}\ }\textbf {\bibinfo {volume}
  {77}},\ \bibinfo {pages} {062319} (\bibinfo {year} {2008})}\BibitemShut
  {NoStop}%
\bibitem [{\citenamefont {Marian}\ and\ \citenamefont
  {Marian}(2012)}]{Marian:Marian:2012}%
  \BibitemOpen
  \bibfield  {author} {\bibinfo {author} {\bibfnamefont {P.}~\bibnamefont
  {Marian}}\ and\ \bibinfo {author} {\bibfnamefont {T.~A.}\ \bibnamefont
  {Marian}},\ }\href {https://doi.org/10.1103/PhysRevA.86.022340} {\bibfield
  {journal} {\bibinfo  {journal} {Phys. Rev. A}\ }\textbf {\bibinfo {volume}
  {86}},\ \bibinfo {pages} {022340} (\bibinfo {year} {2012})}\BibitemShut
  {NoStop}%
\bibitem [{\citenamefont {Lundeen}\ \emph {et~al.}(2011)\citenamefont
  {Lundeen}, \citenamefont {Sutherland}, \citenamefont {Patel}, \citenamefont
  {Stewart},\ and\ \citenamefont {Bamber}}]{Lundeen:Sutherland:2011}%
  \BibitemOpen
  \bibfield  {author} {\bibinfo {author} {\bibfnamefont {J.~S.}\ \bibnamefont
  {Lundeen}}, \bibinfo {author} {\bibfnamefont {B.}~\bibnamefont {Sutherland}},
  \bibinfo {author} {\bibfnamefont {A.}~\bibnamefont {Patel}}, \bibinfo
  {author} {\bibfnamefont {C.}~\bibnamefont {Stewart}},\ and\ \bibinfo {author}
  {\bibfnamefont {C.}~\bibnamefont {Bamber}},\ }\href
  {https://doi.org/10.1038/nature10120} {\bibfield  {journal} {\bibinfo
  {journal} {Nature}\ }\textbf {\bibinfo {volume} {474}},\ \bibinfo {pages}
  {188} (\bibinfo {year} {2011})}\BibitemShut {NoStop}%
\bibitem [{\citenamefont {Chrapkiewicz}\ \emph {et~al.}(2016)\citenamefont
  {Chrapkiewicz}, \citenamefont {Jachura}, \citenamefont {Banaszek},\ and\
  \citenamefont {Wasilewski}}]{Chrapkiewicz:Radoslaw:2016}%
  \BibitemOpen
  \bibfield  {author} {\bibinfo {author} {\bibfnamefont {R.}~\bibnamefont
  {Chrapkiewicz}}, \bibinfo {author} {\bibfnamefont {M.}~\bibnamefont
  {Jachura}}, \bibinfo {author} {\bibfnamefont {K.}~\bibnamefont {Banaszek}},\
  and\ \bibinfo {author} {\bibfnamefont {W.}~\bibnamefont {Wasilewski}},\
  }\href {https://doi.org/10.1038/nphoton.2016.129} {\bibfield  {journal}
  {\bibinfo  {journal} {Nature Photonics}\ }\textbf {\bibinfo {volume} {10}},\
  \bibinfo {pages} {576} (\bibinfo {year} {2016})}\BibitemShut {NoStop}%
\bibitem [{\citenamefont {Menssen}\ \emph {et~al.}(2017)\citenamefont
  {Menssen}, \citenamefont {Jones}, \citenamefont {Metcalf}, \citenamefont
  {Tichy}, \citenamefont {Barz}, \citenamefont {Kolthammer},\ and\
  \citenamefont {Walmsley}}]{Menessen:Jones:2017}%
  \BibitemOpen
  \bibfield  {author} {\bibinfo {author} {\bibfnamefont {A.~J.}\ \bibnamefont
  {Menssen}}, \bibinfo {author} {\bibfnamefont {A.~E.}\ \bibnamefont {Jones}},
  \bibinfo {author} {\bibfnamefont {B.~J.}\ \bibnamefont {Metcalf}}, \bibinfo
  {author} {\bibfnamefont {M.~C.}\ \bibnamefont {Tichy}}, \bibinfo {author}
  {\bibfnamefont {S.}~\bibnamefont {Barz}}, \bibinfo {author} {\bibfnamefont
  {W.~S.}\ \bibnamefont {Kolthammer}},\ and\ \bibinfo {author} {\bibfnamefont
  {I.~A.}\ \bibnamefont {Walmsley}},\ }\href
  {https://doi.org/10.1103/PhysRevLett.118.153603} {\bibfield  {journal}
  {\bibinfo  {journal} {Phys. Rev. Lett.}\ }\textbf {\bibinfo {volume} {118}},\
  \bibinfo {pages} {153603} (\bibinfo {year} {2017})}\BibitemShut {NoStop}%
\bibitem [{\citenamefont {Duan}\ \emph {et~al.}(2001)\citenamefont {Duan},
  \citenamefont {Lukin}, \citenamefont {Cirac},\ and\ \citenamefont
  {Zoller}}]{Duan:Lukin:2001}%
  \BibitemOpen
  \bibfield  {author} {\bibinfo {author} {\bibfnamefont {L.-M.}\ \bibnamefont
  {Duan}}, \bibinfo {author} {\bibfnamefont {M.~D.}\ \bibnamefont {Lukin}},
  \bibinfo {author} {\bibfnamefont {J.~I.}\ \bibnamefont {Cirac}},\ and\
  \bibinfo {author} {\bibfnamefont {P.}~\bibnamefont {Zoller}},\ }\href
  {https://doi.org/10.1038/35106500} {\bibfield  {journal} {\bibinfo  {journal}
  {Nature}\ }\textbf {\bibinfo {volume} {414}},\ \bibinfo {pages} {413}
  (\bibinfo {year} {2001})}\BibitemShut {NoStop}%
\bibitem [{\citenamefont {Vallone}\ \emph {et~al.}(2015)\citenamefont
  {Vallone}, \citenamefont {Bacco}, \citenamefont {Dequal}, \citenamefont
  {Gaiarin}, \citenamefont {Luceri}, \citenamefont {Bianco},\ and\
  \citenamefont {Villoresi}}]{Vallone:Bacco:2015}%
  \BibitemOpen
  \bibfield  {author} {\bibinfo {author} {\bibfnamefont {G.}~\bibnamefont
  {Vallone}}, \bibinfo {author} {\bibfnamefont {D.}~\bibnamefont {Bacco}},
  \bibinfo {author} {\bibfnamefont {D.}~\bibnamefont {Dequal}}, \bibinfo
  {author} {\bibfnamefont {S.}~\bibnamefont {Gaiarin}}, \bibinfo {author}
  {\bibfnamefont {V.}~\bibnamefont {Luceri}}, \bibinfo {author} {\bibfnamefont
  {G.}~\bibnamefont {Bianco}},\ and\ \bibinfo {author} {\bibfnamefont
  {P.}~\bibnamefont {Villoresi}},\ }\href
  {https://link.aps.org/doi/10.1103/PhysRevLett.115.040502} {\bibfield
  {journal} {\bibinfo  {journal} {Phys. Rev. Lett.}\ }\textbf {\bibinfo
  {volume} {115}},\ \bibinfo {pages} {040502} (\bibinfo {year}
  {2015})}\BibitemShut {NoStop}%
\bibitem [{\citenamefont {Liorni}\ \emph {et~al.}(2021)\citenamefont {Liorni},
  \citenamefont {Kampermann},\ and\ \citenamefont
  {Bruß}}]{Liorni:Kampermann:2021}%
  \BibitemOpen
  \bibfield  {author} {\bibinfo {author} {\bibfnamefont {C.}~\bibnamefont
  {Liorni}}, \bibinfo {author} {\bibfnamefont {H.}~\bibnamefont {Kampermann}},\
  and\ \bibinfo {author} {\bibfnamefont {D.}~\bibnamefont {Bruß}},\ }\href
  {https://doi.org/10.1088/1367-2630/abfa63} {\bibfield  {journal} {\bibinfo
  {journal} {New Journal of Physics}\ }\textbf {\bibinfo {volume} {23}},\
  \bibinfo {pages} {053021} (\bibinfo {year} {2021})},\ \bibinfo {note}
  {publisher: IOP Publishing}\BibitemShut {NoStop}%
\bibitem [{\citenamefont {Gündoğan}\ \emph {et~al.}(2021)\citenamefont
  {Gündoğan}, \citenamefont {Sidhu}, \citenamefont {Henderson}, \citenamefont
  {Mazzarella}, \citenamefont {Wolters}, \citenamefont {Oi},\ and\
  \citenamefont {Krutzik}}]{Gundogan:Sidhu:2021}%
  \BibitemOpen
  \bibfield  {author} {\bibinfo {author} {\bibfnamefont {M.}~\bibnamefont
  {Gündoğan}}, \bibinfo {author} {\bibfnamefont {J.~S.}\ \bibnamefont
  {Sidhu}}, \bibinfo {author} {\bibfnamefont {V.}~\bibnamefont {Henderson}},
  \bibinfo {author} {\bibfnamefont {L.}~\bibnamefont {Mazzarella}}, \bibinfo
  {author} {\bibfnamefont {J.}~\bibnamefont {Wolters}}, \bibinfo {author}
  {\bibfnamefont {D.~K.~L.}\ \bibnamefont {Oi}},\ and\ \bibinfo {author}
  {\bibfnamefont {M.}~\bibnamefont {Krutzik}},\ }\href
  {https://doi.org/10.1038/s41534-021-00460-9} {\bibfield  {journal} {\bibinfo
  {journal} {npj Quantum Information}\ }\textbf {\bibinfo {volume} {7}},\
  \bibinfo {pages} {128} (\bibinfo {year} {2021})}\BibitemShut {NoStop}%
\bibitem [{\citenamefont {Mol}\ \emph {et~al.}(2023)\citenamefont {Mol},
  \citenamefont {Esguerra}, \citenamefont {Meister}, \citenamefont {Bruschi},
  \citenamefont {Schell}, \citenamefont {Wolters},\ and\ \citenamefont
  {Wörner}}]{Mol:Esguerra:2023}%
  \BibitemOpen
  \bibfield  {author} {\bibinfo {author} {\bibfnamefont {J.-M.}\ \bibnamefont
  {Mol}}, \bibinfo {author} {\bibfnamefont {L.}~\bibnamefont {Esguerra}},
  \bibinfo {author} {\bibfnamefont {M.}~\bibnamefont {Meister}}, \bibinfo
  {author} {\bibfnamefont {D.~E.}\ \bibnamefont {Bruschi}}, \bibinfo {author}
  {\bibfnamefont {A.~W.}\ \bibnamefont {Schell}}, \bibinfo {author}
  {\bibfnamefont {J.}~\bibnamefont {Wolters}},\ and\ \bibinfo {author}
  {\bibfnamefont {L.}~\bibnamefont {Wörner}},\ }\href
  {https://doi.org/10.1088/2058-9565/acb2f1} {\bibfield  {journal} {\bibinfo
  {journal} {Quantum Science and Technology}\ }\textbf {\bibinfo {volume}
  {8}},\ \bibinfo {pages} {024006} (\bibinfo {year} {2023})}\BibitemShut
  {NoStop}%
\bibitem [{\citenamefont {Okoshi}(2012)}]{Okoshi:2012}%
  \BibitemOpen
  \bibfield  {author} {\bibinfo {author} {\bibfnamefont {T.}~\bibnamefont
  {Okoshi}},\ }\href {https://doi.org/10.1016/B978-0-12-525260-7.X5001-X}
  {\emph {\bibinfo {title} {Optical fibers}}}\ (\bibinfo  {publisher}
  {Elsevier},\ \bibinfo {year} {2012})\BibitemShut {NoStop}%
\bibitem [{\citenamefont {Boaron}\ \emph {et~al.}(2018)\citenamefont {Boaron},
  \citenamefont {Boso}, \citenamefont {Rusca}, \citenamefont {Vulliez},
  \citenamefont {Autebert}, \citenamefont {Caloz}, \citenamefont {Perrenoud},
  \citenamefont {Gras}, \citenamefont {Bussi\`eres}, \citenamefont {Li},
  \citenamefont {Nolan}, \citenamefont {Martin},\ and\ \citenamefont
  {Zbinden}}]{Boaron:Boso:2018}%
  \BibitemOpen
  \bibfield  {author} {\bibinfo {author} {\bibfnamefont {A.}~\bibnamefont
  {Boaron}}, \bibinfo {author} {\bibfnamefont {G.}~\bibnamefont {Boso}},
  \bibinfo {author} {\bibfnamefont {D.}~\bibnamefont {Rusca}}, \bibinfo
  {author} {\bibfnamefont {C.}~\bibnamefont {Vulliez}}, \bibinfo {author}
  {\bibfnamefont {C.}~\bibnamefont {Autebert}}, \bibinfo {author}
  {\bibfnamefont {M.}~\bibnamefont {Caloz}}, \bibinfo {author} {\bibfnamefont
  {M.}~\bibnamefont {Perrenoud}}, \bibinfo {author} {\bibfnamefont
  {G.}~\bibnamefont {Gras}}, \bibinfo {author} {\bibfnamefont {F.}~\bibnamefont
  {Bussi\`eres}}, \bibinfo {author} {\bibfnamefont {M.-J.}\ \bibnamefont {Li}},
  \bibinfo {author} {\bibfnamefont {D.}~\bibnamefont {Nolan}}, \bibinfo
  {author} {\bibfnamefont {A.}~\bibnamefont {Martin}},\ and\ \bibinfo {author}
  {\bibfnamefont {H.}~\bibnamefont {Zbinden}},\ }\href
  {https://doi.org/10.1103/PhysRevLett.121.190502} {\bibfield  {journal}
  {\bibinfo  {journal} {Phys. Rev. Lett.}\ }\textbf {\bibinfo {volume} {121}},\
  \bibinfo {pages} {190502} (\bibinfo {year} {2018})}\BibitemShut {NoStop}%
\bibitem [{\citenamefont {Briegel}\ \emph {et~al.}(1998)\citenamefont
  {Briegel}, \citenamefont {D\"ur}, \citenamefont {Cirac},\ and\ \citenamefont
  {Zoller}}]{Briegel:Duer:1998}%
  \BibitemOpen
  \bibfield  {author} {\bibinfo {author} {\bibfnamefont {H.-J.}\ \bibnamefont
  {Briegel}}, \bibinfo {author} {\bibfnamefont {W.}~\bibnamefont {D\"ur}},
  \bibinfo {author} {\bibfnamefont {J.~I.}\ \bibnamefont {Cirac}},\ and\
  \bibinfo {author} {\bibfnamefont {P.}~\bibnamefont {Zoller}},\ }\href
  {https://doi.org/10.1103/PhysRevLett.81.5932} {\bibfield  {journal} {\bibinfo
   {journal} {Phys. Rev. Lett.}\ }\textbf {\bibinfo {volume} {81}},\ \bibinfo
  {pages} {5932} (\bibinfo {year} {1998})}\BibitemShut {NoStop}%
\bibitem [{\citenamefont {Ruihong}\ and\ \citenamefont
  {Ying}(2019)}]{Ruihong:Ying:2019}%
  \BibitemOpen
  \bibfield  {author} {\bibinfo {author} {\bibfnamefont {Q.}~\bibnamefont
  {Ruihong}}\ and\ \bibinfo {author} {\bibfnamefont {M.}~\bibnamefont {Ying}},\
  }\href {https://doi.org/10.1088/1742-6596/1237/5/052032} {\bibfield
  {journal} {\bibinfo  {journal} {Journal of Physics: Conference Series}\
  }\textbf {\bibinfo {volume} {1237}},\ \bibinfo {pages} {052032} (\bibinfo
  {year} {2019})},\ \bibinfo {note} {publisher: IOP Publishing}\BibitemShut
  {NoStop}%
\bibitem [{\citenamefont {Lu}\ \emph {et~al.}(2022)\citenamefont {Lu},
  \citenamefont {Cao}, \citenamefont {Peng},\ and\ \citenamefont
  {Pan}}]{Lu:Cao:2022}%
  \BibitemOpen
  \bibfield  {author} {\bibinfo {author} {\bibfnamefont {C.-Y.}\ \bibnamefont
  {Lu}}, \bibinfo {author} {\bibfnamefont {Y.}~\bibnamefont {Cao}}, \bibinfo
  {author} {\bibfnamefont {C.-Z.}\ \bibnamefont {Peng}},\ and\ \bibinfo
  {author} {\bibfnamefont {J.-W.}\ \bibnamefont {Pan}},\ }\href
  {https://doi.org/10.1103/RevModPhys.94.035001} {\bibfield  {journal}
  {\bibinfo  {journal} {Rev. Mod. Phys.}\ }\textbf {\bibinfo {volume} {94}},\
  \bibinfo {pages} {035001} (\bibinfo {year} {2022})}\BibitemShut {NoStop}%
\bibitem [{\citenamefont {Liao}\ \emph {et~al.}(2017)\citenamefont {Liao},
  \citenamefont {Cai}, \citenamefont {Liu}, \citenamefont {Zhang},
  \citenamefont {Li}, \citenamefont {Ren}, \citenamefont {Yin}, \citenamefont
  {Shen}, \citenamefont {Cao}, \citenamefont {Li}, \citenamefont {Li},
  \citenamefont {Chen} \emph {et~al.}}]{Liao:Cai:2017}%
  \BibitemOpen
  \bibfield  {author} {\bibinfo {author} {\bibfnamefont {S.-K.}\ \bibnamefont
  {Liao}}, \bibinfo {author} {\bibfnamefont {W.-Q.}\ \bibnamefont {Cai}},
  \bibinfo {author} {\bibfnamefont {W.-Y.}\ \bibnamefont {Liu}}, \bibinfo
  {author} {\bibfnamefont {L.}~\bibnamefont {Zhang}}, \bibinfo {author}
  {\bibfnamefont {Y.}~\bibnamefont {Li}}, \bibinfo {author} {\bibfnamefont
  {J.-G.}\ \bibnamefont {Ren}}, \bibinfo {author} {\bibfnamefont
  {J.}~\bibnamefont {Yin}}, \bibinfo {author} {\bibfnamefont {Q.}~\bibnamefont
  {Shen}}, \bibinfo {author} {\bibfnamefont {Y.}~\bibnamefont {Cao}}, \bibinfo
  {author} {\bibfnamefont {Z.-P.}\ \bibnamefont {Li}}, \bibinfo {author}
  {\bibfnamefont {F.-Z.}\ \bibnamefont {Li}}, \bibinfo {author} {\bibfnamefont
  {X.-W.}\ \bibnamefont {Chen}}, \emph {et~al.},\ }\href
  {https://doi.org/10.1038/nature23655} {\bibfield  {journal} {\bibinfo
  {journal} {Nature}\ }\textbf {\bibinfo {volume} {549}},\ \bibinfo {pages}
  {43} (\bibinfo {year} {2017})}\BibitemShut {NoStop}%
\bibitem [{\citenamefont {Calderaro}\ \emph {et~al.}(2018)\citenamefont
  {Calderaro}, \citenamefont {Agnesi}, \citenamefont {Dequal}, \citenamefont
  {Vedovato}, \citenamefont {Schiavon}, \citenamefont {Santamato},
  \citenamefont {Luceri}, \citenamefont {Bianco}, \citenamefont {Vallone},\
  and\ \citenamefont {Villoresi}}]{Calderaro:Agnesi:2018}%
  \BibitemOpen
  \bibfield  {author} {\bibinfo {author} {\bibfnamefont {L.}~\bibnamefont
  {Calderaro}}, \bibinfo {author} {\bibfnamefont {C.}~\bibnamefont {Agnesi}},
  \bibinfo {author} {\bibfnamefont {D.}~\bibnamefont {Dequal}}, \bibinfo
  {author} {\bibfnamefont {F.}~\bibnamefont {Vedovato}}, \bibinfo {author}
  {\bibfnamefont {M.}~\bibnamefont {Schiavon}}, \bibinfo {author}
  {\bibfnamefont {A.}~\bibnamefont {Santamato}}, \bibinfo {author}
  {\bibfnamefont {V.}~\bibnamefont {Luceri}}, \bibinfo {author} {\bibfnamefont
  {G.}~\bibnamefont {Bianco}}, \bibinfo {author} {\bibfnamefont
  {G.}~\bibnamefont {Vallone}},\ and\ \bibinfo {author} {\bibfnamefont
  {P.}~\bibnamefont {Villoresi}},\ }\href
  {http://dx.doi.org/10.1088/2058-9565/aaefd4} {\bibfield  {journal} {\bibinfo
  {journal} {Quantum Science and Technology}\ }\textbf {\bibinfo {volume}
  {4}},\ \bibinfo {pages} {015012} (\bibinfo {year} {2018})}\BibitemShut
  {NoStop}%
\bibitem [{\citenamefont {Agnesi}\ \emph {et~al.}(2019)\citenamefont {Agnesi},
  \citenamefont {Calderaro}, \citenamefont {Dequal}, \citenamefont {Vedovato},
  \citenamefont {Schiavon}, \citenamefont {Santamato}, \citenamefont {Luceri},
  \citenamefont {Bianco}, \citenamefont {Vallone},\ and\ \citenamefont
  {Villoresi}}]{Agnesi:Calderaro:2019}%
  \BibitemOpen
  \bibfield  {author} {\bibinfo {author} {\bibfnamefont {C.}~\bibnamefont
  {Agnesi}}, \bibinfo {author} {\bibfnamefont {L.}~\bibnamefont {Calderaro}},
  \bibinfo {author} {\bibfnamefont {D.}~\bibnamefont {Dequal}}, \bibinfo
  {author} {\bibfnamefont {F.}~\bibnamefont {Vedovato}}, \bibinfo {author}
  {\bibfnamefont {M.}~\bibnamefont {Schiavon}}, \bibinfo {author}
  {\bibfnamefont {A.}~\bibnamefont {Santamato}}, \bibinfo {author}
  {\bibfnamefont {V.}~\bibnamefont {Luceri}}, \bibinfo {author} {\bibfnamefont
  {G.}~\bibnamefont {Bianco}}, \bibinfo {author} {\bibfnamefont
  {G.}~\bibnamefont {Vallone}},\ and\ \bibinfo {author} {\bibfnamefont
  {P.}~\bibnamefont {Villoresi}},\ }\href
  {https://doi.org/10.1364/JOSAB.36.000B59} {\bibfield  {journal} {\bibinfo
  {journal} {J. Opt. Soc. Am. B}\ }\textbf {\bibinfo {volume} {36}},\ \bibinfo
  {pages} {B59} (\bibinfo {year} {2019})}\BibitemShut {NoStop}%
\bibitem [{\citenamefont {Sidhu}\ \emph {et~al.}(2021)\citenamefont {Sidhu},
  \citenamefont {Joshi}, \citenamefont {Gündoğan}, \citenamefont {Brougham},
  \citenamefont {Lowndes}, \citenamefont {Mazzarella}, \citenamefont {Krutzik},
  \citenamefont {Mohapatra}, \citenamefont {Dequal}, \citenamefont {Vallone},
  \citenamefont {Villoresi}, \citenamefont {Ling}, \citenamefont {Jennewein}
  \emph {et~al.}}]{Sidu:Joshi:2021}%
  \BibitemOpen
  \bibfield  {author} {\bibinfo {author} {\bibfnamefont {J.~S.}\ \bibnamefont
  {Sidhu}}, \bibinfo {author} {\bibfnamefont {S.~K.}\ \bibnamefont {Joshi}},
  \bibinfo {author} {\bibfnamefont {M.}~\bibnamefont {Gündoğan}}, \bibinfo
  {author} {\bibfnamefont {T.}~\bibnamefont {Brougham}}, \bibinfo {author}
  {\bibfnamefont {D.}~\bibnamefont {Lowndes}}, \bibinfo {author} {\bibfnamefont
  {L.}~\bibnamefont {Mazzarella}}, \bibinfo {author} {\bibfnamefont
  {M.}~\bibnamefont {Krutzik}}, \bibinfo {author} {\bibfnamefont
  {S.}~\bibnamefont {Mohapatra}}, \bibinfo {author} {\bibfnamefont
  {D.}~\bibnamefont {Dequal}}, \bibinfo {author} {\bibfnamefont
  {G.}~\bibnamefont {Vallone}}, \bibinfo {author} {\bibfnamefont
  {P.}~\bibnamefont {Villoresi}}, \bibinfo {author} {\bibfnamefont
  {A.}~\bibnamefont {Ling}}, \bibinfo {author} {\bibfnamefont {T.}~\bibnamefont
  {Jennewein}}, \emph {et~al.},\ }\href {https://doi.org/10.1049/qtc2.12015}
  {\bibfield  {journal} {\bibinfo  {journal} {IET Quantum Communication}\
  }\textbf {\bibinfo {volume} {2}},\ \bibinfo {pages} {182} (\bibinfo {year}
  {2021})},\ \bibinfo {note} {publisher: The Institution of Engineering and
  Technology}\BibitemShut {NoStop}%
\bibitem [{\citenamefont {Cuomo}\ \emph {et~al.}(2020)\citenamefont {Cuomo},
  \citenamefont {Caleffi},\ and\ \citenamefont
  {Cacciapuoti}}]{Cuomo:Caleffi:2020}%
  \BibitemOpen
  \bibfield  {author} {\bibinfo {author} {\bibfnamefont {D.}~\bibnamefont
  {Cuomo}}, \bibinfo {author} {\bibfnamefont {M.}~\bibnamefont {Caleffi}},\
  and\ \bibinfo {author} {\bibfnamefont {A.~S.}\ \bibnamefont {Cacciapuoti}},\
  }\href {https://doi.org/10.1049/iet-qtc.2020.0002} {\bibfield  {journal}
  {\bibinfo  {journal} {IET Quantum Communication}\ }\textbf {\bibinfo {volume}
  {1}},\ \bibinfo {pages} {3} (\bibinfo {year} {2020})},\ \bibinfo {note}
  {publisher: The Institution of Engineering and Technology}\BibitemShut
  {NoStop}%
\bibitem [{\citenamefont {Pirandola}\ \emph {et~al.}(2020)\citenamefont
  {Pirandola}, \citenamefont {Andersen}, \citenamefont {Banchi}, \citenamefont
  {Berta}, \citenamefont {Bunandar}, \citenamefont {Colbeck}, \citenamefont
  {Englund}, \citenamefont {Gehring}, \citenamefont {Lupo}, \citenamefont
  {Ottaviani}, \citenamefont {Pereira}, \citenamefont {Razavi}, \citenamefont
  {Shaari}, \citenamefont {Tomamichel}, \citenamefont {Usenko}, \citenamefont
  {Vallone}, \citenamefont {Villoresi},\ and\ \citenamefont
  {Wallden}}]{Pirandola:Andersen:2020}%
  \BibitemOpen
  \bibfield  {author} {\bibinfo {author} {\bibfnamefont {S.}~\bibnamefont
  {Pirandola}}, \bibinfo {author} {\bibfnamefont {U.~L.}\ \bibnamefont
  {Andersen}}, \bibinfo {author} {\bibfnamefont {L.}~\bibnamefont {Banchi}},
  \bibinfo {author} {\bibfnamefont {M.}~\bibnamefont {Berta}}, \bibinfo
  {author} {\bibfnamefont {D.}~\bibnamefont {Bunandar}}, \bibinfo {author}
  {\bibfnamefont {R.}~\bibnamefont {Colbeck}}, \bibinfo {author} {\bibfnamefont
  {D.}~\bibnamefont {Englund}}, \bibinfo {author} {\bibfnamefont
  {T.}~\bibnamefont {Gehring}}, \bibinfo {author} {\bibfnamefont
  {C.}~\bibnamefont {Lupo}}, \bibinfo {author} {\bibfnamefont {C.}~\bibnamefont
  {Ottaviani}}, \bibinfo {author} {\bibfnamefont {J.~L.}\ \bibnamefont
  {Pereira}}, \bibinfo {author} {\bibfnamefont {M.}~\bibnamefont {Razavi}},
  \bibinfo {author} {\bibfnamefont {J.~S.}\ \bibnamefont {Shaari}}, \bibinfo
  {author} {\bibfnamefont {M.}~\bibnamefont {Tomamichel}}, \bibinfo {author}
  {\bibfnamefont {V.~C.}\ \bibnamefont {Usenko}}, \bibinfo {author}
  {\bibfnamefont {G.}~\bibnamefont {Vallone}}, \bibinfo {author} {\bibfnamefont
  {P.}~\bibnamefont {Villoresi}},\ and\ \bibinfo {author} {\bibfnamefont
  {P.}~\bibnamefont {Wallden}},\ }\href {https://doi.org/10.1364/AOP.361502}
  {\bibfield  {journal} {\bibinfo  {journal} {Adv. Opt. Photon.}\ }\textbf
  {\bibinfo {volume} {12}},\ \bibinfo {pages} {1012} (\bibinfo {year}
  {2020})}\BibitemShut {NoStop}%
\bibitem [{\citenamefont {Bennett}\ and\ \citenamefont
  {Brassard}(2014)}]{Bennet:Brassard:1984}%
  \BibitemOpen
  \bibfield  {author} {\bibinfo {author} {\bibfnamefont {C.~H.}\ \bibnamefont
  {Bennett}}\ and\ \bibinfo {author} {\bibfnamefont {G.}~\bibnamefont
  {Brassard}},\ }\href {https://doi.org/10.1016/j.tcs.2014.05.025} {\bibfield
  {journal} {\bibinfo  {journal} {Theoretical Aspects of Quantum Cryptography
  – celebrating 30 years of BB84}\ }\textbf {\bibinfo {volume} {560}},\
  \bibinfo {pages} {7} (\bibinfo {year} {2014})}\BibitemShut {NoStop}%
\bibitem [{\citenamefont {Bennett}\ \emph {et~al.}(1992)\citenamefont
  {Bennett}, \citenamefont {Brassard},\ and\ \citenamefont
  {Mermin}}]{Bennet:Brassard:1992}%
  \BibitemOpen
  \bibfield  {author} {\bibinfo {author} {\bibfnamefont {C.~H.}\ \bibnamefont
  {Bennett}}, \bibinfo {author} {\bibfnamefont {G.}~\bibnamefont {Brassard}},\
  and\ \bibinfo {author} {\bibfnamefont {N.~D.}\ \bibnamefont {Mermin}},\
  }\href {https://doi.org/10.1103/PhysRevLett.68.557} {\bibfield  {journal}
  {\bibinfo  {journal} {Phys. Rev. Lett.}\ }\textbf {\bibinfo {volume} {68}},\
  \bibinfo {pages} {557} (\bibinfo {year} {1992})}\BibitemShut {NoStop}%
\bibitem [{\citenamefont {Shor}\ and\ \citenamefont
  {Preskill}(2000)}]{Shor:Preskill:2000}%
  \BibitemOpen
  \bibfield  {author} {\bibinfo {author} {\bibfnamefont {P.~W.}\ \bibnamefont
  {Shor}}\ and\ \bibinfo {author} {\bibfnamefont {J.}~\bibnamefont
  {Preskill}},\ }\href {https://doi.org/10.1103/PhysRevLett.85.441} {\bibfield
  {journal} {\bibinfo  {journal} {Phys. Rev. Lett.}\ }\textbf {\bibinfo
  {volume} {85}},\ \bibinfo {pages} {441} (\bibinfo {year} {2000})}\BibitemShut
  {NoStop}%
\bibitem [{\citenamefont {Renner}(2005)}]{Renner:2005}%
  \BibitemOpen
  \bibfield  {author} {\bibinfo {author} {\bibfnamefont {R.}~\bibnamefont
  {Renner}},\ }\emph {\bibinfo {title} {Security of Quantum Key
  Distribution}},\ \href@noop {} {Ph.D. thesis},\ \bibinfo  {school} {{ETH
  Zurich}} (\bibinfo {year} {2005}),\ \bibinfo {note} {available at
  http://arxiv.org/abs/quant-ph/0512258}\BibitemShut {NoStop}%
\bibitem [{\citenamefont {Wootters}\ and\ \citenamefont
  {Zurek}(1982)}]{Wootters:Zurek:1982}%
  \BibitemOpen
  \bibfield  {author} {\bibinfo {author} {\bibfnamefont {W.~K.}\ \bibnamefont
  {Wootters}}\ and\ \bibinfo {author} {\bibfnamefont {W.~H.}\ \bibnamefont
  {Zurek}},\ }\href {https://doi.org/10.1038/299802a0} {\bibfield  {journal}
  {\bibinfo  {journal} {Nature}\ }\textbf {\bibinfo {volume} {299}},\ \bibinfo
  {pages} {802} (\bibinfo {year} {1982})}\BibitemShut {NoStop}%
\bibitem [{\citenamefont {Dieks}(1982)}]{Dieks:1982}%
  \BibitemOpen
  \bibfield  {author} {\bibinfo {author} {\bibfnamefont {D.}~\bibnamefont
  {Dieks}},\ }\href
  {https://doi.org/https://doi.org/10.1016/0375-9601(82)90084-6} {\bibfield
  {journal} {\bibinfo  {journal} {Physics Letters A}\ }\textbf {\bibinfo
  {volume} {92}},\ \bibinfo {pages} {271} (\bibinfo {year} {1982})}\BibitemShut
  {NoStop}%
\bibitem [{\citenamefont {Ghirardi}(2013)}]{Ghirardi:2013}%
  \BibitemOpen
  \bibfield  {author} {\bibinfo {author} {\bibfnamefont {G.}~\bibnamefont
  {Ghirardi}},\ }in\ \href {https://doi.org/10.5772/56429} {\emph {\bibinfo
  {booktitle} {Advances in Quantum Mechanics}}},\ \bibinfo {editor} {edited by\
  \bibinfo {editor} {\bibfnamefont {P.}~\bibnamefont {Bracken}}}\ (\bibinfo
  {publisher} {IntechOpen},\ \bibinfo {address} {Rijeka},\ \bibinfo {year}
  {2013})\ Chap.~\bibinfo {chapter} {24}\BibitemShut {NoStop}%
\bibitem [{\citenamefont {Ekert}(1991)}]{Ekert:1991}%
  \BibitemOpen
  \bibfield  {author} {\bibinfo {author} {\bibfnamefont {A.~K.}\ \bibnamefont
  {Ekert}},\ }\href {https://doi.org/10.1103/PhysRevLett.67.661} {\bibfield
  {journal} {\bibinfo  {journal} {Phys. Rev. Lett.}\ }\textbf {\bibinfo
  {volume} {67}},\ \bibinfo {pages} {661} (\bibinfo {year} {1991})}\BibitemShut
  {NoStop}%
\bibitem [{\citenamefont {Gisin}\ \emph {et~al.}(2002)\citenamefont {Gisin},
  \citenamefont {Ribordy}, \citenamefont {Tittel},\ and\ \citenamefont
  {Zbinden}}]{Gisin:Ribordy:Tittel:2002}%
  \BibitemOpen
  \bibfield  {author} {\bibinfo {author} {\bibfnamefont {N.}~\bibnamefont
  {Gisin}}, \bibinfo {author} {\bibfnamefont {G.}~\bibnamefont {Ribordy}},
  \bibinfo {author} {\bibfnamefont {W.}~\bibnamefont {Tittel}},\ and\ \bibinfo
  {author} {\bibfnamefont {H.}~\bibnamefont {Zbinden}},\ }\href
  {https://doi.org/10.1103/RevModPhys.74.145} {\bibfield  {journal} {\bibinfo
  {journal} {Rev. Mod. Phys.}\ }\textbf {\bibinfo {volume} {74}},\ \bibinfo
  {pages} {145} (\bibinfo {year} {2002})}\BibitemShut {NoStop}%
\bibitem [{\citenamefont {Barzel}\ \emph {et~al.}(2022)\citenamefont {Barzel},
  \citenamefont {Bruschi}, \citenamefont {Schell},\ and\ \citenamefont
  {Lämmerzahl}}]{Barzel:Bruschi:2022}%
  \BibitemOpen
  \bibfield  {author} {\bibinfo {author} {\bibfnamefont {R.}~\bibnamefont
  {Barzel}}, \bibinfo {author} {\bibfnamefont {D.~E.}\ \bibnamefont {Bruschi}},
  \bibinfo {author} {\bibfnamefont {A.~W.}\ \bibnamefont {Schell}},\ and\
  \bibinfo {author} {\bibfnamefont {C.}~\bibnamefont {Lämmerzahl}},\
  }\bibfield  {journal} {\bibinfo  {journal} {Physical Review D}\ }\textbf
  {\bibinfo {volume} {105}},\ \href
  {https://doi.org/10.1103/physrevd.105.105016} {10.1103/physrevd.105.105016}
  (\bibinfo {year} {2022})\BibitemShut {NoStop}%
\bibitem [{\citenamefont {Vilasini}\ \emph {et~al.}(2019)\citenamefont
  {Vilasini}, \citenamefont {Portmann},\ and\ \citenamefont {del
  Rio}}]{Vilasini:Portmann:2019}%
  \BibitemOpen
  \bibfield  {author} {\bibinfo {author} {\bibfnamefont {V.}~\bibnamefont
  {Vilasini}}, \bibinfo {author} {\bibfnamefont {C.}~\bibnamefont {Portmann}},\
  and\ \bibinfo {author} {\bibfnamefont {L.}~\bibnamefont {del Rio}},\ }\href
  {https://doi.org/10.1088/1367-2630/ab0e3b} {\bibfield  {journal} {\bibinfo
  {journal} {New Journal of Physics}\ }\textbf {\bibinfo {volume} {21}},\
  \bibinfo {pages} {043057} (\bibinfo {year} {2019})},\ \bibinfo {note}
  {publisher: IOP Publishing}\BibitemShut {NoStop}%
\bibitem [{\citenamefont {Xu}\ \emph {et~al.}(2020)\citenamefont {Xu},
  \citenamefont {Ma}, \citenamefont {Zhang}, \citenamefont {Lo},\ and\
  \citenamefont {Pan}}]{Xu:Ma:2020}%
  \BibitemOpen
  \bibfield  {author} {\bibinfo {author} {\bibfnamefont {F.}~\bibnamefont
  {Xu}}, \bibinfo {author} {\bibfnamefont {X.}~\bibnamefont {Ma}}, \bibinfo
  {author} {\bibfnamefont {Q.}~\bibnamefont {Zhang}}, \bibinfo {author}
  {\bibfnamefont {H.-K.}\ \bibnamefont {Lo}},\ and\ \bibinfo {author}
  {\bibfnamefont {J.-W.}\ \bibnamefont {Pan}},\ }\href
  {https://doi.org/10.1103/RevModPhys.92.025002} {\bibfield  {journal}
  {\bibinfo  {journal} {Rev. Mod. Phys.}\ }\textbf {\bibinfo {volume} {92}},\
  \bibinfo {pages} {025002} (\bibinfo {year} {2020})}\BibitemShut {NoStop}%
\bibitem [{\citenamefont {Tan}\ \emph {et~al.}(2008)\citenamefont {Tan},
  \citenamefont {Erkmen}, \citenamefont {Giovannetti}, \citenamefont {Guha},
  \citenamefont {Lloyd}, \citenamefont {Maccone}, \citenamefont {Pirandola},\
  and\ \citenamefont {Shapiro}}]{Tan:Erkmen:2008}%
  \BibitemOpen
  \bibfield  {author} {\bibinfo {author} {\bibfnamefont {S.-H.}\ \bibnamefont
  {Tan}}, \bibinfo {author} {\bibfnamefont {B.~I.}\ \bibnamefont {Erkmen}},
  \bibinfo {author} {\bibfnamefont {V.}~\bibnamefont {Giovannetti}}, \bibinfo
  {author} {\bibfnamefont {S.}~\bibnamefont {Guha}}, \bibinfo {author}
  {\bibfnamefont {S.}~\bibnamefont {Lloyd}}, \bibinfo {author} {\bibfnamefont
  {L.}~\bibnamefont {Maccone}}, \bibinfo {author} {\bibfnamefont
  {S.}~\bibnamefont {Pirandola}},\ and\ \bibinfo {author} {\bibfnamefont
  {J.~H.}\ \bibnamefont {Shapiro}},\ }\href
  {https://doi.org/10.1103/PhysRevLett.101.253601} {\bibfield  {journal}
  {\bibinfo  {journal} {Phys. Rev. Lett.}\ }\textbf {\bibinfo {volume} {101}},\
  \bibinfo {pages} {253601} (\bibinfo {year} {2008})}\BibitemShut {NoStop}%
\bibitem [{\citenamefont {Liu}\ \emph {et~al.}(2022)\citenamefont {Liu},
  \citenamefont {Wen}, \citenamefont {Tian}, \citenamefont {Jing},\ and\
  \citenamefont {Wang}}]{Liu:Wen:2022}%
  \BibitemOpen
  \bibfield  {author} {\bibinfo {author} {\bibfnamefont {Q.}~\bibnamefont
  {Liu}}, \bibinfo {author} {\bibfnamefont {C.}~\bibnamefont {Wen}}, \bibinfo
  {author} {\bibfnamefont {Z.}~\bibnamefont {Tian}}, \bibinfo {author}
  {\bibfnamefont {J.}~\bibnamefont {Jing}},\ and\ \bibinfo {author}
  {\bibfnamefont {J.}~\bibnamefont {Wang}},\ }\href
  {https://doi.org/10.1103/PhysRevA.105.062428} {\bibfield  {journal} {\bibinfo
   {journal} {Phys. Rev. A}\ }\textbf {\bibinfo {volume} {105}},\ \bibinfo
  {pages} {062428} (\bibinfo {year} {2022})}\BibitemShut {NoStop}%
\bibitem [{\citenamefont {\ifmmode~\dot{Z}\else \.{Z}\fi{}ukowski}\ \emph
  {et~al.}(1993)\citenamefont {\ifmmode~\dot{Z}\else \.{Z}\fi{}ukowski},
  \citenamefont {Zeilinger}, \citenamefont {Horne},\ and\ \citenamefont
  {Ekert}}]{Zukowski:Zeilinger:1993}%
  \BibitemOpen
  \bibfield  {author} {\bibinfo {author} {\bibfnamefont {M.}~\bibnamefont
  {\ifmmode~\dot{Z}\else \.{Z}\fi{}ukowski}}, \bibinfo {author} {\bibfnamefont
  {A.}~\bibnamefont {Zeilinger}}, \bibinfo {author} {\bibfnamefont {M.~A.}\
  \bibnamefont {Horne}},\ and\ \bibinfo {author} {\bibfnamefont {A.~K.}\
  \bibnamefont {Ekert}},\ }\href {https://doi.org/10.1103/PhysRevLett.71.4287}
  {\bibfield  {journal} {\bibinfo  {journal} {Phys. Rev. Lett.}\ }\textbf
  {\bibinfo {volume} {71}},\ \bibinfo {pages} {4287} (\bibinfo {year}
  {1993})}\BibitemShut {NoStop}%
\bibitem [{\citenamefont {Sen(De)}\ \emph {et~al.}(2005)\citenamefont
  {Sen(De)}, \citenamefont {Sen}, \citenamefont {Brukner}, \citenamefont
  {Bu\ifmmode~\check{z}\else \v{z}\fi{}ek},\ and\ \citenamefont
  {\ifmmode~\dot{Z}\else \.{Z}\fi{}ukowski}}]{Sen:Sen:2005}%
  \BibitemOpen
  \bibfield  {author} {\bibinfo {author} {\bibfnamefont {A.}~\bibnamefont
  {Sen(De)}}, \bibinfo {author} {\bibfnamefont {U.}~\bibnamefont {Sen}},
  \bibinfo {author} {\bibfnamefont {i.~c.~v.}\ \bibnamefont {Brukner}},
  \bibinfo {author} {\bibfnamefont {V.}~\bibnamefont {Bu\ifmmode~\check{z}\else
  \v{z}\fi{}ek}},\ and\ \bibinfo {author} {\bibfnamefont {M.}~\bibnamefont
  {\ifmmode~\dot{Z}\else \.{Z}\fi{}ukowski}},\ }\href
  {https://doi.org/10.1103/PhysRevA.72.042310} {\bibfield  {journal} {\bibinfo
  {journal} {Phys. Rev. A}\ }\textbf {\bibinfo {volume} {72}},\ \bibinfo
  {pages} {042310} (\bibinfo {year} {2005})}\BibitemShut {NoStop}%
\bibitem [{\citenamefont {Degen}\ \emph {et~al.}(2017)\citenamefont {Degen},
  \citenamefont {Reinhard},\ and\ \citenamefont {Cappellaro}}]{Degen2017}%
  \BibitemOpen
  \bibfield  {author} {\bibinfo {author} {\bibfnamefont {C.~L.}\ \bibnamefont
  {Degen}}, \bibinfo {author} {\bibfnamefont {F.}~\bibnamefont {Reinhard}},\
  and\ \bibinfo {author} {\bibfnamefont {P.}~\bibnamefont {Cappellaro}},\
  }\href@noop {} {\bibfield  {journal} {\bibinfo  {journal} {Reviews of modern
  physics}\ }\textbf {\bibinfo {volume} {89}},\ \bibinfo {pages} {035002}
  (\bibinfo {year} {2017})}\BibitemShut {NoStop}%
\bibitem [{\citenamefont {Peters}\ \emph {et~al.}(2001)\citenamefont {Peters},
  \citenamefont {Chung},\ and\ \citenamefont {Chu}}]{Peters:Chung:2001}%
  \BibitemOpen
  \bibfield  {author} {\bibinfo {author} {\bibfnamefont {A.}~\bibnamefont
  {Peters}}, \bibinfo {author} {\bibfnamefont {K.~Y.}\ \bibnamefont {Chung}},\
  and\ \bibinfo {author} {\bibfnamefont {S.}~\bibnamefont {Chu}},\ }\href
  {https://doi.org/10.1088/0026-1394/38/1/4} {\bibfield  {journal} {\bibinfo
  {journal} {Metrologia}\ }\textbf {\bibinfo {volume} {38}},\ \bibinfo {pages}
  {25} (\bibinfo {year} {2001})}\BibitemShut {NoStop}%
\bibitem [{\citenamefont {Yang}\ and\ \citenamefont
  {Zhang}(2018)}]{Yang:Zhang:2018}%
  \BibitemOpen
  \bibfield  {author} {\bibinfo {author} {\bibfnamefont {S.}~\bibnamefont
  {Yang}}\ and\ \bibinfo {author} {\bibfnamefont {G.}~\bibnamefont {Zhang}},\
  }\href {https://doi.org/10.1088/1361-6501/aad732} {\bibfield  {journal}
  {\bibinfo  {journal} {Measurement Science and Technology}\ }\textbf {\bibinfo
  {volume} {29}},\ \bibinfo {pages} {102001} (\bibinfo {year}
  {2018})}\BibitemShut {NoStop}%
\bibitem [{\citenamefont {Kaltenbaek}\ \emph {et~al.}(2016)\citenamefont
  {Kaltenbaek}, \citenamefont {Aspelmeyer}, \citenamefont {Barker},
  \citenamefont {Bassi}, \citenamefont {Bateman}, \citenamefont {Bongs},
  \citenamefont {Bose}, \citenamefont {Braxmaier}, \citenamefont {Brukner},
  \citenamefont {Christophe}, \citenamefont {Chwalla}, \citenamefont {Cohadon},
  \citenamefont {Cruise} \emph {et~al.}}]{Kaltenbaek:Aspelmeyer:2016}%
  \BibitemOpen
  \bibfield  {author} {\bibinfo {author} {\bibfnamefont {R.}~\bibnamefont
  {Kaltenbaek}}, \bibinfo {author} {\bibfnamefont {M.}~\bibnamefont
  {Aspelmeyer}}, \bibinfo {author} {\bibfnamefont {P.~F.}\ \bibnamefont
  {Barker}}, \bibinfo {author} {\bibfnamefont {A.}~\bibnamefont {Bassi}},
  \bibinfo {author} {\bibfnamefont {J.}~\bibnamefont {Bateman}}, \bibinfo
  {author} {\bibfnamefont {K.}~\bibnamefont {Bongs}}, \bibinfo {author}
  {\bibfnamefont {S.}~\bibnamefont {Bose}}, \bibinfo {author} {\bibfnamefont
  {C.}~\bibnamefont {Braxmaier}}, \bibinfo {author} {\bibfnamefont
  {v.}~\bibnamefont {Brukner}}, \bibinfo {author} {\bibfnamefont
  {B.}~\bibnamefont {Christophe}}, \bibinfo {author} {\bibfnamefont
  {M.}~\bibnamefont {Chwalla}}, \bibinfo {author} {\bibfnamefont {P.-F.}\
  \bibnamefont {Cohadon}}, \bibinfo {author} {\bibfnamefont {A.~M.}\
  \bibnamefont {Cruise}}, \emph {et~al.},\ }\href
  {https://doi.org/10.1140/epjqt/s40507-016-0043-7} {\bibfield  {journal}
  {\bibinfo  {journal} {EPJ Quantum Technology}\ }\textbf {\bibinfo {volume}
  {3}},\ \bibinfo {pages} {5} (\bibinfo {year} {2016})}\BibitemShut {NoStop}%
\bibitem [{\citenamefont {Amaro-Seoane}\ \emph {et~al.}(2017)\citenamefont
  {Amaro-Seoane}, \citenamefont {Audley}, \citenamefont {Babak}, \citenamefont
  {Baker}, \citenamefont {Barausse}, \citenamefont {Bender}, \citenamefont
  {Berti}, \citenamefont {Binetruy}, \citenamefont {Born}, \citenamefont
  {Bortoluzzi}, \citenamefont {Camp}, \citenamefont {Caprini}, \citenamefont
  {Cardoso} \emph {et~al.}}]{Amaro-Seoane:Audley:2017}%
  \BibitemOpen
  \bibfield  {author} {\bibinfo {author} {\bibfnamefont {P.}~\bibnamefont
  {Amaro-Seoane}}, \bibinfo {author} {\bibfnamefont {H.}~\bibnamefont
  {Audley}}, \bibinfo {author} {\bibfnamefont {S.}~\bibnamefont {Babak}},
  \bibinfo {author} {\bibfnamefont {J.}~\bibnamefont {Baker}}, \bibinfo
  {author} {\bibfnamefont {E.}~\bibnamefont {Barausse}}, \bibinfo {author}
  {\bibfnamefont {P.}~\bibnamefont {Bender}}, \bibinfo {author} {\bibfnamefont
  {E.}~\bibnamefont {Berti}}, \bibinfo {author} {\bibfnamefont
  {P.}~\bibnamefont {Binetruy}}, \bibinfo {author} {\bibfnamefont
  {M.}~\bibnamefont {Born}}, \bibinfo {author} {\bibfnamefont {D.}~\bibnamefont
  {Bortoluzzi}}, \bibinfo {author} {\bibfnamefont {J.}~\bibnamefont {Camp}},
  \bibinfo {author} {\bibfnamefont {C.}~\bibnamefont {Caprini}}, \bibinfo
  {author} {\bibfnamefont {V.}~\bibnamefont {Cardoso}}, \emph {et~al.},\ }\href
  {https://doi.org/10.48550/ARXIV.1702.00786} {\bibinfo {title} {Laser
  interferometer space antenna}} (\bibinfo {year} {2017})\BibitemShut {NoStop}%
\bibitem [{\citenamefont {El-Neaj}\ \emph {et~al.}(2020)\citenamefont
  {El-Neaj}, \citenamefont {Alpigiani}, \citenamefont {Amairi-Pyka},
  \citenamefont {Araújo}, \citenamefont {Balaž}, \citenamefont {Bassi},
  \citenamefont {Bathe-Peters}, \citenamefont {Battelier}, \citenamefont
  {Belić}, \citenamefont {Bentine}, \citenamefont {Bernabeu}, \citenamefont
  {Bertoldi}, \citenamefont {Bingham}, \citenamefont {Blas}, \citenamefont
  {Bolpasi} \emph {et~al.}}]{El-Neaj:Alpigiani:2020}%
  \BibitemOpen
  \bibfield  {author} {\bibinfo {author} {\bibfnamefont {Y.~A.}\ \bibnamefont
  {El-Neaj}}, \bibinfo {author} {\bibfnamefont {C.}~\bibnamefont {Alpigiani}},
  \bibinfo {author} {\bibfnamefont {S.}~\bibnamefont {Amairi-Pyka}}, \bibinfo
  {author} {\bibfnamefont {H.}~\bibnamefont {Araújo}}, \bibinfo {author}
  {\bibfnamefont {A.}~\bibnamefont {Balaž}}, \bibinfo {author} {\bibfnamefont
  {A.}~\bibnamefont {Bassi}}, \bibinfo {author} {\bibfnamefont
  {L.}~\bibnamefont {Bathe-Peters}}, \bibinfo {author} {\bibfnamefont
  {B.}~\bibnamefont {Battelier}}, \bibinfo {author} {\bibfnamefont
  {A.}~\bibnamefont {Belić}}, \bibinfo {author} {\bibfnamefont
  {E.}~\bibnamefont {Bentine}}, \bibinfo {author} {\bibfnamefont
  {J.}~\bibnamefont {Bernabeu}}, \bibinfo {author} {\bibfnamefont
  {A.}~\bibnamefont {Bertoldi}}, \bibinfo {author} {\bibfnamefont
  {R.}~\bibnamefont {Bingham}}, \bibinfo {author} {\bibfnamefont
  {D.}~\bibnamefont {Blas}}, \bibinfo {author} {\bibfnamefont {V.}~\bibnamefont
  {Bolpasi}}, \emph {et~al.},\ }\href
  {https://doi.org/10.1140/epjqt/s40507-020-0080-0} {\bibfield  {journal}
  {\bibinfo  {journal} {EPJ Quantum Technology}\ }\textbf {\bibinfo {volume}
  {7}},\ \bibinfo {pages} {6} (\bibinfo {year} {2020})}\BibitemShut {NoStop}%
\bibitem [{\citenamefont {Scholz}\ \emph {et~al.}(2009)\citenamefont {Scholz},
  \citenamefont {Koch}, \citenamefont {Ullmann},\ and\ \citenamefont
  {Benson}}]{Scholz:Koch:2009}%
  \BibitemOpen
  \bibfield  {author} {\bibinfo {author} {\bibfnamefont {M.}~\bibnamefont
  {Scholz}}, \bibinfo {author} {\bibfnamefont {L.}~\bibnamefont {Koch}},
  \bibinfo {author} {\bibfnamefont {R.}~\bibnamefont {Ullmann}},\ and\ \bibinfo
  {author} {\bibfnamefont {O.}~\bibnamefont {Benson}},\ }\href
  {https://doi.org/10.1063/1.3139768} {\bibfield  {journal} {\bibinfo
  {journal} {Applied Physics Letters}\ }\textbf {\bibinfo {volume} {94}},\
  \bibinfo {pages} {201105} (\bibinfo {year} {2009})},\ \Eprint
  {https://arxiv.org/abs/https://doi.org/10.1063/1.3139768}
  {https://doi.org/10.1063/1.3139768} \BibitemShut {NoStop}%
\bibitem [{\citenamefont {Wahl}\ \emph {et~al.}(2013)\citenamefont {Wahl},
  \citenamefont {Röhlicke}, \citenamefont {Rahn}, \citenamefont {Erdmann},
  \citenamefont {Kell}, \citenamefont {Ahlrichs}, \citenamefont {Kernbach},
  \citenamefont {Schell},\ and\ \citenamefont {Benson}}]{Wahl:Roerhlicke:2013}%
  \BibitemOpen
  \bibfield  {author} {\bibinfo {author} {\bibfnamefont {M.}~\bibnamefont
  {Wahl}}, \bibinfo {author} {\bibfnamefont {T.}~\bibnamefont {Röhlicke}},
  \bibinfo {author} {\bibfnamefont {H.-J.}\ \bibnamefont {Rahn}}, \bibinfo
  {author} {\bibfnamefont {R.}~\bibnamefont {Erdmann}}, \bibinfo {author}
  {\bibfnamefont {G.}~\bibnamefont {Kell}}, \bibinfo {author} {\bibfnamefont
  {A.}~\bibnamefont {Ahlrichs}}, \bibinfo {author} {\bibfnamefont
  {M.}~\bibnamefont {Kernbach}}, \bibinfo {author} {\bibfnamefont {A.~W.}\
  \bibnamefont {Schell}},\ and\ \bibinfo {author} {\bibfnamefont
  {O.}~\bibnamefont {Benson}},\ }\href {https://doi.org/10.1063/1.4795828}
  {\bibfield  {journal} {\bibinfo  {journal} {Review of Scientific
  Instruments}\ }\textbf {\bibinfo {volume} {84}},\ \bibinfo {pages} {043102}
  (\bibinfo {year} {2013})},\ \Eprint
  {https://arxiv.org/abs/https://doi.org/10.1063/1.4795828}
  {https://doi.org/10.1063/1.4795828} \BibitemShut {NoStop}%
\bibitem [{\citenamefont {Altschul}\ \emph {et~al.}(2015)\citenamefont
  {Altschul}, \citenamefont {Bailey}, \citenamefont {Blanchet}, \citenamefont
  {Bongs}, \citenamefont {Bouyer}, \citenamefont {Cacciapuoti}, \citenamefont
  {Capozziello}, \citenamefont {Gaaloul}, \citenamefont {Giulini},
  \citenamefont {Hartwig}, \citenamefont {Iess}, \citenamefont {Jetzer},
  \citenamefont {Landragin} \emph {et~al.}}]{Altschul:Bailey:2015}%
  \BibitemOpen
  \bibfield  {author} {\bibinfo {author} {\bibfnamefont {B.}~\bibnamefont
  {Altschul}}, \bibinfo {author} {\bibfnamefont {Q.~G.}\ \bibnamefont
  {Bailey}}, \bibinfo {author} {\bibfnamefont {L.}~\bibnamefont {Blanchet}},
  \bibinfo {author} {\bibfnamefont {K.}~\bibnamefont {Bongs}}, \bibinfo
  {author} {\bibfnamefont {P.}~\bibnamefont {Bouyer}}, \bibinfo {author}
  {\bibfnamefont {L.}~\bibnamefont {Cacciapuoti}}, \bibinfo {author}
  {\bibfnamefont {S.}~\bibnamefont {Capozziello}}, \bibinfo {author}
  {\bibfnamefont {N.}~\bibnamefont {Gaaloul}}, \bibinfo {author} {\bibfnamefont
  {D.}~\bibnamefont {Giulini}}, \bibinfo {author} {\bibfnamefont
  {J.}~\bibnamefont {Hartwig}}, \bibinfo {author} {\bibfnamefont
  {L.}~\bibnamefont {Iess}}, \bibinfo {author} {\bibfnamefont {P.}~\bibnamefont
  {Jetzer}}, \bibinfo {author} {\bibfnamefont {A.}~\bibnamefont {Landragin}},
  \emph {et~al.},\ }\href {https://doi.org/10.1016/j.asr.2014.07.014}
  {\bibfield  {journal} {\bibinfo  {journal} {Advances in Space Research}\
  }\textbf {\bibinfo {volume} {55}},\ \bibinfo {pages} {501} (\bibinfo {year}
  {2015})}\BibitemShut {NoStop}%
\bibitem [{\citenamefont {Tino}\ \emph {et~al.}(2020)\citenamefont {Tino},
  \citenamefont {Cacciapuoti}, \citenamefont {Capozziello}, \citenamefont
  {Lambiase},\ and\ \citenamefont {Sorrentino}}]{Tino:Cacciapuoti:2020}%
  \BibitemOpen
  \bibfield  {author} {\bibinfo {author} {\bibfnamefont {G.}~\bibnamefont
  {Tino}}, \bibinfo {author} {\bibfnamefont {L.}~\bibnamefont {Cacciapuoti}},
  \bibinfo {author} {\bibfnamefont {S.}~\bibnamefont {Capozziello}}, \bibinfo
  {author} {\bibfnamefont {G.}~\bibnamefont {Lambiase}},\ and\ \bibinfo
  {author} {\bibfnamefont {F.}~\bibnamefont {Sorrentino}},\ }\href
  {https://doi.org/10.1016/j.ppnp.2020.103772} {\bibfield  {journal} {\bibinfo
  {journal} {Progress in Particle and Nuclear Physics}\ }\textbf {\bibinfo
  {volume} {112}},\ \bibinfo {pages} {103772} (\bibinfo {year}
  {2020})}\BibitemShut {NoStop}%
\bibitem [{\citenamefont {Bassi}\ \emph {et~al.}(2022)\citenamefont {Bassi},
  \citenamefont {Cacciapuoti}, \citenamefont {Capozziello}, \citenamefont
  {Dell'Agnello}, \citenamefont {Diamanti}, \citenamefont {Giulini},
  \citenamefont {Iess}, \citenamefont {Jetzer}, \citenamefont {Joshi},
  \citenamefont {Landragin}, \citenamefont {Poncin-Lafitte}, \citenamefont
  {Rasel}, \citenamefont {Roura}, \citenamefont {Salomon},\ and\ \citenamefont
  {Ulbricht}}]{Bassi:Cacciapuoti:2022}%
  \BibitemOpen
  \bibfield  {author} {\bibinfo {author} {\bibfnamefont {A.}~\bibnamefont
  {Bassi}}, \bibinfo {author} {\bibfnamefont {L.}~\bibnamefont {Cacciapuoti}},
  \bibinfo {author} {\bibfnamefont {S.}~\bibnamefont {Capozziello}}, \bibinfo
  {author} {\bibfnamefont {S.}~\bibnamefont {Dell'Agnello}}, \bibinfo {author}
  {\bibfnamefont {E.}~\bibnamefont {Diamanti}}, \bibinfo {author}
  {\bibfnamefont {D.}~\bibnamefont {Giulini}}, \bibinfo {author} {\bibfnamefont
  {L.}~\bibnamefont {Iess}}, \bibinfo {author} {\bibfnamefont {P.}~\bibnamefont
  {Jetzer}}, \bibinfo {author} {\bibfnamefont {S.~K.}\ \bibnamefont {Joshi}},
  \bibinfo {author} {\bibfnamefont {A.}~\bibnamefont {Landragin}}, \bibinfo
  {author} {\bibfnamefont {C.~L.}\ \bibnamefont {Poncin-Lafitte}}, \bibinfo
  {author} {\bibfnamefont {E.}~\bibnamefont {Rasel}}, \bibinfo {author}
  {\bibfnamefont {A.}~\bibnamefont {Roura}}, \bibinfo {author} {\bibfnamefont
  {C.}~\bibnamefont {Salomon}},\ and\ \bibinfo {author} {\bibfnamefont
  {H.}~\bibnamefont {Ulbricht}},\ }\href
  {https://doi.org/10.1038/s41526-022-00229-0} {\bibfield  {journal} {\bibinfo
  {journal} {npj Microgravity}\ }\textbf {\bibinfo {volume} {8}},\ \bibinfo
  {pages} {49} (\bibinfo {year} {2022})}\BibitemShut {NoStop}%
\bibitem [{\citenamefont {Zych}\ and\ \citenamefont
  {Brukner}(2018)}]{Zych:Brukner:2015}%
  \BibitemOpen
  \bibfield  {author} {\bibinfo {author} {\bibfnamefont {M.}~\bibnamefont
  {Zych}}\ and\ \bibinfo {author} {\bibfnamefont {{\v C}.}~\bibnamefont
  {Brukner}},\ }\href {https://doi.org/10.1038/s41567-018-0197-6} {\bibfield
  {journal} {\bibinfo  {journal} {Nature Physics}\ }\textbf {\bibinfo {volume}
  {14}},\ \bibinfo {pages} {1027} (\bibinfo {year} {2018})}\BibitemShut
  {NoStop}%
\bibitem [{\citenamefont {Gasbarri}\ \emph {et~al.}(2021)\citenamefont
  {Gasbarri}, \citenamefont {Belenchia}, \citenamefont {Carlesso},
  \citenamefont {Donadi}, \citenamefont {Bassi}, \citenamefont {Kaltenbaek},
  \citenamefont {Paternostro},\ and\ \citenamefont
  {Ulbricht}}]{Gasbarri:Belenchia:2021}%
  \BibitemOpen
  \bibfield  {author} {\bibinfo {author} {\bibfnamefont {G.}~\bibnamefont
  {Gasbarri}}, \bibinfo {author} {\bibfnamefont {A.}~\bibnamefont {Belenchia}},
  \bibinfo {author} {\bibfnamefont {M.}~\bibnamefont {Carlesso}}, \bibinfo
  {author} {\bibfnamefont {S.}~\bibnamefont {Donadi}}, \bibinfo {author}
  {\bibfnamefont {A.}~\bibnamefont {Bassi}}, \bibinfo {author} {\bibfnamefont
  {R.}~\bibnamefont {Kaltenbaek}}, \bibinfo {author} {\bibfnamefont
  {M.}~\bibnamefont {Paternostro}},\ and\ \bibinfo {author} {\bibfnamefont
  {H.}~\bibnamefont {Ulbricht}},\ }\href
  {https://doi.org/10.1038/s42005-021-00656-7} {\bibfield  {journal} {\bibinfo
  {journal} {Communications Physics}\ }\textbf {\bibinfo {volume} {4}},\
  \bibinfo {pages} {155} (\bibinfo {year} {2021})}\BibitemShut {NoStop}%
\bibitem [{\citenamefont {Belenchia}\ \emph {et~al.}(2022)\citenamefont
  {Belenchia}, \citenamefont {Carlesso}, \citenamefont {Ömer Bayraktar},
  \citenamefont {Dequal}, \citenamefont {Derkach}, \citenamefont {Gasbarri},
  \citenamefont {Herr}, \citenamefont {Li}, \citenamefont {Rademacher},
  \citenamefont {Sidhu}, \citenamefont {Oi}, \citenamefont {Seidel},
  \citenamefont {Kaltenbaek} \emph {et~al.}}]{Belenchia:Carlesso:2022}%
  \BibitemOpen
  \bibfield  {author} {\bibinfo {author} {\bibfnamefont {A.}~\bibnamefont
  {Belenchia}}, \bibinfo {author} {\bibfnamefont {M.}~\bibnamefont {Carlesso}},
  \bibinfo {author} {\bibnamefont {Ömer Bayraktar}}, \bibinfo {author}
  {\bibfnamefont {D.}~\bibnamefont {Dequal}}, \bibinfo {author} {\bibfnamefont
  {I.}~\bibnamefont {Derkach}}, \bibinfo {author} {\bibfnamefont
  {G.}~\bibnamefont {Gasbarri}}, \bibinfo {author} {\bibfnamefont
  {W.}~\bibnamefont {Herr}}, \bibinfo {author} {\bibfnamefont {Y.~L.}\
  \bibnamefont {Li}}, \bibinfo {author} {\bibfnamefont {M.}~\bibnamefont
  {Rademacher}}, \bibinfo {author} {\bibfnamefont {J.}~\bibnamefont {Sidhu}},
  \bibinfo {author} {\bibfnamefont {D.~K.}\ \bibnamefont {Oi}}, \bibinfo
  {author} {\bibfnamefont {S.~T.}\ \bibnamefont {Seidel}}, \bibinfo {author}
  {\bibfnamefont {R.}~\bibnamefont {Kaltenbaek}}, \emph {et~al.},\ }\href
  {https://doi.org/10.1016/j.physrep.2021.11.004} {\bibfield  {journal}
  {\bibinfo  {journal} {Physics Reports}\ }\textbf {\bibinfo {volume} {951}},\
  \bibinfo {pages} {1} (\bibinfo {year} {2022})}\BibitemShut {NoStop}%
\bibitem [{\citenamefont {Yamaguchi}\ \emph {et~al.}(2012)\citenamefont
  {Yamaguchi}, \citenamefont {Okamoto},\ and\ \citenamefont
  {Mahboob}}]{Yamaguchi:Hajime:2012}%
  \BibitemOpen
  \bibfield  {author} {\bibinfo {author} {\bibfnamefont {H.}~\bibnamefont
  {Yamaguchi}}, \bibinfo {author} {\bibfnamefont {H.}~\bibnamefont {Okamoto}},\
  and\ \bibinfo {author} {\bibfnamefont {I.}~\bibnamefont {Mahboob}},\ }\href
  {https://doi.org/10.1143/APEX.5.014001} {\bibfield  {journal} {\bibinfo
  {journal} {Applied Physics Express}\ }\textbf {\bibinfo {volume} {5}},\
  \bibinfo {pages} {014001} (\bibinfo {year} {2012})}\BibitemShut {NoStop}%
\bibitem [{\citenamefont {Tam}\ and\ \citenamefont {Hu}(1989)}]{Tam:Hu:1989}%
  \BibitemOpen
  \bibfield  {author} {\bibinfo {author} {\bibfnamefont {C.~K.~W.}\
  \bibnamefont {Tam}}\ and\ \bibinfo {author} {\bibfnamefont {F.~Q.}\
  \bibnamefont {Hu}},\ }\href {https://doi.org/10.1017/S0022112089001370}
  {\bibfield  {journal} {\bibinfo  {journal} {Journal of Fluid Mechanics}\
  }\textbf {\bibinfo {volume} {203}},\ \bibinfo {pages} {51–76} (\bibinfo
  {year} {1989})}\BibitemShut {NoStop}%
\bibitem [{\citenamefont {Lou}\ \emph {et~al.}(2012)\citenamefont {Lou},
  \citenamefont {Dai},\ and\ \citenamefont {Wosinski}}]{Lou:Daoxin:2012}%
  \BibitemOpen
  \bibfield  {author} {\bibinfo {author} {\bibfnamefont {F.}~\bibnamefont
  {Lou}}, \bibinfo {author} {\bibfnamefont {D.}~\bibnamefont {Dai}},\ and\
  \bibinfo {author} {\bibfnamefont {L.}~\bibnamefont {Wosinski}},\ }\href
  {https://doi.org/10.1364/OL.37.003372} {\bibfield  {journal} {\bibinfo
  {journal} {Opt. Lett.}\ }\textbf {\bibinfo {volume} {37}},\ \bibinfo {pages}
  {3372} (\bibinfo {year} {2012})}\BibitemShut {NoStop}%
\bibitem [{\citenamefont {Bellini}\ \emph {et~al.}(2014)\citenamefont
  {Bellini}, \citenamefont {Ludhova}, \citenamefont {Ranucci},\ and\
  \citenamefont {Villante}}]{Bellini:Ludhova:2014}%
  \BibitemOpen
  \bibfield  {author} {\bibinfo {author} {\bibfnamefont {G.}~\bibnamefont
  {Bellini}}, \bibinfo {author} {\bibfnamefont {L.}~\bibnamefont {Ludhova}},
  \bibinfo {author} {\bibfnamefont {G.}~\bibnamefont {Ranucci}},\ and\ \bibinfo
  {author} {\bibfnamefont {F.~L.}\ \bibnamefont {Villante}},\ }\href
  {https://doi.org/10.1155/2014/191960} {\bibfield  {journal} {\bibinfo
  {journal} {Advances in High Energy Physics}\ }\textbf {\bibinfo {volume}
  {2014}},\ \bibinfo {pages} {1} (\bibinfo {year} {2014})},\ \bibinfo {note}
  {publisher: Hindawi Limited}\BibitemShut {NoStop}%
\bibitem [{\citenamefont {Salas}\ \emph {et~al.}(2021)\citenamefont {Salas},
  \citenamefont {Forero}, \citenamefont {Gariazzo}, \citenamefont
  {Mart{\'\i}nez-Mirav{\'e}}, \citenamefont {Mena}, \citenamefont {Ternes},
  \citenamefont {T{\'o}rtola},\ and\ \citenamefont
  {Valle}}]{Salas:Forero:2020}%
  \BibitemOpen
  \bibfield  {author} {\bibinfo {author} {\bibfnamefont {P.~F.~d.}\
  \bibnamefont {Salas}}, \bibinfo {author} {\bibfnamefont {D.~V.}\ \bibnamefont
  {Forero}}, \bibinfo {author} {\bibfnamefont {S.}~\bibnamefont {Gariazzo}},
  \bibinfo {author} {\bibfnamefont {P.}~\bibnamefont
  {Mart{\'\i}nez-Mirav{\'e}}}, \bibinfo {author} {\bibfnamefont
  {O.}~\bibnamefont {Mena}}, \bibinfo {author} {\bibfnamefont {C.~A.}\
  \bibnamefont {Ternes}}, \bibinfo {author} {\bibfnamefont {M.}~\bibnamefont
  {T{\'o}rtola}},\ and\ \bibinfo {author} {\bibfnamefont {J.~W.~F.}\
  \bibnamefont {Valle}},\ }\bibfield  {journal} {\bibinfo  {journal} {Journal
  of High Energy Physics}\ }\textbf {\bibinfo {volume} {2021}},\ \href
  {https://doi.org/10.1007/jhep02(2021)071} {10.1007/jhep02(2021)071} (\bibinfo
  {year} {2021}),\ \bibinfo {note} {publisher: Springer Science and Business
  Media LLC}\BibitemShut {NoStop}%
\bibitem [{\citenamefont {Do}\ \emph {et~al.}(2019)\citenamefont {Do},
  \citenamefont {Hees}, \citenamefont {Ghez}, \citenamefont {Martinez},
  \citenamefont {Chu}, \citenamefont {Jia}, \citenamefont {Sakai},
  \citenamefont {Lu}, \citenamefont {Gautam}, \citenamefont {O’Neil},
  \citenamefont {Becklin}, \citenamefont {Morris}, \citenamefont {Matthews}
  \emph {et~al.}}]{Do:Hees:2019}%
  \BibitemOpen
  \bibfield  {author} {\bibinfo {author} {\bibfnamefont {T.}~\bibnamefont
  {Do}}, \bibinfo {author} {\bibfnamefont {A.}~\bibnamefont {Hees}}, \bibinfo
  {author} {\bibfnamefont {A.}~\bibnamefont {Ghez}}, \bibinfo {author}
  {\bibfnamefont {G.~D.}\ \bibnamefont {Martinez}}, \bibinfo {author}
  {\bibfnamefont {D.~S.}\ \bibnamefont {Chu}}, \bibinfo {author} {\bibfnamefont
  {S.}~\bibnamefont {Jia}}, \bibinfo {author} {\bibfnamefont {S.}~\bibnamefont
  {Sakai}}, \bibinfo {author} {\bibfnamefont {J.~R.}\ \bibnamefont {Lu}},
  \bibinfo {author} {\bibfnamefont {A.~K.}\ \bibnamefont {Gautam}}, \bibinfo
  {author} {\bibfnamefont {K.~K.}\ \bibnamefont {O’Neil}}, \bibinfo {author}
  {\bibfnamefont {E.~E.}\ \bibnamefont {Becklin}}, \bibinfo {author}
  {\bibfnamefont {M.~R.}\ \bibnamefont {Morris}}, \bibinfo {author}
  {\bibfnamefont {K.}~\bibnamefont {Matthews}}, \emph {et~al.},\ }\href
  {https://doi.org/10.1126/science.aav8137} {\bibfield  {journal} {\bibinfo
  {journal} {Science}\ }\textbf {\bibinfo {volume} {365}},\ \bibinfo {pages}
  {664} (\bibinfo {year} {2019})},\ \Eprint
  {https://arxiv.org/abs/https://www.science.org/doi/pdf/10.1126/science.aav8137}
  {https://www.science.org/doi/pdf/10.1126/science.aav8137} \BibitemShut
  {NoStop}%
\bibitem [{\citenamefont {Gutiérrez}\ and\ \citenamefont
  {Ramos-Chernenko}(2022)}]{Gutierrez:Ramos-Chernenko:2022}%
  \BibitemOpen
  \bibfield  {author} {\bibinfo {author} {\bibfnamefont {C.~M.}\ \bibnamefont
  {Gutiérrez}}\ and\ \bibinfo {author} {\bibfnamefont {N.}~\bibnamefont
  {Ramos-Chernenko}},\ }\href {https://doi.org/10.3847/1538-4357/ac5a59}
  {\bibfield  {journal} {\bibinfo  {journal} {The Astrophysical Journal}\
  }\textbf {\bibinfo {volume} {929}},\ \bibinfo {pages} {29} (\bibinfo {year}
  {2022})}\BibitemShut {NoStop}%
\bibitem [{\citenamefont {Mosquera~Cuesta}\ and\ \citenamefont
  {Salim}(2004)}]{Mosquera:Salim:2004}%
  \BibitemOpen
  \bibfield  {author} {\bibinfo {author} {\bibfnamefont {H.~J.}\ \bibnamefont
  {Mosquera~Cuesta}}\ and\ \bibinfo {author} {\bibfnamefont {J.~M.}\
  \bibnamefont {Salim}},\ }\href
  {https://doi.org/10.1111/j.1365-2966.2004.08375.x} {\bibfield  {journal}
  {\bibinfo  {journal} {Monthly Notices of the Royal Astronomical Society}\
  }\textbf {\bibinfo {volume} {354}},\ \bibinfo {pages} {L55} (\bibinfo {year}
  {2004})},\ \bibinfo {note} {\_eprint:
  https://academic.oup.com/mnras/article-pdf/354/4/L55/3610635/354-4-L55.pdf}\BibitemShut
  {NoStop}%
\bibitem [{\citenamefont {Degnan}(1993)}]{Degnan:1993}%
  \BibitemOpen
  \bibfield  {author} {\bibinfo {author} {\bibfnamefont {J.~J.}\ \bibnamefont
  {Degnan}},\ }in\ \href {https://doi.org/10.1029/GD025p0133} {\emph {\bibinfo
  {booktitle} {Contributions of {Space} {Geodesy} to {Geodynamics}:
  {Technology}}}},\ \bibinfo {series and number} {Geodynamics {Series}}\
  (\bibinfo {year} {1993})\ pp.\ \bibinfo {pages} {133--162}\BibitemShut
  {NoStop}%
\bibitem [{\citenamefont {Saeed}\ \emph {et~al.}(2020)\citenamefont {Saeed},
  \citenamefont {Elzanaty}, \citenamefont {Almorad}, \citenamefont {Dahrouj},
  \citenamefont {Al-Naffouri},\ and\ \citenamefont
  {Alouini}}]{Saeed:Elzanaty:2020}%
  \BibitemOpen
  \bibfield  {author} {\bibinfo {author} {\bibfnamefont {N.}~\bibnamefont
  {Saeed}}, \bibinfo {author} {\bibfnamefont {A.}~\bibnamefont {Elzanaty}},
  \bibinfo {author} {\bibfnamefont {H.}~\bibnamefont {Almorad}}, \bibinfo
  {author} {\bibfnamefont {H.}~\bibnamefont {Dahrouj}}, \bibinfo {author}
  {\bibfnamefont {T.~Y.}\ \bibnamefont {Al-Naffouri}},\ and\ \bibinfo {author}
  {\bibfnamefont {M.-S.}\ \bibnamefont {Alouini}},\ }\href
  {https://doi.org/10.1109/COMST.2020.2990499} {\bibfield  {journal} {\bibinfo
  {journal} {IEEE Communications Surveys \& Tutorials}\ }\textbf {\bibinfo
  {volume} {22}},\ \bibinfo {pages} {1839} (\bibinfo {year}
  {2020})}\BibitemShut {NoStop}%
\bibitem [{\citenamefont {Araniti}\ \emph {et~al.}(2022)\citenamefont
  {Araniti}, \citenamefont {Iera}, \citenamefont {Molinaro}, \citenamefont
  {Pizzi},\ and\ \citenamefont {Rinaldi}}]{Araniti:Iera:2022}%
  \BibitemOpen
  \bibfield  {author} {\bibinfo {author} {\bibfnamefont {G.}~\bibnamefont
  {Araniti}}, \bibinfo {author} {\bibfnamefont {A.}~\bibnamefont {Iera}},
  \bibinfo {author} {\bibfnamefont {A.}~\bibnamefont {Molinaro}}, \bibinfo
  {author} {\bibfnamefont {S.}~\bibnamefont {Pizzi}},\ and\ \bibinfo {author}
  {\bibfnamefont {F.}~\bibnamefont {Rinaldi}},\ }\href
  {https://doi.org/10.1109/JIOT.2021.3115160} {\bibfield  {journal} {\bibinfo
  {journal} {IEEE Internet of Things Journal}\ }\textbf {\bibinfo {volume}
  {9}},\ \bibinfo {pages} {14876} (\bibinfo {year} {2022})}\BibitemShut
  {NoStop}%
\bibitem [{\citenamefont {Bloser}\ \emph {et~al.}(2022)\citenamefont {Bloser},
  \citenamefont {Murphy}, \citenamefont {Fiore},\ and\ \citenamefont
  {Perkins}}]{Bloser:Murphy:2022}%
  \BibitemOpen
  \bibfield  {author} {\bibinfo {author} {\bibfnamefont {P.}~\bibnamefont
  {Bloser}}, \bibinfo {author} {\bibfnamefont {D.}~\bibnamefont {Murphy}},
  \bibinfo {author} {\bibfnamefont {F.}~\bibnamefont {Fiore}},\ and\ \bibinfo
  {author} {\bibfnamefont {J.}~\bibnamefont {Perkins}},\ }\bibinfo {title}
  {Cubesats for gamma-ray astronomy},\ in\ \href
  {https://doi.org/10.1007/978-981-16-4544-0_53-1} {\emph {\bibinfo {booktitle}
  {Handbook of X-ray and Gamma-ray Astrophysics}}},\ \bibinfo {editor} {edited
  by\ \bibinfo {editor} {\bibfnamefont {C.}~\bibnamefont {Bambi}}\ and\
  \bibinfo {editor} {\bibfnamefont {A.}~\bibnamefont {Santangelo}}}\ (\bibinfo
  {publisher} {Springer Nature Singapore},\ \bibinfo {address} {Singapore},\
  \bibinfo {year} {2022})\ pp.\ \bibinfo {pages} {1--33}\BibitemShut {NoStop}%
\bibitem [{\citenamefont {Moore}\ \emph {et~al.}(2016)\citenamefont {Moore},
  \citenamefont {Woods}, \citenamefont {Caspi},\ and\ \citenamefont
  {Mason}}]{Moore:Woods:2016}%
  \BibitemOpen
  \bibfield  {author} {\bibinfo {author} {\bibfnamefont {C.~S.}\ \bibnamefont
  {Moore}}, \bibinfo {author} {\bibfnamefont {T.~N.}\ \bibnamefont {Woods}},
  \bibinfo {author} {\bibfnamefont {A.}~\bibnamefont {Caspi}},\ and\ \bibinfo
  {author} {\bibfnamefont {J.~P.}\ \bibnamefont {Mason}},\ }in\ \href
  {https://doi.org/10.1117/12.2231945} {\emph {\bibinfo {booktitle} {Space
  Telescopes and Instrumentation 2016: Ultraviolet to Gamma Ray}}},\ Vol.\
  \bibinfo {volume} {9905},\ \bibinfo {editor} {edited by\ \bibinfo {editor}
  {\bibfnamefont {J.-W.~A.}\ \bibnamefont {den Herder}}, \bibinfo {editor}
  {\bibfnamefont {T.}~\bibnamefont {Takahashi}},\ and\ \bibinfo {editor}
  {\bibfnamefont {M.}~\bibnamefont {Bautz}}},\ \bibinfo {organization}
  {International Society for Optics and Photonics}\ (\bibinfo  {publisher}
  {SPIE},\ \bibinfo {year} {2016})\ p.\ \bibinfo {pages} {990509}\BibitemShut
  {NoStop}%
\bibitem [{\citenamefont {Shkolnik}(2018)}]{Shkolnik:2018}%
  \BibitemOpen
  \bibfield  {author} {\bibinfo {author} {\bibfnamefont {E.~L.}\ \bibnamefont
  {Shkolnik}},\ }\href {https://doi.org/10.1038/s41550-018-0438-8} {\bibfield
  {journal} {\bibinfo  {journal} {Nature Astronomy}\ }\textbf {\bibinfo
  {volume} {2}},\ \bibinfo {pages} {374} (\bibinfo {year} {2018})}\BibitemShut
  {NoStop}%
\bibitem [{\citenamefont {{Bowman, D. M.}}\ \emph {et~al.}(2022)\citenamefont
  {{Bowman, D. M.}}, \citenamefont {{Vandenbussche, B.}}, \citenamefont {{Sana,
  H.}}, \citenamefont {{Tkachenko, A.}}, \citenamefont {{Raskin, G.}},
  \citenamefont {{Delabie, T.}}, \citenamefont {{Vandoren, B.}}, \citenamefont
  {{Royer, P.}}, \citenamefont {{Garcia, S.}}, \citenamefont {{Van Reeth,
  T.}},\ and\ \citenamefont {{the CubeSpec
  Collaboration}}}]{Bowman:Vandenbussche:2022}%
  \BibitemOpen
  \bibfield  {author} {\bibinfo {author} {\bibnamefont {{Bowman, D. M.}}},
  \bibinfo {author} {\bibnamefont {{Vandenbussche, B.}}}, \bibinfo {author}
  {\bibnamefont {{Sana, H.}}}, \bibinfo {author} {\bibnamefont {{Tkachenko,
  A.}}}, \bibinfo {author} {\bibnamefont {{Raskin, G.}}}, \bibinfo {author}
  {\bibnamefont {{Delabie, T.}}}, \bibinfo {author} {\bibnamefont {{Vandoren,
  B.}}}, \bibinfo {author} {\bibnamefont {{Royer, P.}}}, \bibinfo {author}
  {\bibnamefont {{Garcia, S.}}}, \bibinfo {author} {\bibnamefont {{Van Reeth,
  T.}}},\ and\ \bibinfo {author} {\bibnamefont {{the CubeSpec
  Collaboration}}},\ }\href {https://doi.org/10.1051/0004-6361/202142375}
  {\bibfield  {journal} {\bibinfo  {journal} {A\&A}\ }\textbf {\bibinfo
  {volume} {658}},\ \bibinfo {pages} {A96} (\bibinfo {year}
  {2022})}\BibitemShut {NoStop}%
\bibitem [{\citenamefont {Oi}\ \emph {et~al.}(2017)\citenamefont {Oi},
  \citenamefont {Ling}, \citenamefont {Vallone}, \citenamefont {Villoresi},
  \citenamefont {Greenland}, \citenamefont {Kerr}, \citenamefont {Macdonald},
  \citenamefont {Weinfurter}, \citenamefont {Kuiper}, \citenamefont {Charbon},\
  and\ \citenamefont {et~al.}}]{Oi:Ling:2017}%
  \BibitemOpen
  \bibfield  {author} {\bibinfo {author} {\bibfnamefont {D.~K.}\ \bibnamefont
  {Oi}}, \bibinfo {author} {\bibfnamefont {A.}~\bibnamefont {Ling}}, \bibinfo
  {author} {\bibfnamefont {G.}~\bibnamefont {Vallone}}, \bibinfo {author}
  {\bibfnamefont {P.}~\bibnamefont {Villoresi}}, \bibinfo {author}
  {\bibfnamefont {S.}~\bibnamefont {Greenland}}, \bibinfo {author}
  {\bibfnamefont {E.}~\bibnamefont {Kerr}}, \bibinfo {author} {\bibfnamefont
  {M.}~\bibnamefont {Macdonald}}, \bibinfo {author} {\bibfnamefont
  {H.}~\bibnamefont {Weinfurter}}, \bibinfo {author} {\bibfnamefont
  {H.}~\bibnamefont {Kuiper}}, \bibinfo {author} {\bibfnamefont
  {E.}~\bibnamefont {Charbon}},\ and\ \bibinfo {author} {\bibnamefont
  {et~al.}},\ }\href {http://dx.doi.org/10.1140/epjqt/s40507-017-0060-1}
  {\bibfield  {journal} {\bibinfo  {journal} {EPJ Quantum Technology}\ }\textbf
  {\bibinfo {volume} {4}} (\bibinfo {year} {2017})}\BibitemShut {NoStop}%
\bibitem [{\citenamefont {Joshi}\ \emph {et~al.}(2018)\citenamefont {Joshi},
  \citenamefont {Pienaar}, \citenamefont {Ralph}, \citenamefont {Cacciapuoti},
  \citenamefont {McCutcheon}, \citenamefont {Rarity}, \citenamefont
  {Giggenbach}, \citenamefont {Lim}, \citenamefont {Makarov}, \citenamefont
  {Fuentes}, \citenamefont {Scheidl}, \citenamefont {Beckert}, \citenamefont
  {Bourennane}, \citenamefont {Bruschi} \emph {et~al.}}]{Joshi:Pneaar:2018}%
  \BibitemOpen
  \bibfield  {author} {\bibinfo {author} {\bibfnamefont {S.~K.}\ \bibnamefont
  {Joshi}}, \bibinfo {author} {\bibfnamefont {J.}~\bibnamefont {Pienaar}},
  \bibinfo {author} {\bibfnamefont {T.~C.}\ \bibnamefont {Ralph}}, \bibinfo
  {author} {\bibfnamefont {L.}~\bibnamefont {Cacciapuoti}}, \bibinfo {author}
  {\bibfnamefont {W.}~\bibnamefont {McCutcheon}}, \bibinfo {author}
  {\bibfnamefont {J.}~\bibnamefont {Rarity}}, \bibinfo {author} {\bibfnamefont
  {D.}~\bibnamefont {Giggenbach}}, \bibinfo {author} {\bibfnamefont {J.~G.}\
  \bibnamefont {Lim}}, \bibinfo {author} {\bibfnamefont {V.}~\bibnamefont
  {Makarov}}, \bibinfo {author} {\bibfnamefont {I.}~\bibnamefont {Fuentes}},
  \bibinfo {author} {\bibfnamefont {T.}~\bibnamefont {Scheidl}}, \bibinfo
  {author} {\bibfnamefont {E.}~\bibnamefont {Beckert}}, \bibinfo {author}
  {\bibfnamefont {M.}~\bibnamefont {Bourennane}}, \bibinfo {author}
  {\bibfnamefont {D.~E.}\ \bibnamefont {Bruschi}}, \emph {et~al.},\ }\href
  {https://doi.org/10.1088/1367-2630/aac58b} {\bibfield  {journal} {\bibinfo
  {journal} {New Journal of Physics}\ }\textbf {\bibinfo {volume} {20}},\
  \bibinfo {pages} {063016} (\bibinfo {year} {2018})}\BibitemShut {NoStop}%
\bibitem [{\citenamefont {Mazzarella}\ \emph {et~al.}(2020)\citenamefont
  {Mazzarella}, \citenamefont {Lowe}, \citenamefont {Lowndes}, \citenamefont
  {Joshi}, \citenamefont {Greenland}, \citenamefont {McNeil}, \citenamefont
  {Mercury}, \citenamefont {Macdonald}, \citenamefont {Rarity},\ and\
  \citenamefont {Oi}}]{Mazzarella:Lowe:2020}%
  \BibitemOpen
  \bibfield  {author} {\bibinfo {author} {\bibfnamefont {L.}~\bibnamefont
  {Mazzarella}}, \bibinfo {author} {\bibfnamefont {C.}~\bibnamefont {Lowe}},
  \bibinfo {author} {\bibfnamefont {D.}~\bibnamefont {Lowndes}}, \bibinfo
  {author} {\bibfnamefont {S.~K.}\ \bibnamefont {Joshi}}, \bibinfo {author}
  {\bibfnamefont {S.}~\bibnamefont {Greenland}}, \bibinfo {author}
  {\bibfnamefont {D.}~\bibnamefont {McNeil}}, \bibinfo {author} {\bibfnamefont
  {C.}~\bibnamefont {Mercury}}, \bibinfo {author} {\bibfnamefont
  {M.}~\bibnamefont {Macdonald}}, \bibinfo {author} {\bibfnamefont
  {J.}~\bibnamefont {Rarity}},\ and\ \bibinfo {author} {\bibfnamefont
  {D.~K.~L.}\ \bibnamefont {Oi}},\ }\bibfield  {journal} {\bibinfo  {journal}
  {Cryptography}\ }\textbf {\bibinfo {volume} {4}},\ \href
  {https://doi.org/10.3390/cryptography4010007} {10.3390/cryptography4010007}
  (\bibinfo {year} {2020})\BibitemShut {NoStop}%
\end{thebibliography}%

\newpage
\appendix

\end{document}